\documentclass[a4paper,11pt]{book}

\usepackage{amsmath}
\usepackage{amsfonts}
\usepackage{amssymb}
\usepackage{slashed}
\usepackage{a4wide}
\usepackage{graphicx}
\usepackage{slashed}
\usepackage[dvips]{feynmp}
\usepackage{picins}

\usepackage{mathbbol}
\usepackage{wick}
\usepackage{xcolor}

\usepackage{fancyheadings}
\usepackage{overpic}
\usepackage[nottoc]{tocbibind}

\usepackage{fancyheadings}
\usepackage[a4paper,inner=35mm,outer=25mm,nofoot,bmargin=3cm,tmargin=35mm,headheight=1cm,headsep=10mm,footnotesep=10mm]{geometry}

\newcommand{\units}[1]{\ensuremath{\,\mathrm{#1}}}

\newcommand{\bra}[1]{\left\langle #1 \right|}
\newcommand{\ket}[1]{\left| #1 \right\rangle}
\newcommand{\braket}[1]{\left\langle #1 \right\rangle}
\newcommand{\ul}[1]{\underline{#1}}
\newcommand{\mmax}{\text{max}}
\newcommand{\mmin}{\text{min}}
\newcommand{\Tr}{\mathop{\mathrm{Tr}}}
\newcommand{\elll}{l}

\newcommand{\note}[2]{\emph{#1} #2}
\newcommand{\crossout}[1]{\setbox0\hbox{#1} \hbox to \wd0{\rlap{\rule[.6ex]{\wd0}{.5pt}}\box0}} 
\newcommand\topalignbox[1]{\leavevmode\vtop{\vskip-1.5ex\hbox{#1}}} 
\newcommand{\Small}[1]{{\fontsize{9pt}{9pt} \selectfont #1}}

\newcommand{\terminol}[1]{\textit{\textbf{#1}}}

\newlength{\savearraycolsep} 

\begin{document}

\pagestyle{empty}

\begin{titlepage}

\vspace{3cm}

\begin{center}
 \underline{$\left.\hspace{14cm} \right.$} \vspace{1cm}
 
\Huge{\bf{ Hadron Masses: \\ Lattice QCD and \\ Chiral Effective Field Theory}}
\vskip 2cm \large
\end{center}


\def\TUM#1{%
\textcolor{gray}{
\dimen1=#1\dimen1=.1143\dimen1%
\dimen2=#1\dimen2=.419\dimen2%
\dimen3=#1\dimen3=.0857\dimen3%
\dimen4=\dimen1\advance\dimen4 by\dimen2%
\setbox0=\vbox{\hrule width\dimen3 height\dimen1 depth0pt\vskip\dimen2}%
\setbox1=\vbox{\hrule width\dimen1 height\dimen4 depth0pt}%
\setbox2=\vbox{\hrule width\dimen3 height\dimen1 depth0pt}%
\setbox3=\hbox{\copy0\copy1\copy0\copy1\box2\copy1\copy0\copy1\box0\box1}%
\leavevmode\vbox{\box3}}}
\def\oTUM#1{%
\dimen1=#1\dimen1=.1143\dimen1%
\dimen2=#1\dimen2=.419\dimen2%
\dimen3=#1\dimen3=.0857\dimen3%
\dimen0=#1\dimen0=.018\dimen0%
\dimen4=\dimen1\advance\dimen4 by-\dimen0%
\setbox1=\vbox{\hrule width\dimen0 height\dimen4 depth0pt}%
\advance\dimen4 by\dimen2%
\setbox8=\vbox{\hrule width\dimen0 height\dimen4 depth0pt}%
\advance\dimen4 by-\dimen2\advance\dimen4 by-\dimen0%
\setbox4=\vbox{\hrule width\dimen4 height\dimen0 depth0pt}%
\advance\dimen4 by\dimen1\advance\dimen4 by\dimen3%
\setbox6=\vbox{\hrule width\dimen4 height\dimen0 depth0pt}%
\advance\dimen4 by\dimen3\advance\dimen4 by\dimen0%
\setbox9=\vbox{\hrule width\dimen4 height\dimen0 depth0pt}%
\advance\dimen4 by\dimen1%
\setbox7=\vbox{\hrule width\dimen4 height\dimen0 depth0pt}%
\dimen4=\dimen3%
\setbox5=\vbox{\hrule width\dimen4 height\dimen0 depth0pt}%
\advance\dimen4 by-\dimen0%
\setbox2=\vbox{\hrule width\dimen4 height\dimen0 depth0pt}%
\dimen4=\dimen2\advance\dimen4 by\dimen0%
\setbox3=\vbox{\hrule width\dimen0 height\dimen4 depth0pt}%
\setbox0=\vbox{\hbox{\box9\lower\dimen2\copy3\lower\dimen2\copy5%
\lower\dimen2\copy3\box7}\kern-\dimen2\nointerlineskip%
\hbox{\raise\dimen2\box1\raise\dimen2\box2\copy3\copy4\copy3%
\raise\dimen2\copy5\copy3\box6\copy3\raise\dimen2\copy5\copy3\copy4\copy3%
\raise\dimen2\box5\box3\box4\box8}}%
\leavevmode\box0}


\begin{center}
\vskip 1cm {\Large Diploma Thesis by} 
\\ \vskip 0.3cm {\Large {\sc Bernhard Musch}} \\
\vskip 1.5cm { {\Large December 2005}}
\vskip 1 cm
\vskip 2.5 cm
{{\Large Technische Universit\"at M\"unchen} \\ \vspace{0.2cm} {\Large Physik-Department} \\ \vspace{0.2cm}
{\Large T39 (Prof.~Dr.~Wolfram Weise)}}
\end{center}

\vspace{2.0cm}

\begin{center}
  \underline{$\left.\hspace{14cm} \right.$}\\
  \vspace{0.5cm}
  \oTUM{2.3cm} \hspace{9.0cm} \raisebox{-0.6cm}{\includegraphics[width=2cm]{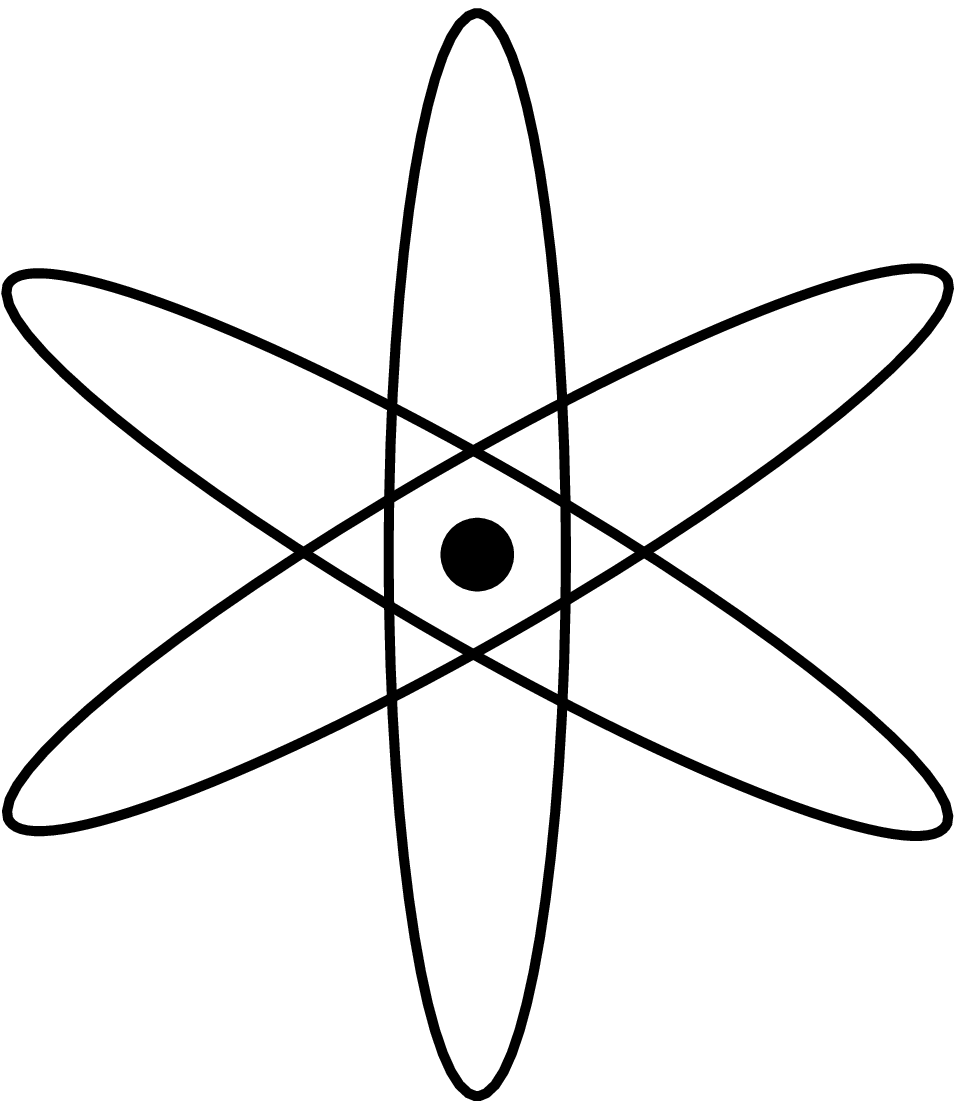}}
\end{center}

\end{titlepage}


$\phantom{+}$\\
\clearpage
\newpage

\pagestyle{fancyplain}

\renewcommand{\chaptermark}[1]{\markboth{#1}{}}
\renewcommand{\sectionmark}[1]{\markright{\thesection\ #1}{}}
\lhead[\fancyplain{}{\thepage}]
	{\fancyplain{}{\rightmark}}
\rhead[\fancyplain{}{\leftmark}]
	{\fancyplain{}{\thepage}}
\cfoot{}

\pagenumbering{arabic}
\setcounter{page}{1}

\tableofcontents
\newpage

\chapter{Introduction}


\terminol{Quantum Chromodynamics} (\terminol{QCD}) describes the interactions of \terminol{quarks} and \nobreak\terminol{gluons}, particles which carry a charge called \terminol{color}. For small momentum transfer, quantum fluctuations of interactions between color charged particles become so important that they cannot be treated any more as small corrections. In fact, the interactions become so strong that on larger scales ($\gtrsim 1\units{fm}$), quarks and gluons are never observed individually, but are confined in \terminol{hadrons}, composite particles which have no net color. \terminol{Nucleons} (neutrons and protons), the constituents of atomic nuclei, belong to the class of hadrons. Analyzing their inner structure in terms of QCD with the usual method of \terminol{perturbation theory} fails due to the strong interactions \cite{TW01}. A different approach is needed. In the long run, it is vital to describe hadrons quantitatively using Standard Model QCD. It helps us to gain accuracy in the interpretation of low- and high-energy experiments, adds to our understanding of nuclear physics and plays a role in the attempt to go beyond the Standard Model. \par

One idea is to set up an \terminol{Effective Field Theory} (\terminol{EFT}), like \terminol{Chiral Perturbation Theory} (\terminol{$\chi$PT}). Instead of quarks and gluons, the theory is formulated in terms of the composite particles (\terminol{effective degrees of freedom}) that are thought to be physically relevant in the regime of interest, e.g., pions and nucleons. A systematic approach guarantees that the symmetry features of QCD are preserved in $\chi$PT. A disadvantage is the infinite number of interaction terms and coupling parameters (\terminol{low energy constants}, or \terminol{LECs}) that enter $\chi$PT. Among other things, they encode the form factors, i.e. the shape and size of the composite particles. In practice, one works with an expansion in small parameters. Then, only a finite number of interaction terms yield significant contributions. The numerical values of the coupling parameters are a priori unknown. Usually they are determined by fitting the parametric expressions of a set of observables to empirical data. Once they are known, predictions for observables inaccessible to experiment are feasible.
However, the necessity to make use of empirical data conflicts with the wish to calculate hadron properties \terminol{ab initio}, i.e., from Standard Model parameters only.

This is the point where numerical \terminol{Lattice Field Theory} calculations become useful. These calculations evaluate the dynamics of quark and gluon fields numerically in a four dimensional cube of space-time. Continuous fields are approximated by the field values specified at lattice points inside the cube. This is a good approximation as long as the cube is larger than the hadrons and the lattice is finer than structural sizes within hadrons. Thus by putting the fields on the lattice, we have a method that permits ab-initio calculations of masses, form factors and other properties of hadrons. Since calculations within Quantum Field Theory require in principle the consideration of all possible field configurations, lattice calculations become numerically very expensive. In order to obtain results on modern computers, the up- and down-quark masses are set to much higher values than they are in physical reality. Simulations at the true quark masses still seem to be a long way off.
Thus one needs some kind of extrapolation to lower quark masses in order to extract \emph{physical} quantities.

The obvious means for extrapolations of lattice QCD results is chiral perturbation theory, which offers quark mass dependent formulas for observables in the low energy domain.

The prime examples of quantities that can be calculated using lattice QCD are the pion mass and the nucleon mass.\footnote{At this stage, one can neglect the comparatively small difference between neutron and proton mass due to \terminol{isospin breaking}.}
Experiments yield for the pion mass $m_\pi^\text{phys}$ and the nucleon mass $m_N^\text{phys}$ realized in nature
\begin{equation*}
	m_\pi^\text{phys} \approx 0.14\units{GeV}\ , \qquad m_N^\text{phys} \approx 0.94\units{GeV} 
	\end{equation*}
With highly accurate experimental results available, why does one engage in complicated calculations for these two observables? Firstly, they can show us how well lattice extra\-polations really work and what obstacles to look out for. Secondly, it is possible to extract information from the quark mass dependence that is difficult or impossible to measure in experiments. Last but not least, it would be a brilliant scientific feat to demonstrate that the nucleon mass really follows from the Standard Model.

Lattice data with quark masses low enough to attempt extrapolation using $\chi$PT have become available only recently and are still scarce.
Promising results have been published in reference \cite{PHW04}, showing a successful fit to nucleon mass data from the lattice.
Not all information entering this fit comes from lattice calculations. Some of the LECs have been fixed to values obtained from empirical data. Additionally, the authors included the nucleon mass at the physical point as input.

My task in this diploma thesis is to discuss the reliability, possible obstacles and the future perspective of chiral extrapolations, with particular focus on the nucleon mass. How large are the statistical uncertainties of the fit in ref. \cite{PHW04}? Do we observe convergence of the perturbative series? Are the assumptions made well founded? Are the extracted LECs compatible with empirical information? Will it be possible any time soon to perform the fit with information purely from the lattice? 
This work cannot give final solutions, but it can present tools and strategies for the future, when more data become available.

The following four chapters are intended to give novices a quick-start introduction to the mechanisms and terminology of the theoretical fields that have been brought together in this work: Chiral Perturbation Theory, Lattice Field Theory and error analysis. Furthermore, chapter 4 describes the algorithms used for the statistical analysis. 

Chapter 6 constitutes the principal part and is devoted to fitting $\chi$PT to two-flavor lattice data for the nucleon mass. It starts out with a discussion of the existing fit of ref. \cite{PHW04}. A simple numerical experiment reveals important insight about the strategy needed to perform ab-initio extrapolations. In an attempt to match our results to empirical pion-nucleon scattering measurements, we stumble over discrepancies, but we resolve the issue by taking effects of the delta resonance into account. Finally, use of a finite volume correction formula extends the selection of lattice data usable for our fit.

The pion mass is a crucial input variable in many chiral extrapolations. Its precise dependence on the quark mass has yet to be clarified. Chapter 7 presents a step in that direction, making use of a simultaneous fit to pion and kaon mass from three-flavor lattice data. The chapter also demonstrates the transfer of three-flavor results to the two-flavor framework.

Finally, chapter 8 gives a brief summary of my conclusions.

\chapter{Basics of Relativistic Baryon Chiral Perturbation Theory}

This chapter introduces the basic concepts of $\chi$PT. 
The focus rests on conceptual understanding and issues relevant for application.

\section{The QCD Lagrangian}
 
\newcommand{\zf}{\mathrm{f}}
\newcommand{\zg}{\mathrm{g}}
\newcommand{\zh}{\mathrm{h}}
\newcommand{\zF}{\mathrm{F}}
\newcommand{\zG}{\mathrm{G}}
\newcommand{\zH}{\mathrm{H}}

\newcommand{\za}{\mathrm{a}}
\newcommand{\zb}{\mathrm{b}}
\newcommand{\zc}{\mathrm{c}}
\newcommand{\zA}{\mathrm{A}}
\newcommand{\zB}{\mathrm{B}}
\newcommand{\zC}{\mathrm{C}}

\newcommand{\ef}{\mathrm{f}}
\newcommand{\ce}{\mathrm{c}}
\newcommand{\Ce}{\mathrm{C}}
\newcommand{\Ef}{\mathrm{F}}

\newcommand{\zi}{\mathrm{i}}
\newcommand{\zj}{\mathrm{j}}
\newcommand{\zk}{\mathrm{k}}

Chiral Perturbation Theory is an Effective Field Theory for the basic gauge invariant QCD Lagrangian
\begin{equation}
	\mathcal{L}_\text{QCD}\ =\ \bar q_{\zf \za}\ i \slashed D_{\za \zb}\ q_{\zf \zb} 
	\ -\ \bar q_{\zf \za}\ m_{\zf \zg}\ q_{\zg \za}
	\ -\ \frac{1}{4} \mathcal G_{\mu \nu\, \zA}\ \mathcal G^{\mu \nu}_{\zA} 
	\label{eq-LagrangeanQCD}
	\end{equation}
with flavor indices $\zf,\zg=\{u,d,s,...\}$ for the $n_\zf$ quark flavors, color indices $\za,\zb = 1..3$, Lorentz indices $\mu, \nu, ...$ and indices $\zA,\zB,\zC=1..8$ numbering the generators of the color group. Einstein's summation convention is implied.
The Dirac spinors $q_{\zf \za}=q_{\zf \za}(x)$ represent the quarks. The QCD Lagrangian is set up in such a way that it is invariant under local SU(3) gauge transformations 
\begin{equation}
	q_{\zf \za}(x)\ \rightarrow\ W_{\za \zb}(x)\  q_{\zf \zb}(x) 
	\label{eq-gaugetrafo}
	\end{equation}
for any (space dependent) unitary unimodular $3 \times 3$ color matrix $W_{\za \zb}(x)$. This is achieved by introducing the covariant derivative 
\begin{equation}
	D_{\mu\, \za \zb}\ q_{\zf \zb} \equiv \partial_\mu q_{\zf \za} - i g \frac{\lambda_{\za\zb,\zA}}{2} \mathcal{A}_{\mu\,\zA}\ q_{\zf \zb}
	\end{equation}
where $\lambda_{\za\zb,\zA}$ for $\zA=1..8$ are the generators of SU(3), the Gell-Mann matrices. $g$ is the strong coupling constant, and the bosonic gauge fields $\mathcal{A}_{\mu\,\zA}=\mathcal{A}_{\mu\,\zA}(x)$ describe the gluons. Their transformation behavior is fixed by the requirement of gauge invariance. The propagation of the gluons is assured by the gauge invariant kinetic gluon term involving the field strength tensors
\begin{equation}
	\mathcal{G}_{\mu \nu\, \zA} \equiv \partial_\mu \mathcal{A}_{\nu\, \zA} - \partial_\nu \mathcal{A}_{\mu\, \zA} + g\ f_{\zA\zB\zC}\; \mathcal{A}_{\mu\, \zB}\; \mathcal{A}_{\nu\, \zC}
	\end{equation}
where $f_{\zA\zB\zC}$ are the structure constants of the color group SU(3). In a perturbative picture, the term involving the structure constants is responsible for interaction vertices of three or four gluon lines. In the low energy domain, these gluonic self-interactions abound, to the point where the perturbative picture becomes inappropriate altogether. \par

In practice, not all quark flavors need to be considered in our Lagrangian (\ref{eq-LagrangeanQCD}). For our applications, energies are far too low to excite states involving the heavy quarks charm, bottom or top. They can therefore be ``\terminol{integrated out}''. In first approximation, they only cause slight shifts in the values of the quark masses $m_{\zf\zg}$ and the coupling $g$. A theory with three flavors ``up'', ``down'' and ``strange'' has $n_\zf=3$ and is therefore typically called an SU(3) theory. For SU(2) theories with $n_\zf=2$, even the strange quark is neglected.\par

The quark mass matrix $m_{\zf \zg}$ is chosen diagonal:
\begin{equation}
	n_\zf=2\ :  \qquad
	m = \begin{pmatrix} m_u & \\ & m_d \end{pmatrix}
\end{equation}
\begin{equation}
	n_\zf=3\ : \qquad 
	m = \begin{pmatrix} m_u & & \\ & m_d & \\ & & m_s \end{pmatrix}
\end{equation}
Since we are not interested in isospin breaking effects, we always set $m_u = m_d \equiv \hat{m}$.

The QCD Lagrangian is a real Lorentz scalar and even under the discrete symmetries C, P and T.

\section{Chiral Symmetry}

The more we know about mutual symmetry transformation properties of the EFT's elementary degrees of freedom, the less terms are allowed in the effective Lagrangian. This motivates further inquiries about the symmetries of the theory. We concentrate on flavor symmetries. For the purpose of clarity, we omit the color indices $\za,\zb,\zc$.

We define left- and right-handed (\terminol{chiral})  
components of the
quark fields 
$q_{\zf}^L=\frac{1}{2}(1-\gamma_5) q_{\zf}$ and
$q_{\zf}^R=\frac{1}{2}(1+\gamma_5) q_{\zf}$ .
In terms of these fields, the Lagrangian (\ref{eq-LagrangeanQCD}) appears as
\setlength{\savearraycolsep}{\arraycolsep} \setlength{\arraycolsep}{5pt}
\renewcommand{\arraystretch}{1.6}
\begin{equation}
	\begin{array}{lccc}
	\mathcal{L}_\text{QCD}\  = &
		\phantom{-} \overline{q_{\zf}^L}\ i \slashed D\ q_{\zf}^L & 
		+\ \overline{q_{\zf}^R}\ i \slashed D\ q_{\zf}^R \\
		& -\ \overline{q_{\zf}^L}\ m_{\zf\zg}\ q_{\zg}^R 
		& -\ \overline{q_{\zf}^R}\ m_{\zf\zg}\ q_{\zg}^L 
		& -\ \frac{1}{4} \mathcal G_{\mu \nu\, \zA}\ \mathcal G^{\mu \nu}_\zA 
		\label{eq-ChiralQCD}
	\end{array}
	\end{equation}
\setlength{\arraycolsep}{\savearraycolsep}
\renewcommand{\arraystretch}{1.0}
\\
So for $m_{\zf \zg}=0$, called the \terminol{chiral limit}, the QCD Lagrangian (\ref{eq-LagrangeanQCD}) is 
invariant under global transformations in flavor space 
\begin{equation}
	q_{\zf}^L\ \rightarrow L_{\zf\zg}\ q_{\zg}^L \ , \qquad
	q_{\zf}^R\ \rightarrow R_{\zf\zg}\ q_{\zg}^R
	\label{eq-qarktrafo}
	\end{equation}
The transformation matrices $L$ and $R$ must be unitary. 
If $L=R=V$, we will speak of a \terminol{vector transformation}, while the case $L^\dagger=R=A$ shall be called an \terminol{axial transformation}.
We observe that the QCD Lagrangian in the chiral limit exhibits a global $\mathrm{U}(n_\zf)_L \otimes \mathrm{U}(n_\zf)_R$ symmetry.
The mass term breaks this symmetry \terminol{explicitly}. 

The group $\mathrm{SU}(n_\zf)$ has $n_\zF \equiv n_\zf^2-1$  generators $\lambda_{\zf\zg,\zF}$ , where $\zF$ is running from $1$ to $n_\zF$. It can be extended to  $\mathrm{U}(n_\zf)$ with an extra $\mathrm{U}(1)$ generator $\lambda_{\zf\zg,0}:=\delta_{\zf\zg}$. Using this notation, the $2 n_\ef^2$ conserved symmetry currents read
\begin{equation}
	{j^L}^{\mu}_{\zF} = \overline{ q^L_{\zf}}\ \gamma^\mu \frac{\lambda_{\zf\zg,\zF}}{2}\ q^L_{\zg} \, \qquad
	{j^R}^{\mu}_{\zF} = \overline{ q^R_{\zf}}\ \gamma^\mu \frac{\lambda_{\ef\zg,\zF}}{2}\ q^R_{\zg} 
	\end{equation}
It is useful to define the following linear combinations \\
\renewcommand{\arraystretch}{1.8}
\begin{tabular}{ll}
	$\displaystyle {j^V}^{\mu}_{\zF} = {j^R}^{\mu}_{\zF} + {j^L}^{\mu}_{\zF} = \bar q_{\zf}\ \gamma^\mu \frac{\lambda_{\zf\zg,\zF}}{2}\ q_{\zg}$ & 
	vector current  \\
	$\displaystyle {j^A}^{\mu}_{\zF} = {j^R}^{\mu}_{\zF} - {j^L}^{\mu}_{\zF} = \bar q_{\zf}\ \gamma^\mu \gamma_5 \frac{\lambda_{\zf\zg,\zF}}{2}\ q_{\zg}$ &
	axial vector current 
	\end{tabular} 
\renewcommand{\arraystretch}{1} \vspace{10pt} \\
The corresponding charges are
\begin{equation}
	Q_{\zF}^V(t) = \int d^3 x\ {j^V}^0_{\zF}(\mathbf{x},t)
	\quad \text{and} \quad
	Q_{\zF}^A(t) = \int d^3 x\ {j^A}^0_{\zF}(\mathbf{x},t)
\end{equation}
\par 
The vector charges generate vector transformations, which form a subgroup called $\mathrm{U}(1)_V\otimes\mathrm{SU}(n_\zf)_V$. Axial charges can be used to generate infinitesimal axial transformations. However, the axial transformations do \emph{not} form a group, because commutators of axial charges do not yield pure axial charges again. Even though, one speaks of an $\mathrm{U}(1)_A\otimes\mathrm{SU}(n_\zf)_A$ symmetry. Thus in total we say that the Lagrangian in the chiral limit exhibits a $\mathrm{U}(1)_V\otimes\mathrm{SU}(n_\zf)_V$ and a $\mathrm{U}(1)_A\otimes\mathrm{SU}(n_\zf)_A$ symmetry.

Only a part of the above symmetries remain valid for our EFT, even in the chiral limit. Firstly, the $\mathrm{U}(1)_A$-symmetry is not preserved by quantization (\terminol{axial anomaly}), such that the QCD Hamilton operator $H^0_\text{QCD}$ for massless quarks is only invariant under $\mathrm{SU}(n_\zf)_V \otimes \mathrm{U}(1)_V$ and $\mathrm{SU}(n_\zf)_A$ transformations.
Secondly, empirical facts strongly suggest that the vacuum state does not share the Hamiltonian's invariance under $\mathrm{SU}(n_\zf)_A$.  This phenomenon is called \terminol{spontaneous symmetry breaking}. \par

\section{Spontaneous Symmetry Breaking}
\label{sec-symmbreak}

Candidates for vacua in the spectrum of eigenstates of $H^0_\text{QCD}$ are states $\ket{\Omega_i}$, for which the momentum eigenvalue zero is discrete. This set of states can be chosen orthonormal. The rest of the Hilbert space can be described by an orthonormal set of states $\ket{N,\vec{p}}$ carrying three-momenta $\vec{p}$. The vacuum may be a linear combination of states belonging to a degenerate subset of the $\ket{\Omega_i}$. \par
According to an argument found in ref. \cite{Wei2}, local operators do not permit transitions among the $\ket{\Omega_i}$ in infinite volume. To see this, consider two Hermitian operators at equal times $A(\vec{x})$ and $B(0)$. Now
\begin{eqnarray}
	\bra{\Omega_j} A(\vec{x}) B(0) \ket{\Omega_i} 
	& = &\sum_k \bra{\Omega_j} A(0) \ket{\Omega_k} \bra{\Omega_k} B(0) \ket{\Omega_i} \nonumber \\
	& & + \int d^3 p \sum_N \bra{\Omega_j} A(0) \ket{N,\vec{p}} \bra{N,\vec{p}} B(0) \ket{\Omega_i} 
		e^{-i \vec{p} \vec{x}} \nonumber \\
	& \xrightarrow[|\vec{x}| \rightarrow \infty]{} &
		\sum_k \bra{\Omega_j} A(0) \ket{\Omega_k} \bra{\Omega_k} B(0) \ket{\Omega_i}
\end{eqnarray}
(where we have assumed that the $\vec{p}$-dependent matrix elements are Lebesque-integrable). Causality requires the equal-time commutator $[A(\vec{x}),B(0)]$ to vanish. Therefore 
\begin{eqnarray}
	0 & = &\bra{\Omega_j} [A(\vec{x}), B(0)] \ket{\Omega_i} \nonumber \\ 
	& \xrightarrow[|\vec{x}| \rightarrow \infty]{} &
		\sum_k \bra{\Omega_j} A(0) \ket{\Omega_k} \bra{\Omega_k} B(0) \ket{\Omega_i}
		- \bra{\Omega_j} B(0) \ket{\Omega_k} \bra{\Omega_k} A(0) \ket{\Omega_i}
	\end{eqnarray}
The right hand side tells us, that with respect to the vacuum states, any two local operators commute. They can therefore all be simultaneously diagonalized, i.e. the $\ket{\Omega_i}$ can be chosen in such a way that
\begin{equation}
	\bra{\Omega_j} A(0) \ket{\Omega_i} \propto \delta_{ji}
	\qquad \text{for all local operators }A(0)
	\label{eq-diagvacua}
	\end{equation}
In particular, a small symmetry breaking perturbation built out of local operators (such as the mass term) will be diagonal with respect to the vacuum states, can break degeneracies and can thus select one of the vacuum states $\ket{\Omega_0}$ as the one realized in nature. So we see that the vacuum is in general not a linear combination of the $\ket{\Omega_i}$ that forms a singlet under the symmetries of the Hamiltonian, but rather a state which is stable with respect to local symmetry breaking perturbations. \par

According to \cite{VW84}, the QCD ground state in the chiral limit \emph{must} be invariant under $\mathrm{SU}(n_\zf)_V \otimes \mathrm{U}(1)_V$ (\terminol{isospin}). However, the possibility remains, that the state(s) of lowest energy are not invariant under $\mathrm{SU}(n_\zf)_A$. Indeed, observations indicate that this is the case.

Operators with non-trivial transformation behavior under unbroken symmetries must have the vacuum expectation value zero. Otherwise their expectation value would change under symmetry transformations - a contradiction to the invariance of the vacuum. Operators invariant under all unbroken symmetries, however, may take non-zero vacuum expectation values.  An example is the scalar quark density operator $\bar{q}_{\zf}(x) q_{\zf}(x)$. It is invariant under the unbroken symmetries $\mathrm{U}(n_\zf)_V$. We switch to matrix-vector notation with respect to the indices $\zf$,$\zg$,... and drop position arguments:
\begin{equation}
	\bar{q} \equiv \left(\;\bar{q}_1(x),...,\bar{q}_{n_\zf}(x)\;\right)\ , \qquad
	     q  \equiv \left(\;      q_1(x),...,      q_{n_\zf}(x)\;\right)^T 
	\end{equation}
Under infinitesimal axial rotation $L^\dagger = R \equiv A = 1 + i \epsilon_\zF \lambda_\zF / 2$, the scalar quark density $\bar{q} q$ develops pseudoscalar components of the form $\bar{q} \gamma^5 \lambda_{\zF} q$  $(\zF=1..n_\zF)$ :
\begin{equation}
	\bar{q} q \rightarrow \overline{q^L} A A q^R + \overline{q^R} A^\dagger A^\dagger q^L 
	= \bar{q} q + i \epsilon_F\, (\overline{q^L} \lambda_F q^R - \overline{q^R} \lambda_F q^L)
	= \bar{q} q + i \epsilon_F\, \overline{q} \gamma^5 \lambda_F q
	\end{equation}
The latter transform as vectors under vector rotations $L=R=V=1+i \epsilon_\zG \lambda_\zG/2$ :
\begin{align}
	\bar{q} \gamma^5 \lambda_{\zF} q & \rightarrow 
	\overline{q^L} V^\dagger \lambda_{\zF} V q^R - \overline{q^R} V^\dagger \lambda_{\zF} V q^L 
	= \bar{q} \gamma^5 \lambda_{\zF} q 
	+ \frac{i}{2} \epsilon_G \ \overline{q} \gamma^5 [ \lambda_{\zF}, \lambda_{\zG} ]  q \nonumber \\
	& = \bar{q} \gamma^5 \lambda_{\zF} q 
	- f_{\zF \zG \zH} \epsilon_G \ \overline{q} \gamma^5 \lambda_{\zH} q
	\end{align}
and must therefore possess vanishing vacuum expectation values. Together with a numerical estimate \footnote{$\overline{\mathrm{MS}}$, scale $2\units{GeV}$, see e.g. \cite{Giu01}}, we write down the \terminol{vacuum alignment conditions} as
\begin{equation}
	\bra{\Omega_0} \bar{q} q \ket{\Omega_0} / n_\zf \approx (270 \units{GeV})^3 \ ,\qquad
	\bra{\Omega_0} \bar{q} \gamma^5 \lambda_\zF q \ket{\Omega_0} = 0 , \quad F=1..(n_\zf^2-1)
	\end{equation}

In a coordinate system with the different di-quark densities as the axes, the vacuum alignment condition can be thought of as an arrow, starting at the origin and pointing to the di-quark density realized in the QCD vacuum. Here the length of the arrow is the value of the \terminol{scalar quark condensate} $\braket{\bar{q} q}$. It would be zero if the symmetry were unbroken, which is why it is called an \terminol{order parameter}. In principle, we could choose a different vacuum alignment, shifting the non-zero vacuum expectation values to other di-quark densities. Then, however, we would also have to modify our definition of vector and axial rotations, because the vacuum must be invariant under vector rotations. (It should be noted that the di-quark condensates are not the only quantities that characterize the vacuum. Spontaneous symmetry breaking could even be realized with $\braket{\bar{q} q}=0$, because other objects, such as four-quark condensates, may independently exhibit nonzero values.)

\parpic[l][l]{
	\includegraphics[width=.4\textwidth]{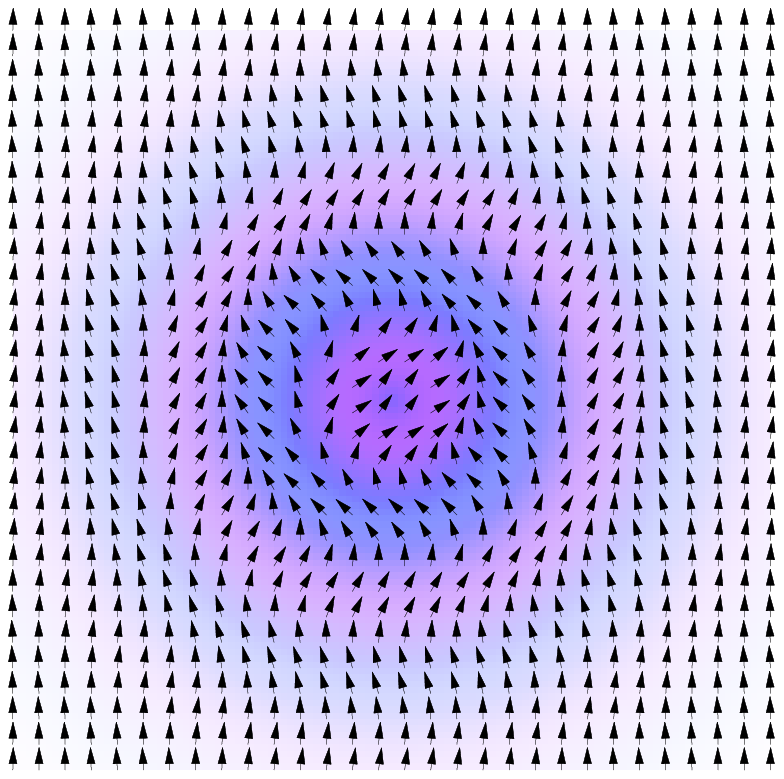} 
	}
Excitations of the vacuum can modify the alignment locally. If we imagine di-quark density arrows of the kind discussed above at every point in space, we would see parallel arrows everywhere except in the region of the excitation, see left inset. A misalignment causes tension, which leads to a wavelike propagation and spreading of the misalignment. The more gradual the misalignment, i.e. the larger the wave length, the less tension energy. In the limit of an infinite wave length, the energy of the misalignment goes to zero - the dispersion relation starts out at the origin \cite{Georgi}. Thus local vacuum misalignments manifest themselves as massless particles, the so called Goldstone Bosons. Their existence in a spontaneously broken field theory can be shown quite generally \cite{Wei2}. They play an important role in our theory, because due to their masslessness, we expect them to dominate the low energy behavior of QCD. In reality, with explicit symmetry breaking interactions present, the 
Goldstone Bosons are identified as the lightest particles in the spectrum of QCD. For $n_\zf=2$, these are the pions $\pi^+$, $\pi^-$, $\pi^0$. For $n_\zf=3$, one includes additionally light excitations involving strangeness: $K^+$, $K^-$, $K^0$, $\bar K^0$ and $\eta$.

\section{The Goldstone Boson Field}
\label{sec-gobofields}

To explore the dynamics of such vacuum misalignments, we must single out local axial rotations from the fields. We write down the quark fields in the form
\begin{equation}
	q^L(x) = \tilde L(x) \begin{pmatrix} 0 \\ \vdots \\ 0 \\ q^L_0(x) \end{pmatrix} , \qquad
	q^R(x) = \tilde R(x) \begin{pmatrix} 0 \\ \vdots \\ 0 \\ q^R_0(x) \end{pmatrix} 
	\label{eq-rotquark}
	\end{equation}
with unitary $n_\zf \times n_\zf$ flavor rotation matrices $\tilde L(x)$, $\tilde R(x)$. These can be expressed as products of a unitary axial rotation $u(x)$ and a unitary vector rotation $v(x)$:
\begin{equation}
	\tilde L(x) = u(x)^\dagger \; v(x)\; , \qquad \tilde R(x) = u(x) \; v(x)
	\label{eq-chiraltrafodecomp}
	\end{equation}
Multiplying the second equation from the right with the Hermitian conjugate of the first equation gives
\begin{equation}
	\tilde R(x)\; \tilde L(x)^\dagger = u(x)\; u(x) \equiv U(x)
	\end{equation}
For $U(x)$ we can read off a simple transformation law under global chiral rotations $L$, $R$:
\begin{equation}	
	q^L(x) \rightarrow L\; q^L(x)\;, \quad 
	q^R(x) \rightarrow R\; q^R(x)\; \quad \Rightarrow \quad
	U(x) \rightarrow R\; U(x)\; L^\dagger
	\end{equation}
Equation (\ref{eq-rotquark}) may now be reexpressed as
\begin{eqnarray}
	q^L(x) & = & u(x)^\dagger v(x) \begin{pmatrix} 0 \\ \vdots \\ 0 \\ q^L_0(x) \end{pmatrix}\ \equiv\ u(x)^\dagger \tilde q^L(x), \\
	q^R(x) & = & u(x)\; v(x) \begin{pmatrix} 0 \\ \vdots \\ 0 \\ q^R_0(x) \end{pmatrix}\ \equiv\ u(x)\; \tilde q^R(x) 
	\label{eq-singleoutu}
	\end{eqnarray}
The new quark fields $\tilde q(x)$ are constrained. They encode less degrees of freedom, because the degrees of freedom corresponding to local axial rotations have been taken out and put into $u(x)$. We parametrize $U(x)$, and along with it $u(x)$, in terms of real fields $\phi_{\zF}$:
\begin{equation}
	U(x) = \exp\left(i \frac{\lambda_{\zF}\; \phi_{\zF}(x)}{f_\pi^0} \right) 
	\end{equation}
The ground state corresponds to $U(x)=\Eins_{n_\zf}$ and $\phi_{\zF}(x)=0$. The meaning of the normalization constant $f_\pi^0$ will become clear later. The fields $\phi_{\zF}(x)$ excite Goldstone Bosons. For example, in an SU(2) theory, the pion fields are mapped to linear combinations of the $\phi_{\zF}$ in the following way:
\begin{equation}
	\lambda_{\zF}\; \phi_{\zF} = 
	\begin{pmatrix}
		\phi_3 & \phi_1 - i \phi_2 \\
		\phi_1 - i \phi_2 & - \phi_3 \\
	\end{pmatrix} = 
	\begin{pmatrix}
		\pi^0 & \sqrt{2} \pi^+ \\
		\sqrt{2} \pi^- & -\pi^0
	\end{pmatrix}
	\end{equation}

A parity transformation $P$ exchanges right- and left-handed quark fields. From our representation (\ref{eq-singleoutu}), it follows that $u$ transforms into its Hermitian conjugate under parity. For the Goldstone Boson fields $\phi_\zF$ in turn, this implies a change of sign:
\begin{equation}
	P\, \phi_\zF(t,\vec x)\, P = - \phi_\zF(t,-\vec x)
	\end{equation}
Thus Goldstone Bosons have negative parity. They form an isospin vector, because under infinitesimal vector transformations $L=R=V=1+i \epsilon_\zG \lambda_\zG/2$, we find
\begin{equation}
	U \rightarrow V U V^\dagger \qquad \Rightarrow \qquad
	\phi_\zF \rightarrow \phi_\zF -  f_{\zF \zG \zH}\, \epsilon_\zG\, \phi_\zH
	\end{equation}

We could now rewrite the Lagrangian in terms of the constrained quark Fermion fields $\tilde q(x)$ and the above parametrization of $u(x)$ in terms of Bosonic fields $\phi_{(F)}$. We are not going to do this, however, because we are interested in the low energy limit of QCD dynamics, where quarks are confined and are not the observed degrees of freedom. Instead, we start building an \emph{effective} theory.

\section{Chiral Perturbation Theory for the Goldstone Bosons}

The general idea of an Effective Field Theory is to introduce fields for all excitations that are in resonance in the process under study. The fields come in multiplets of known behavior under symmetry transformations. In our framework, the theory contains the Goldstone Bosons (pions,~...) and the lightest baryon multiplet (proton, neutron,~...). Formally, the Lagrangian contains all interaction terms compatible with the underlying symmetries of QCD, where each term contains a different coupling constant. These couplings implicitly encode the rest of the spectrum. In $\chi$PT, the underlying symmetries are those in the chiral limit. Quark masses are introduced as perturbations using a trick: The quark mass matrix $m_{\zf \zg}$ is interpreted as a constant external field with a special transformation behavior. 

We start out by building a Lagrangian in the limit of vanishing quark masses with Goldstone Boson fields only. Thus the only fields we take over from our discussion of the QCD Lagrangian are those characterizing the local perturbations of the vacuum alignment: the fields collected in $U(x)$. The most general, chirally invariant, effective Lagrangian density with the minimal number of derivatives \cite{Scher} is
\begin{equation}
	\mathcal{L}_0^{(2)} = \frac{(f_\pi^0)^2}{4} \Tr \left( \partial_\mu U \partial^\mu U \right)
	\label{eq-L20}
	\end{equation}
To perform actual calculations with this Lagrangian, $U$ is expanded in terms of the Goldstone Boson fields $\Phi_\zF$. The prefactor $(f_\pi^0)^2/4$ in $\mathcal{L}_0^{(2)}$ makes sure that first term in the expansion is $\frac{1}{2} \partial_\mu \phi_{\zF}\, \partial^\mu \phi_{\zF}$, the standard kinetic term. Why we start out with terms with minimal number of derivatives will become clear when power counting is introduced in section \ref{sec-powercnt}. One always seeks to find a \terminol{minimal} Lagrangian. For example, a term of the form $\Tr \left( \partial_\mu \partial^\mu U \; U \right)$ turns out to produce the same equations of motion as the Lagrangian \ref{eq-L20}, and can therefore be ignored. Note that we are not able to find a chirally invariant expression in terms of $U$ fields that would produce mass terms of the form $\phi_{\zF} M_{\zF \zG} \phi_{\zG}$ for the Goldstone Bosons: $\Tr{U}$ is not invariant, $\Tr{U U^\dagger}=n_\zf$ is constant, etc. This is the way it should be; the symmetry properties of the Lagrangian automatically require the Goldstone Bosons to be massless.

Already at this stage, we can read off physical implications. For example, by applying Noether's theorem the axial current comes out as \cite{Scher}
\begin{equation}
	{j^A}^{\mu}_{\zF} 
	= -i \frac{(f_\pi^0)^2}{4} \Tr \left( \lambda_{\zF} [ U, \partial^\mu U^\dagger ] \right)
	= - f_\pi^0\; \partial^\mu \phi_{\zF} + ...
	\label{eq-effaxialcurr}
	\end{equation}
\begin{equation} 
	\Rightarrow \qquad
	\bra{\Omega_0} {j^A}^{\mu}_{\zF} \ket{\phi_{\zF}(p)} \approx
	-f^0_\pi\; \partial^\mu e^{-i p \cdot x}\; \delta_{\zF \zG} =
	i\; p^\mu\; f^0_\pi\; e^{-i p \cdot x}\; \delta_{\zF \zG}
	\end{equation}
The axial current, an operator that induces a local misalignment of the axial degrees of freedom, can create Goldstone Bosons out of the vacuum! This is a feature of spontaneous symmetry breaking. In fact, the matrix element of a broken symmetry current sandwiched between the vacuum and the corresponding Goldstone Boson state is necessarily non-zero. It is now clear why $f_\pi^0$ is called the \terminol{pion decay constant in the chiral limit}. The pion decay constant $f_\pi^\text{phys}$ measurable in experiments differs from $f_\pi^0$ only by corrections originating in the finite quark masses.

This reminds us again that the theory we have set up so far would only apply if quarks were massless. There is a way to introduce quark masses without having to give up the tight constraints of chiral symmetry. The trick is to interpret the quark mass matrix $m_{\zf \zg}$ as an external field with the transformation behavior
\begin{equation}	
	q^L(x) \rightarrow L\; q^L(x)\;, \quad 
	q^R(x) \rightarrow R\; q^R(x)\; \quad \Rightarrow \quad
	m \rightarrow R\; m\; L^\dagger
	\end{equation}
Then obviously the QCD Lagrangian eq. (\ref{eq-ChiralQCD}) is still chirally invariant. Where does the quark mass matrix show up in the effective theory? Every combination of $U$ and $m$ fields is allowed that transforms as a singlet under the symmetries of the Lagrangian. At lowest order in $m$, we can add a term of the form
\begin{equation}
	\mathcal{L}^{(2)}_\text{s.b.} = \frac{(f^0_\pi)^2\; B}{2} \Tr \left( m U^\dagger + U m^\dagger \right)
	\label{eq-L2sb}
	\end{equation}
where $B$ is a low energy constant. As such its value is not predicted by the effective theory itself. $B$ is related to the quark condensate, which can be seen by comparing the vacuum expectation value of the mass term in the effective theory with the mass term in the QCD Lagrangian:
\begin{equation}
	\braket{ \mathcal{L}^{(2)}_\text{s.b.} } 
	\quad \mathop{=}^{\braket{U}=\Eins} \quad
	\frac{(f^0_\pi)^2\; B}{2}\; \Tr (m+m^\dagger) \quad \mathop{=}^{m=m^\dagger} \quad
	(f^0_\pi)^2\; B\; \Tr (m) 
	\label{eq-massexpecteff}
	\end{equation}
\begin{equation}
	\braket{ - \bar{q}\, m\, q } = 
	- m_{\zf \zg}\, \braket{ \bar{q}_{\zf} q_{\zg} } =
	- m_{\zf \zg}\, \frac{1}{n_\zf}\, \lambda_{\zf \zg, 0}\, \braket{ \bar{q} q } =
	- \frac{1}{n_\zf}\, \Tr(m)\, \braket{ \bar{q} q }
	\label{eq-massexpectQCD}
	\end{equation}
In the last line we have made use of $\braket{\bar{q}\, \lambda_\zF\, q } = 0$ for $\zF=1..n_\zF$, which follows from the fact that these objects form a vector under $\mathrm{SU}(n_\zf)_V$.
Equating eq. (\ref{eq-massexpecteff}) and (\ref{eq-massexpectQCD}) yields
\begin{equation} \boxed{
	(f_\pi^0)^2\; B = - \frac{\braket{\bar{q} q}}{n_\zf} 
	} \label{eq-GMORcondensate} \end{equation}
We read off the leading order Goldstone Boson masses from the terms of second order in the $\phi_{\zF}$, i.e. we bring the Lagrangian into the form $\mathcal{L}^{(2)}_\text{s.b.} = ... + \phi_{\zF}\, \overline{m}_{\zF \zG}\, \phi_\zG + ...$ . In our case $\overline{m}_{\zF \zG}$ is already diagonal, and by going back to the definitions of the particle fields one arrives at leading order particle masses \cite{Scher}

\renewcommand{\arraystretch}{1.4}
\begin{equation} \boxed{
	\begin{array}{lll}
	\overline{m}_\pi^2 & = 2\, B\, \hat{m} \qquad & \text{for } n_\zf = 2,3\\
	\overline{m}_K^2 & = B\, ( \hat{m} + m_s ) \qquad & \text{for } n_\zf = 3 \\
	\overline{m}_\eta^2 & = \frac{2}{3}\, B\, ( \hat{m} + 2 m_s ) \qquad & \text{for } n_\zf = 3
	\end{array}
	} \label{eq-GMOR} \end{equation}
\renewcommand{\arraystretch}{1.0}

These equations, referred to as the \terminol{Gell-Mann-, Oakes- and Renner- relations} (\terminol{GMOR}), are of central importance to our discussion. The fact that the different subtypes of pions $\pi^0$, $\pi^\pm$, and kaons $K^\pm$,$K^0$, $\bar K$ come out at equal masses is a consequence of neglecting differences in $m_u$ and $m_d$ on the one hand and ignoring the electromagnetic self-energy contribution on the other hand.

\section{Adding Baryons}

For baryons, we need in the SU(2) theory Fermion fields for the proton-neutron doublet, and in the SU(3) case Fermion fields transforming as an octet. For simplicity, let us consider the SU(2) case, where we have a doublet of Fermionic baryon fields
\begin{equation}
	\Psi = \begin{pmatrix} p \\ n \end{pmatrix}
	\end{equation}
We can look at the quark fields in section \ref{sec-gobofields} for inspiration. Remember that we have completely moved the axial degrees of freedom of our theory into the fields $u(x)$. So the excitations produced by $\Psi(x)$ should be devoid of axial degrees of freedom. For the purpose of constructing the Lagrangian, however, it is also helpful to introduce ``unconstrained'' baryon fields $\psi(x)$. In analogy to eq. (\ref{eq-singleoutu}), we decompose the fields into left- and right-handed components and write
\begin{equation}
	\psi^L \equiv u^\dagger \Psi^L \qquad \psi^R \equiv u \Psi^R
	\end{equation}
We assign the usual transformation behavior to the unconstrained fields
\begin{equation}
	\psi^L(x) \rightarrow L\, \psi^L(x) \qquad \psi^R(x) \rightarrow R\, \psi^R(x)
	\end{equation}
Looking at equations (\ref{eq-chiraltrafodecomp}) and (\ref{eq-singleoutu}), we see that the axially constrained baryon fields $\Psi^L(x)$, $\Psi^R(x)$ and thus $\Psi(x)$, like the constrained quark fields, transform as isovectors
\begin{equation}
	L = R = V \qquad \Rightarrow \qquad
	\Psi(x) \rightarrow V\, \Psi(x)
	\end{equation}
but remain invariant under pure axial transformations $L^\dagger=R=A$. Thus $\Psi(x)$ meets our requirements as a representation of the baryon doublet.\footnote{The explicit transformation behavior of $\Psi(x)$ is complicated and non-global due to a dependence on $u(x)$ \cite{Scher}.}  
From the chirally invariant structures
\begin{align}
	\overline{\psi^R}\, U\, \psi^L & = \overline{\Psi^R}\, \Psi^L &
	\overline{\psi^L}\, \slashed{\partial} \psi^L 
		& = \overline{\Psi^L}\, \slashed{\partial} \Psi^L 
		+ \overline{\Psi^L}\, u \slashed{\partial} u^\dagger\, \Psi^L \nonumber \\ 
	\overline{\psi^L}\, U^\dagger\, \psi^R & = \overline{\Psi^L}\, \Psi^R &
	\overline{\psi^R}\, \slashed{\partial} \psi^R 
		& = \overline{\Psi^R}\, \slashed{\partial} \Psi^R 
		+ \overline{\Psi^R}\, u^\dagger \slashed{\partial} u\, \Psi^R 
	\end{align}
we assemble the \terminol{lowest order} Lagrangian of pion-nucleon interaction: 
\begin{equation}
	\mathcal{L}^{(1)}_{\pi N} 
	= i \overline{\Psi} \left( \slashed \partial 
	+ \frac{1}{2}( u \slashed{\partial} u^\dagger + u^\dagger \slashed{\partial} u) 
	+ (g_A^0) \frac{\gamma^5}{2}(u \slashed{\partial} u^\dagger - u^\dagger \slashed{\partial} u) 
	\right) \Psi
	+ m_0 \, \overline{\Psi} \, \Psi
	\label{eq-L1piN}
	\end{equation}
(A term $\gamma^5 \, \overline{\Psi} \, \Psi$ is forbidden, because the Lagrangian must be of positive parity). There are two LECs in this Lagrangian: the nucleon mass in the chiral limit $m_0$,\footnote{The existence of a finite nucleon mass $m_0$ in the chiral limit is related to the \emph{trace anomaly}, and can be shown to originate to a large part from gluonic field energy \cite{TW01}.} and the axial-vector coupling constant in the chiral limit $g_A^0$. The latter owes its name to the form of the axial-vector current of the nucleon
\begin{equation}
	\bra{N(p)} {j^A}^{\mu}_{\zF} \ket{N(p)} = \overline{u}(p) g_A \gamma^\mu \gamma_5 \frac{\lambda_{\zF}}{2} u(p)
	\end{equation}
For massless quarks, $g_A$ assumes its chiral limit value $g_A^0$ \cite{Scher}.

Spontaneous symmetry breaking can account for the absence of \terminol{parity doublets} in the spectrum. We can now understand how. Without spontaneous symmetry breaking, it would be the operators $\psi^\dagger$ which would generate the nucleons. Consider a nucleon state $\ket{\mathcal{N},+} \equiv \psi^\dagger \ket{0} = \psi^{L \dagger} + \psi^{R \dagger} \ket{0}$ of positive parity. An infinitesimal axial rotation, e.g. $\hat A = \exp(2 i \epsilon Q^A_1)$, transforms  $\psi$ according to
\begin{equation}
	\psi^L \rightarrow \psi^L - i \epsilon \lambda_1 \psi^L \qquad
	\psi^R \rightarrow \psi^R + i \epsilon \lambda_1 \psi^R 
	\end{equation}
but leaves the vacuum $\ket{0}$ invariant by definition. Thus the transformation introduces a nucleon component $\ket{\mathcal{N}',-} \equiv \lambda_1 ( \psi^{L \dagger} - \psi^{R \dagger}) \ket{0}$ of opposite parity. Since Lagrangian and Hamiltonian are invariant under axial transformations, $\ket{\mathcal{N},+}$ and $\ket{\mathcal{N}',-}$ have the same energy and can both be excited. So in a world symmetric under $\mathrm{SU}(n_\zf)_A$, one would expect to find nucleon states of both parities at the same energy in the spectrum. This is not what we observe.
 
Due to spontaneous symmetry breaking we have a different picture of the world. What we call nucleons are excitations of the QCD vacuum by the constrained nucleon operators: 
$\ket{N,+} \equiv \Psi \ket{\Omega_0} = \Psi^{L \dagger} + \Psi^{R \dagger} \ket{\Omega_0}$. 
Attributing positive parity to the vacuum $\ket{\Omega_0}$, the nucleon $\ket{N,+}$ has positive parity as well.
An axial rotation like $\hat{A}$ leaves $\Psi$ invariant per definition.\footnote{In our new picture, $\hat{A}$ acts on the Goldstone Boson field $U(x)$, as can be seen from the axial current eq. (\ref{eq-effaxialcurr}) in the effective theory.} However, it does not leave the vacuum $\ket{\Omega_0}$ invariant. In the chiral limit, $\hat{A} \ket{\Omega_0} \equiv \ket{\Omega_1}$ has the same energy as $\ket{\Omega_0}$ and still has the discrete momentum eigenvalue 0. It must be another vacuum state. Then an axially rotated nucleon state looks like
\begin{equation}
	\ket{N'} \equiv \hat{A} \ket{N,+} = \Psi \hat{A} \ket{\Omega_0} = \Psi \ket{\Omega_1}
	\end{equation}
It is an excitation of a \emph{different vacuum}. In eq. (\ref{eq-diagvacua}), we saw that no local interaction operator can mediate the transition between different vacuum states. Therefore, $\ket{N'}$ cannot be excited in nature. Spontaneous symmetry breaking constrains the spectrum. The axial degrees of freedom of states can only be modified \emph{locally}, giving rise to Goldstone Bosons rather than axially rotated particle states.

\section{Higher Orders and Power Counting}
\label{sec-powercnt}

In the spirit of effective field theories, we should add to our Lagrangian all terms which share the symmetries of the QCD Lagrangian. An example is the purely mesonic term
\begin{equation}
	\hat{O}_4 \equiv L_4 \Tr \left( \partial_\mu U \partial^\mu U^\dagger \right)\, \Tr \left( m U^\dagger + U m^\dagger \right)
	\label{eq-highordsample}
	\end{equation}
with a low energy constant $L_4$ in front. Obviously, there are infinitely many such terms.
 
Weinberg's power counting scheme offers a way to assign a formal order to a Feynman diagram. We expect diagrams of higher order to be suppressed. The amplitude $\mathcal{M}(p_1,p_2,...,m)$ of the diagram depends on ``small'' external momenta $p_i$ and quark masses $m$. Formally, the expansion of $\chi$PT is performed around vanishing quark masses (chiral limit) and vanishing momentum transfer (low energy limit), so we analyze the order of a diagram by substituting scaled momenta $p_i \rightarrow t p_i$ and quadratically scaled quark masses $m \rightarrow t^2 m$. Counting the quark masses quadratically is inspired by the GMOR relations, which tell us that the leading order contribution to the Goldstone Boson masses is proportional to the quark masses squared. Then from
\begin{equation}
	\mathcal{M}(t p_1, t p_2,...,t^2 m) = t^D\;\mathcal{M}(p_1,p_2,...,m) + \mathcal{O}(t^{D+1})
	\label{eq-scalemomenta}
	\end{equation}
we read off the \terminol{chiral dimension} $D$ of the diagram. The meson momenta of in- and out-states are small quantities, because they vanish in the low energy limit of the chiral limit. They appear directly as arguments $p_i$ of $\mathcal{M}$ in eq. (\ref{eq-scalemomenta}). External baryon momenta $P_j$ cannot be treated as small quantities. We should rather decompose them according to $P_j = m_0 \cdot v + p_j$, where $v$ is a timelike unit four-vector, e.g. the four-velocity of the center of mass. The $p_j$ are again four-momenta that can be treated as ``small'', and appear as arguments of $\mathcal{M}$ in (\ref{eq-scalemomenta}).
 
To simplify notation, the Lagrangian is organized in the following way
\begin{equation}
	\mathcal{L} = \mathcal{L}^{(1)}_{\pi N} 
	+ \mathcal{L}^{(2)}_{\pi \pi} 
	+ \mathcal{L}^{(2)}_{\pi N} 
	+ \mathcal{L}^{(3)}_{\pi N}
	+ \mathcal{L}^{(4)}_{\pi \pi}
	+ \mathcal{L}^{(4)}_{\pi N}
	+ ...
	\end{equation}
Contributions $\mathcal{L}_{\pi \pi}$ contain terms made up of Goldstone Boson fields only, while the contributions $\mathcal{L}_{\pi N}$ contain interactions with baryons. The superscript indicates the chiral dimension of the terms. 

For example, the term $\hat{O}_4$ from (\ref{eq-highordsample}) is part of $\mathcal{L}^{(4)}_{\pi \pi}$, because a tree level amplitude $\mathcal{M} = \bra{\text{out}} \hat{O}_4 \ket{\text{in}}$ is proportional to the quark masses $m$ (yielding two chiral dimensions) and to two external momenta (yielding one chiral dimension each). The two external momenta appear in the amplitude because of the derivatives of Goldstone Boson fields in the $\hat{O}_4$ vertex.

We have already specified the lowest order contributions $\mathcal{L}^{(1)}_{\pi N}$ and $\mathcal{L}^{(2)}_{\pi \pi} = \mathcal{L}^{(2)}_0 + \mathcal{L}^{(2)}_\text{s.b.}$ of the Lagrangian in eq. (\ref{eq-L1piN}), (\ref{eq-L20}) and (\ref{eq-L2sb}). At higher orders, the Lagrangian becomes quite long, and the number of independent LECs explodes \cite{Bij99}, \cite{FettThes}. Here, we quote the terms needed for an SU(2) calculation of the nucleon mass to order $p^4$ \cite{PHW04}. We make use of the following abbreviations
\begin{align}
	\chi & \equiv 2 B m \mathop{=}^{(\ref{eq-GMOR})} \Eins_2 \; \overline{m}_\pi^2 &
	\chi_\pm & \equiv u^\dagger \chi u^\dagger \pm u \chi^\dagger u \\
	D_\mu A & \equiv \partial_\mu A + \frac{1}{2} \left( u^\dagger \partial_\mu u + u \partial_\mu u^\dagger \right) A &
	u_\mu & \equiv i \left( u^\dagger \partial_\mu u - u \partial_\mu u^\dagger \right) 
\end{align}
where $A$ is any object that transforms like $\Psi$, e.g., $D_\nu \Psi$.
From the Lagrangian, the terms displayed below contribute to the nucleon mass:
\begin{eqnarray}
	\mathcal{L}_{\pi \pi}^{(2)} &=& \frac{(f_\pi^0)^2}{4} \Tr \left( \partial_\mu U \partial^\mu U \right) + \frac{(f^0_\pi)^2}{4} \Tr \left( \chi U^\dagger + U \chi^\dagger \right) \nonumber \\
	{\cal{L}}_N^{(1)}&=&\bar{\Psi}\,(i\gamma_\mu D^{\mu}-m_0)\,\Psi+\frac{1}{2}\,g_A^0\, \bar{\Psi}\,\gamma_\mu \gamma_5 u^{\mu}\, \Psi\,, \nonumber\\
	{\cal{L}}_{N\pi}^{(2)}&=&c_1\,{\rm{Tr}}(\chi_{+}) \bar{\Psi} \Psi-\frac{c_2}{4m_0^2}\,{\rm{Tr}}(u_\mu u_\nu)\,(\bar{\Psi} D^{\mu}D^{\nu} \Psi+ {\rm{h.c.}})+\frac{c_3}{2}\,{\rm{Tr}}(u_\mu u^{\mu})\, \bar{\Psi}\Psi+... \nonumber\\
	{\cal{L}}_{N\pi}^{(4)}&=&e_{38}\,({\rm{Tr}}(\chi_{+}))^2 \bar{\Psi}\Psi+\frac{e_{115}}{4}\,{\rm{Tr}}(\chi_{+}^2-\chi_{-}^2)\bar{\Psi}\Psi\nonumber\\&&-\frac{e_{116}}{4}\,\left({\rm{Tr}}(\chi_{-}^2)-({\rm{Tr}}(\chi_{-}))^2+{\rm{Tr}}(\chi_{+}^2)-({\rm{Tr}}(\chi_{+}))^2\right) \bar{\Psi}\Psi +\dots
\end{eqnarray}
$\mathcal{L}_{N \pi}^{(3)}$ does not contribute. $c_1$, $c_2$, $c_3$, $e_{38}$, $e_{115}$, $e_{116}$ are new LECs. 
 
One can derive a formula to quickly read off the chiral dimension of a diagram \cite{Scher}. Let $N_L$ be the number of independent loop momenta, $I_N$ be the number of internal nucleon lines, $N_{\pi \pi}^{(2n)}$ be the number of vertices from $\mathcal{L}^{(2n)}_{\pi \pi}$ and $N_{\pi N}^{(n)}$ be the number of vertices from $\mathcal{L}^{(n)}_{\pi N}$ in the diagram, then
\begin{equation}
	D = 2 N_L + I_N + 2 + \sum_{n=1}^{\infty} 2 (n-1) N_{\pi \pi}^{(2n)} + \sum_{n=1}^\infty (n-2)N_{\pi N}^{(n)}
	\label{eq-powercnt}
	\end{equation}
Obviously diagrams with loops are automatically of higher order. However, for this nice power counting scheme to be applicable, renormalization in the baryon sector has to be handled with special care, see section \ref{sec-infreg}.

Equipped with a power counting rule we can calculate amplitudes in perturbation theory to a given \terminol{chiral order} $p^D$, where $p$ symbolizes ``small'' momenta or masses.

\section{Propagators}

We read off the lowest-order Feynman propagators from the free field parts of the Lagrangian, e.g., for a pion of momentum $k$
\begin{equation}
	i S^\pi_F(k) = \frac{i}{k^2 - m_\pi^2 + i 0^+}
	\label{eq-pionprob}
	\end{equation}
and a nucleon of momentum $P$
\begin{equation}
	i S^N_F(P) = \frac{i}{\slashed P - m_0 + i 0^+}
	\label{eq-nucprop}
	\end{equation}
Using these propagators for internal lines, we now have almost all the tools at hand to calculate arbitrary diagrams within $\chi$PT, which is done as in any other Quantum Field Theory.
	
\section{Renormalization}
\label{sec-renorm}

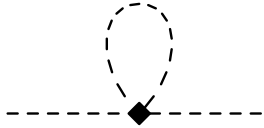
\begin{figure}
	\centering

	\unitlength=1mm
	\begin{fmffile}{mesonloop}
	\fmfset{arrow_len}{3mm}
	\fmfset{arrow_ang}{15}

		\begin{fmfgraph*}(35,13)
		\fmfleft{l}
		\fmfright{r}
		\fmfforce{(0.0w,0.0h)}{l}
		\fmfforce{(1.0w,0.0h)}{r}
		\fmfforce{(0.5w,0.0h)}{d}
		\fmf{dashes}{l,d,r}
		\fmf{dashes,right=90,tension=.8,l.side=left}{d,d}
		\fmfv{d.shape=diamond,d.filled=full,d.size=4thick}{d}
		\end{fmfgraph*}

	\end{fmffile}
	
	\caption{Example of a meson loop diagram.}
	\label{fig-mesonloop}
\end{figure}

Let us take a look at one of the diagrams resulting from an interaction term of the Lagrangian in SU(2) 
\begin{equation}
	\bra{\phi_\zF(p)} \mathcal{T} \mathcal{L}^{(2)}_{s.b.} \ket{\phi_\zF(p')} = 
	... + \frac{B\, \Tr(m)}{24\, (f_\pi^0)^2}
	\wick[u]{222}{<1{\bra{\phi_\zF(p)}}\ >1{\phi_\zG}\, 
	<3{\phi_\zH}\, >3{\phi_\zH}\,
	<2{\phi_\zG}\,  >2{\ket{\phi_\zF(p')}}} + ...
	\label{eq-sampleloopamp}
\end{equation}
$\wick[u]{1}{<3{\phi_\zH} >3{\phi_\zH}}$ is a sum of meson loops -- here pions, see fig. \ref{fig-mesonloop}.
Thus an amplitude involving the diagram above contains a term proportional to
\begin{equation}
	\int \frac{d^4 k}{(2 \pi)^4} \frac{i}{k^2 - \overline{m}_\pi^2 + i 0^+} \equiv I_\pi 
	\label{eq-pionloopintegral}
	\end{equation}
The integral over all four-momenta $k$ is not defined, because it diverges for large $k$. The problem can be solved using dimensional regularization, which replaces the four-dimensional integral by a $(4+\epsilon)$-dimensional one \footnote{How these integrals are defined and calculated can be found in the literature \cite{Muta87,Scher}.}:
\begin{align}
	I_\pi(\epsilon,\lambda)\ & \equiv\ \frac{1}{\lambda^\epsilon} \int \frac{d^{4+\epsilon} k}{(2 \pi)^{4+\epsilon}} \frac{i}{k^2 - \overline{m}_\pi^2 + i 0^+} \nonumber \\
	& = 2 \overline{m}_\pi^{2} \left( \frac{\overline{m}_\pi}{\lambda} \right)^\epsilon \frac{1}{(4\pi)^2} \left(  \frac {1}{\epsilon} - \frac{1}{2}(\ln 4\pi - \gamma_E + 1)\right) + \mathcal{O}(\epsilon)
	\end{align}

$\lambda$ has unit mass and is needed to give $I_\pi(\epsilon,\lambda)$ the correct, $\epsilon$ independent units. It is called \terminol{renormalization scale}. In principle, $\lambda$ can be chosen arbitrarily. 

The resulting value of the integral has a pole proportional to $1/\epsilon$. If the LECs were all finite, the divergence $1/\epsilon$ would end up in the final results for physical observables. This problem can be cured if we allow the LECs to have poles at $\epsilon=0$.
In our case $I_\pi(\epsilon \rightarrow 0,\lambda) \propto \overline m_\pi^2$, so in accordance with the power counting rule (\ref{eq-powercnt}) our diagram is of chiral order $p^4$. If we equip the LECs in $\mathcal{L}_{\pi\pi}^{(4)}$ with appropriate divergences, tree graph contributions from $\mathcal{L}_{\pi\pi}^{(4)}$ can cancel out our loop divergence. Thus a set of terms in the Lagrangian acts as \terminol{counterterms}, balancing the infinities of the loop integral. Since an effective theory already contains all terms conforming to the symmetries, structures able to act as counter terms are available from the start.
The finite remainder of a LEC, carrying information of physical relevance, is expressed in terms of a renormalized coupling constant. 

Let us look at our situation in mesonic $\chi$PT. Here, we can apply the following prescription for the one-loop renormalization of a low energy constant $L_i$, now formally $\epsilon$- and $\lambda$-dependent:
\begin{equation}
	L_i(\epsilon,\lambda) \equiv L^r_i(\lambda) + C_i\, \bar{\lambda}(\epsilon) + \mathcal{O}(\epsilon)
	\label{eq-LECrunning}
	\end{equation}
where $C_i$ is a fixed constant which encodes the strength of the pole, and
\begin{equation}
	\bar{\lambda}(\epsilon) \equiv \frac{1}{(4 \pi)^2} \left( \frac{1}{\epsilon} - \frac{1}{2}( \ln(4\pi) - \gamma_E + 1) \right) 
	\label{eq-LECsing}
	\end{equation}
carries the singularity.
The \terminol{renormalized} coupling constants $L^r_i(\lambda)$ are finite. These are the values that can actually be extracted from fits to experiment or lattice calculations. For our discussion it is important to keep in mind that renormalized LECs determined in different analyses are only comparable when the renormalization scale $\lambda$ is the same. Demanding $\lambda$-independence of the final amplitude in leading order of $\epsilon$ yields the transition rule from one scale $\lambda_1$ to another $\lambda_2$~:
\begin{equation}
	L_i^r(\lambda_2) = L_i^r(\lambda_1) + \frac{C_i}{(4 \pi)^2} \ln \left( \frac{\lambda_1 }{\lambda_2} \right)
	\end{equation}
Through renormalization of the LECs, in any physical observable, the divergent part of $I_\pi(\epsilon,\lambda)$ will be cancelled. In our example, only the finite, renormalized part
\begin{equation}
	\bar{I}_\pi(\lambda) \equiv I_\pi(\epsilon,\lambda) - 2 \overline{m}_\pi^2 \bar{\lambda}(\epsilon) + \mathcal{O}(\epsilon)
	= \frac{2 \overline{m}_\pi^2}{(4\pi)^2} \ln \frac{\overline{m}_\pi}{\lambda} 
	\label{eq-renormint}
\end{equation}
enters the final expression. Thus loop graphs give rise to the typical \terminol{chiral logs}.

The chiral log $\ln(\overline{m}_\pi/\lambda)$ vanishes if we set $\lambda = \overline{m}_\pi$. This is a convenient choice \emph{as long as the pion mass remains fixed throughout the discussion}. 
In the SU(2) mesonic theory, where $\overline{m}_\pi$ is the only mass scale, one therefore defines for the LECs $\elll_i$ of $\mathcal{L}_{\pi \pi}^{(4)}$
\begin{equation}
	\overline{\elll}_i \equiv \frac{2(4\pi)^2}{\gamma_i} \elll_i^r(\overline{m}_\pi)
	\qquad \text{such that} \quad
	\elll_i^r(\lambda) = \frac{\gamma_i}{2(4\pi)^2} \left( \overline{\elll}_i + \ln \left(\frac{\overline{m}_\pi^2}{\lambda^2} \right) \right)
	\end{equation}
The $\gamma_i$ correspond to the $C_i$ in eq. (\ref{eq-LECrunning}).
In terms of the $\overline{\elll}_i$, formulae for mesonic amplitudes are free of chiral logs, as though renormalization had never been an issue. However, the LECs are now implicitly $\overline{m}_\pi$-dependent. For our purposes, this is not acceptable, because for us $\overline{m}_\pi$ encodes the quark masses via GMOR. We want to study quark mass dependence of observables, keeping the LECs fixed.

The integral (\ref{eq-pionloopintegral}) was originally undefined. Renormalization assigns a value $\bar{I}_\pi(\lambda)$ to it that captures the dependence of the integral on the external \emph{variables} in the valid regime of the theory. The method of choosing such a value is not unique, so that various renormalization schemes at arbitrary renormalization scales describe the same physics (up to higher order effects). Checking whether divergences and scale dependences cancel in the final result offers a way of validating internal consistency of the renormalization scheme. 

The loop graph discussed in this section is of a particularly good nature. It results from a term at chiral order $p^2$ and produces a single contribution at a higher order, namely $p^4$. Thus the loop is automatically suppressed in the chiral expansion. This is a pre-requisite if we want to terminate the series at a specific loop order. 

\section{Chiral Scale and Natural Size}
\label{sec-natsize}

Putting (\ref{eq-sampleloopamp}) and (\ref{eq-renormint}) together, the factor $1/(4\pi f_\pi^0)^2$ appears in our final loop amplitude. This factor is generally found in meson loops, and inspires the estimate of the \terminol{chiral scale} 
\begin{equation}
	\Lambda_\chi \approx 4 \pi f_\pi^0
	\end{equation}
This choice of $\Lambda_\chi$ as the \terminol{symmetry breaking scale} and the ``dimensionful parameter that suppresses non-renormizable terms'' was originally motivated in ref. \cite{MG84}, where it also served as a physically sensible cutoff. The value $\Lambda_\chi \approx 1 \units{GeV}$ can be brought into direct connection with empirical observations as well \cite{DGH84}.

Loop contributions are suppressed by a factor of the order $p/\Lambda_\chi$ per chiral order. It is sensible to expect this also for the contributions from structures in the Lagrangian that constitute counterterms of loops. In order to treat all interaction structures on an equal footing, one may come to the belief that, at an appropriate renormalization scale $\lambda$, \emph{all} terms in the Lagrangian should follow this rule. This has implications for the expected magnitude of LECs, and is called the \terminol{natural size} argument. Consider, as an example, leading order tree level contributions in SU(2) from two sample terms 
\begin{align}
	\mathcal{L}_{\pi N}^{(2)} & = c_1\,\Tr(\chi_+) \overline{\Psi} \Psi + \ldots
	= c_1\, 4 \overline{m}_\pi^2 \overline{\Psi} \Psi + \ldots \\
	\mathcal{L}_{\pi N}^{(4)} & = e_{38}\,\left(\Tr(\chi_+)\right)^2 \overline{\Psi}\Psi + \ldots 
	= e_{38}\,16 \overline{m}_\pi^4 \overline{\Psi} \Psi + \ldots
	\end{align}
Identifying $\overline{m}_\pi^2$ in the above expressions with a quantity of order $p^2$, we see that requiring
\begin{equation}
	\mathcal{O}\left( e_{38} \right) = \frac{1}{\Lambda_\chi^2} \mathcal{O}\left( c_1 \right)   
	\end{equation}
establishes ``natural size'' behavior of the term proportional to $e_{38}$. With $c_1 \approx -1 \units{GeV^{-1}}$ and allowing for some variations, we have in general a quick-and-dirty rule of thumb, stating that (in our notation of the Lagrangian) couplings should take values at scales 
\begin{equation*}
	\sim \ -3 \ldots 3 \units{GeV^{-D+1}}
	\end{equation*}
where D is the chiral dimension at which they occur in the Lagrangian.

Going even one step further, we may think of a probability distribution (in the Bayesian sense) according to which we find the LECs distributed. The terms in the Lagrangian are an effective description of QCD, summing up various underlying elementary processes. Thus within this line of thinking, it is sensible to assume the LECs follow a Gaussian distribution. 

Personally, I do not put much confidence in constraints following from natural size considerations.

\section{Infrared Regularization} 
\label{sec-infreg}

Loop diagrams containing baryon lines are less good-natured. It can be shown (e.g., \cite{BL99}) that in general all loop diagrams can be expressed in terms of scalar integrals containing baryon and meson propagators only. As an example, the following typical loop integral needs to be evaluated in the process of calculating the nucleon mass to order $p^3$:
\begin{equation}
	H_{N \pi}(\epsilon,\lambda) \equiv \frac{1}{\lambda^\epsilon} \int \frac{d^{4+\epsilon} k}{(2\pi)^{4+\epsilon}} 
		\ \frac{i}{k^2 - \overline{m}_\pi^2 + i 0^+}
		\ \frac{i}{(P-k)^2 - m_0^2 + i 0^+}
	\end{equation}
where $P$ is the nucleon momentum, with $P^2 - m_0^2 = \mathcal{O}(p)$. 

Calculating such integrals using straightforward dimensional regularization gives results that are \emph{not} of higher order. Consequently the power counting rule (\ref{eq-powercnt}) is not valid, because loops are not suppressed. This means that in principle loop graphs must be summed up to all loop orders at a given chiral order. 
Responsible for this unpleasant behavior is the appearance of the baryon mass $m_0$ in the integrands. It is a constant of the order of $\Lambda_\chi$, and must be treated as a quantity of chiral order $p^0$. 

One way of getting around this problem is the heavy baryon formalism (HB$\chi$PT), corresponding to an expansion of relativistic effects in powers of $1/m_0$ in a suitable inertial frame of reference. The method avoids parameters of order $p^0$ in the propagator, making power counting according to (\ref{eq-powercnt}) possible. However, the convergence properties of the expansion are known to be bad in some important cases \cite{BL99}.

A more recent approach to the problem is called \terminol{infrared regularized} or \terminol{relativistic} baryon chiral perturbation theory, was first described at one-loop order in ref. \cite{BL99} and has been extended to arbitrary loop order \cite{SGS04}. It is a variant of dimensional regularization applied to the original, fully relativistic Lagrangian of $\chi$PT. 

In section \ref{sec-renorm} we had an integrand containing only one fixed Lorentz-invariant quantity, $\overline{m}_\pi$. Setting $\lambda=\overline{m}_\pi$, the loop integral vanished, but then the LECs were implicitly $\overline{m}_\pi$-dependent. The baryon propagator introduces new Lorentz-invariant quantities, one of them being the nucleon mass $m_0$. We would not mind if the renormalization of other LECs depended implicitly on $m_0$. The LECs are fixed properties of the ground state of the QCD Hamiltonian in the chiral limit; the interdependence of the LECs is not an issue here. In contrast to $\overline{m}_\pi$, which encodes the strength of our external quark mass field, $m_0$ is not a \emph{variable}. Can we adapt the $\lambda=\overline{m}_\pi$-trick to the baryon sector? 


Infrared regularization can be thought of as an implementation of this trick in the general case, when both $m_0$ and ``small'' parameters of the order $p$ occur in the integrands. For the integral $H_{N \pi}$ we may introduce dimensionless parameters of order $p$ 
\begin{equation}
	\alpha \equiv \frac{\overline{m}_\pi}{m_0}\ ,\qquad
	\beta \equiv \frac{P^2 - m_0^2}{m_0^2}
	\end{equation}
Working with the Feynman-parametrization, the integral becomes
\begin{equation}
	H_{N \pi}(\epsilon,\lambda) =  \left( \frac{m_0}{\lambda} \right)^{\epsilon} 
	    \frac{\Gamma(-\epsilon/2)}{(4 \pi)^{2+\epsilon/2}}
		 \int_0^1 dz \left( z^2 ( \beta + 1 ) - z ( \beta + \alpha^2 ) + \alpha^2 \right )^{\epsilon/2}
	\label{eq-infredintegralfeynm}
	\end{equation}
Our task is to capture the behavior of the integral for $\alpha, | \beta | \ll 1$. In the limit $\alpha, \beta \rightarrow 0$, the integral does not vanish
\begin{align}
	\mathcal{R}_{N \pi}(\epsilon,\lambda) & \equiv H_{N \pi}(\epsilon,\lambda) \vert_{\alpha,\beta=0} \nonumber \\
	& = \left( \frac{m_0}{\lambda} \right)^{\epsilon} \frac{\Gamma(-\epsilon/2)}{(4 \pi)^{2+\epsilon/2}} \frac{1}{1+\epsilon} 
	  = \left( \frac{m_0}{\lambda} \right)^\epsilon \left( \frac{1}{(4\pi)^2} - 2 \bar \lambda \right) + \mathcal{O}(\epsilon)
	\end{align}
$\mathcal{R}_{N \pi}(\epsilon,\lambda)$ is the culprit in the integral value, spoiling power counting with its contribution of order $p^0$. Yet, it does not depend on the variables of order $p$. Therefore it is called \terminol{infrared regular}. It can easily be absorbed by counterterms. In fact, the form of $\mathcal{R}_{N \pi}(\epsilon,\lambda)$ suggests that it can be canceled by a suitable renormalization of $m_0$ itself. On the other hand, the remaining part of the integral
\begin{equation}
	\mathcal{I}_{N \pi}(\epsilon,\lambda) \equiv H_{N \pi}(\epsilon,\lambda) - \mathcal{R}_{N \pi}(\epsilon,\lambda)
\end{equation} 
turns out to be of order $p$ and respects the power counting rule eq. (\ref{eq-powercnt}).

Now renormalization is performed in two steps. In the first step we renormalize the LECs in such a way that in all loop integrals terms like $\mathcal{R}_{N \pi}(\epsilon,\lambda)$ are completely compensated for by counterterms. In other words we define the renormalized value of our ``culprit'' term to be $\overline{\mathcal{R}}_{N \pi} = 0$. Note however, that now $m_0$ enters our definition of the renormalized coupling constants. Thus we have an $m_0$-dependent renormalization scheme. Power counting is violated, therefore this part of renormalization must be performed to infinite loop order. However, we never need to do it explicitly. From now on, whenever we speak about a LEC $L_i$, we actually refer to its self-consistently renormalized value after this primary implicit renormalization step. Since the renormalized value $\overline{\mathcal{R}}_{N \pi}$ of the integral is $\lambda$-independent, so are the renormalized coupling constants $L_i$. At this stage, the loop integral only contributes with $\mathcal{I}_{N \pi}(\epsilon,\lambda)$. 
This remaining part still carries divergences, which are treated explicitly in a second, mass independent renormalization step, in the same way as in the mesonic sector. 

The choice of $\mathcal{R}_{N \pi}(\epsilon,\lambda)$ above corresponds to the simple method described in ref. \cite{FG03}. In ref. \cite{BL99} Becher and Leutwyler actually propose a (more involved) procedure of determining the infrared regular part. 
\footnote{The infrared regular part $\mathcal{R}_{N \pi}(\epsilon,\lambda)$ of ref. \cite{FG03} could be called ``minimal'', in the sense that only terms violating the power counting are included. In contrast, the infrared regular part $R_{N \pi}(\epsilon,\lambda)$ of ref. \cite{BL99} could be called ``maximal''. Becher and Leutwyler obtain it by modifying integration limits. Alternatively, the integrand in eq. (\ref{eq-infredintegralfeynm}) can be expanded in terms of the small quantities $\alpha$, $\beta$. Interchanging summation and integration and adjusting $\epsilon$ as needed for each term produces a series expansion of this maximal infrared regular part \cite{SGS04}. }
The result is of the form
\begin{equation}
	R_{N \pi}(\epsilon,\lambda) = \left( \frac{m_0}{\lambda} \right)^{\epsilon} 
	\frac{\Gamma(-\epsilon/2)}{(4 \pi)^{2+\epsilon/2}} g(\alpha,\beta,\epsilon) = \mathcal{R}_{N \pi}(\epsilon,\lambda) + \mathcal{O}(\alpha, \beta)
	\end{equation}
Here, $g$ is a function which can be expanded in an ordinary power series in $\alpha$ and $\beta$. Again, $R_{N \pi}(\epsilon,\lambda)$ can be absorbed by counter terms, thanks to the properties of $g$. The remainder $I_{N \pi}(\epsilon,\lambda) = H_{N\pi}(\epsilon,\lambda) - R_{N \pi}(\epsilon,\lambda)$ is called the \terminol{infrared singular} part and respects the power counting rule eq. (\ref{eq-powercnt}). The advantage of the method is its simple relationship to HB$\chi$PT. Expanding an expectation value calculated in this framework in powers of $1/m_0$ and truncating the series at a given order produces the corresponding expression obtained within HB$\chi$PT \cite{BL99}. Thus infrared regularization can be seen as a way of doing HB$\chi$PT to infinite order in $1/m_0$. In practice, infrared regularization offers a way to directly calculate the infrared singular part $I_{N \pi}(\epsilon,\lambda)$. The infrared regular part plays no role in the calculation and can be discarded without ever being evaluated.

In summary, infrared regularization restores a valid power counting. This is important, because otherwise there would be no concept of order in the chiral expansion. If one works with an invalid power counting scheme, going to the next ``order'' changes the renormalization scheme of the LECs, making numerical results obtained at lower order worthless.

\section {Calculating the Nucleon Mass}
\label{sec-nucleonmass}

This diploma thesis makes heavy use of a formula for the nucleon mass calculated within infrared regularized $\chi$PT in isospin symmetric SU(2).
In the chiral limit, the nucleon mass is $m_0$. This is ensured by infrared regularization.
At non-vanishing pion/quark masses, the nucleon mass can be read off from the pole of the nucleon propagator
\footnote{$i \Delta(p)$ is actually a matrix in flavor space. We can ignore this, since we neglect isospin breaking ($m_u = m_d$).}
\begin{equation}
	i \Delta(P) = \int d^4 x\,e^{-i P \cdot x} \bra{\Omega_0} \mathcal{T} \Psi(x) \overline{\Psi}(0) \ket{\Omega_0} 
	\end{equation}
Making use of the nucleon free field propagator, a perturbative expansion yields
\begin{equation}
	i \Delta(P) 
	= \frac{i}{\slashed P - m_0 + i 0^+} 
	\sum_{n=0}^{\infty} \left( -i  \Sigma(P) \frac{i}{\slashed P - m_0 + i 0^+} \right)^n
	= \frac{i}{\slashed P - m_0 - \Sigma(P) + i 0^+}
	\label{eq-nucproppert}
\end{equation}
The \terminol{self-energy} $\Sigma(P)$ is the amplitude of all one-particle-irreducible diagrams of a propagating nucleon. Lorentz invariance requires that $\Sigma(P)$ can be written as
\begin{equation}
	\Sigma(P) = m_0\, g(P^2) - \slashed{P}\, f(P^2) 
	\label{eq-selfendecomp}
	\end{equation}
The nucleon is on the mass shell when $\slashed P - m_0 - \Sigma(P)$ in eq. (\ref{eq-nucproppert}) is a singular Dirac matrix, and at rest when $P = (m_N,0,0,0)^T$. Hence $m_N \gamma^0 \left( 1 + f(m_N^2) \right) - m_0 \left( 1 + g(m_N^2) \right)$ must be singular. The negative solution belongs to the anti-particle and is ignored. Then
\begin{equation}
	m_N \left( 1 + f(m_N^2) \right) = m_0 \left( 1 + g(m_N^2) \right)
	\label{eq-nucmasseq}
	\end{equation}
Thus the nucleon mass is calculated in the following steps:
\begin{itemize}
	\item Evaluate the self energy $\Sigma(p)$ to the desired order in $p$.
	\item Renormalize loop diagrams in $\Sigma(p)$ using infrared regularization.
	\item Decompose $\Sigma(p)$ according to eq. (\ref{eq-selfendecomp}).
	\item Solve eq. (\ref{eq-nucmasseq}) for $m_N$ at the desired order in $p$.
\end{itemize}

Here, we study the nucleon mass formulae up to order $p^4$ given in \cite{PHW04}. Using the Lagrangian as quoted in section \ref{sec-powercnt}, the authors of \cite{PHW04} find the following mass formulae:
\begin{itemize}
	\item to order $p^2$ (\terminol{leading order}, \terminol{LO}):
	\begin{eqnarray}
		m_N^{(\leq 2)}&=&m_0-4c_1 \overline{m}_\pi^2~, \label{eq-massp2} 
		\end{eqnarray}
	\item to order $p^3$, (\terminol{next to leading order}, \terminol{NLO}):
	\begin{eqnarray}
		m_N^{(\leq 3)}&=&m_0-4c_1 \overline{m}_\pi^2+ \left[e_1^{(3)}(\lambda)+\frac{3 (g_A^0)^2}{64 \pi^2 (f_\pi^0)^2 m_0 }(1-2 \ln{\frac{\overline{m}_\pi}{\lambda}})\right]\overline{m}_\pi^4\nonumber\\&& -\frac{3(g_A^0)^2}{16 \pi^2 (f_\pi^0)^2}\, \overline{m}_\pi^3\, \sqrt{1-\frac{\overline{m}_\pi^2}{4 m_0^2}}\arccos\left({-\frac{\overline{m}_\pi}{2 m_0}}\right). \label{eq-massp3} 
	\end{eqnarray} 
	\item to order $p^4$ (\terminol{next to next to leading order}, \terminol{NNLO}), truncated at $\overline{m}_\pi^5$:
	\begin{eqnarray}
		m_N^{(\leq 4\cdot)}
		&=& m_0-4c_1 \overline{m}_\pi^2-\frac{3 (g_A^0)^2}{32 \pi (f_\pi^0)^2 }\overline{m}_\pi^3\nonumber\\
		&+& \left[ e_1^{(4\cdot)}(\lambda)-\frac{3}{64\pi^2 (f_\pi^0)^2}\left(\frac{(g_A^0)^2}{m_0}-\frac{c_2}{2}\right) \right. \nonumber \\
		&-& \left. \frac{3}{32 \pi^2 (f_\pi^0)^2 }\left(\frac{(g_A^0)^2}{m_0} - 8 c_1+c_2+4c_3\right)\ln{\frac{\overline{m}_\pi}{\lambda}}
		\right]\,\overline{m}_\pi^4 \nonumber\\
		&+& \frac{3 (g_A^0)^2}{256 \pi (f_\pi^0)^2 m_0^2} \overline{m}_\pi^5 + {\cal{O}}(\overline{m}_\pi^6)
		\label{eq-massp4}
		\end{eqnarray}
	\end{itemize}

\begin{figure}[tbh]
	\centering

	\unitlength=1mm
	\begin{fmffile}{msd}
	\fmfset{arrow_len}{3mm}
	\fmfset{arrow_ang}{15}

	\begin{tabular}{ccc}

		\begin{fmfgraph*}(35,13)
		\fmfleft{l}
		\fmfright{r}
		\fmfforce{(0.0w,0.0h)}{l}
		\fmfforce{(1.0w,0.0h)}{r}
		\fmfforce{(0.2w,0.0h)}{ol}
		\fmfforce{(0.8w,0.0h)}{or}
		\fmf{fermion}{l,ol}
		\fmf{fermion,label=$N$,l.side=right}{ol,or}
		\fmf{fermion}{or,r}
		\fmf{dashes,left=1,tension=0.5,label=$\pi$,l.side=right}{ol,or}
		\fmfdot{ol,or}
		\end{fmfgraph*}
		&

		\begin{fmfgraph*}(35,13)
		\fmfleft{l}
		\fmfright{r}
		\fmfforce{(0.0w,0.0h)}{l}
		\fmfforce{(1.0w,0.0h)}{r}
		\fmfforce{(0.5w,0.0h)}{d}
		\fmf{fermion}{l,d,r}
		\fmf{dashes,right=90,tension=.8,label=$\pi$,l.side=left}{d,d}
		\fmfv{d.shape=diamond,d.filled=full,d.size=4thick}{d}
		\end{fmfgraph*}
		& 

		\begin{fmfgraph*}(35,13)
		\fmfleft{l}
		\fmfright{r}
		\fmfforce{(0.0w,0.0h)}{l}
		\fmfforce{(1.0w,0.0h)}{r}
		\fmfforce{(0.2w,0.0h)}{ol}
		\fmfforce{(0.8w,0.0h)}{or}
		\fmfforce{(0.5w,0.0h)}{om}
		\fmf{fermion}{l,ol,om,or,r}
		\fmf{fermion,label=$N$,l.side=right}{ol,om,or}
		\fmf{dashes,left=1,tension=0.5,label=$\pi$,l.side=right}{ol,or}
		\fmfdot{ol,or}
		\fmfv{d.shape=diamond,d.filled=full,d.size=4thick}{om}
		\end{fmfgraph*}
		\vspace{14pt} \\
		(a) & (b) & (c)
		\end{tabular}
	\end{fmffile}

	\caption{One-loop graphs of order $p^3$ (a) and $p^4$ (b,c) contributing to the nucleon self-energy. The solid dot denotes a vertex from $\mathcal{L}_{\pi N}^{(1)}$, the diamond a vertex from $\mathcal{L}_{\pi N}^{(2)}$.}
	\label{fig-massdiag}
	\end{figure}
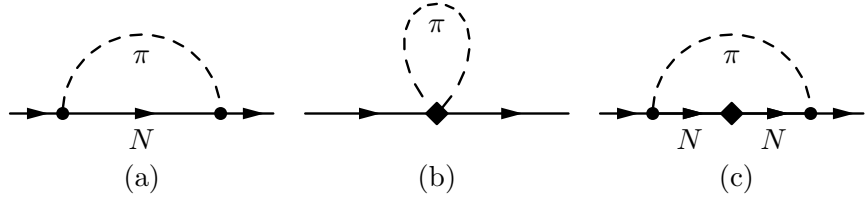

Thanks to the application of the renormalization procedure described above, all these formulae are independent of the renormalization scale $\lambda$. As we have seen, this may require the inclusion of diagrams from higher chiral order to provide appropriate counterterms. Indeed, at order $p^3$, some linear combination of $e_{38}$, $e_{115}$ and $e_{116}$ from $\mathcal{L}_{\pi N}^{(4)}$ is needed to compensate the divergence of the loop in fig. \ref{fig-massdiag} (a). We have called the corresponding combined renormalized coupling constant $e_1^{(3)}(\lambda)$.
At order $p^4$, contributions from $\mathcal{L}_{\pi N}^{(4)}$ are fully included, and result in a term proportional to $e_1 \equiv - 16 e_{38} - 2 e_{115} - 2 e_{116}$, which acts as a counterterm to all three diagrams in fig. \ref{fig-massdiag}. We call its finite contribution $e_1^{(4)}(\lambda)$. The necessity to include counter terms at order $p^6$ would have arisen if the authors of \cite{PHW04} had not chosen to truncate their formula at order $\overline{m}_\pi^5$, which has been checked to be an adequate approximation.

Note that  even though $e_1^{(3)}(\lambda)$ and $e_1^{(4)}(\lambda)$ appear at the same order in the mass formulae, they may belong to different combinations of coupling constants and exhibit a different $\lambda$-dependence. This will be a complication in our convergence analysis, see section \ref{sec-conv}.

\section{Pion-Nucleon Sigma-Term $\sigma_N$}

A quantity of interest closely related to the nucleon mass is the \terminol{pion-nucleon sigma-term}
\begin{equation}
	\sigma_N \equiv \bra{ N}\ m_u \bar u u + m_d \bar d d\ \ket{N}
	\end{equation}
It is a measure of explicit symmetry breaking, giving the contribution of the light quark mass terms to the nucleon mass. 

From the Feynman-Hellmann theorem, also employed in \cite{PHW04}, a reasonably accurate expression for $\sigma_N$ in
$\chi$PT can be obtained via 
\begin{equation}
	\sigma_N \approx m_\pi^2\; \frac{ \partial m_N }{ \partial m_\pi^2 }
	\label{eq-sigmaterm}
	\end{equation}

Thus all we have to do to produce a value of $\sigma_N$ is to form a derivative of our nucleon mass formula at the physical pion mass $m_\pi^\text{phys}$. The result can be compared to values in the literature.

\section{Other Frameworks}

Chiral perturbation theory is a very flexible tool. The choice of fields as well as the method of evaluation can be adapted to the problem under study. 

For an analysis of the quark mass dependence of the axial-vector coupling $g_A$, for example, the authors of ref. \cite{HPW03} find that it is vital to explicitly include the $\Delta(1230)$ nucleon excitation. They utilize HB$\chi$PT with extra fields encoding $\Delta(1230)$, a framework called non-relativistic \terminol{Small Scale Expansion} (\terminol{SSE}) \cite{HHK98}. The expansion parameters, collectively called ``small~scale''~$\epsilon$, now include the mass difference $m_\Delta - m_N \approx 0.234\units{GeV}$ of the $\Delta(1230)$ resonance and the nucleon. 

$\chi$PT can also be adapted to the special needs of lattice extrapolation. For instance, converting meson loop integrals into finite sums can account for the effects generated by the finite box size of the lattice volume, see section \ref{sec-voleff}. Even the side effects of a particular lattice action can be taken into account, like taste violations from staggered Fermions, see e.g. ref. \cite{AuMeson}. 

\chapter{Basics of Lattice Field Theory}

The following chapter aims to give a quick introduction to the principles, terminology, advantages and deficiencies of modern lattice theory useful to know when making use of lattice results. The focus is on clarity rather than on completeness and accuracy. For details, consult refs. \cite{Roth,Le98}.

\section{Philosophy}

Lattice Field Theory is a way of evaluating Quantum Field Theory numerically with ``brute force'' -- but it is more: It is a way of understanding Quantum Field Theory as such. 

The Lagrangian of any continuum Quantum Field Theory contains fields $\phi(x)$ and their derivatives $\partial_\mu \phi(x)$. Remember how a derivative is defined:
\begin{equation}
	\partial_\mu \phi(x) \equiv \lim_{a \rightarrow 0} \frac{ \phi(x + a e_\mu) - \phi(x) }{a} 
	\end{equation}
where $e_\mu$ is a base vector of unit length ($x = x^\mu e_\mu$). In the following, we abbreviate $a e_\mu = \hat \mu$.
The definition of the derivative is not unique. We could equally have chosen the \terminol{central difference}
\begin{equation}
	\partial_\mu \phi(x) 
	\equiv \lim_{a \rightarrow 0} \frac{ \phi(x + \hat \mu) - \phi(x - \hat \mu)}{2a} 
	\end{equation}
At finite $a$, the different representations have certain numerical advantages or disadvantages. For example, for a smooth field $\phi$ that can be represented as a Taylor series around $x$, we have
\begin{equation}
	\partial_\mu \phi(x) = \frac{ \phi(x + \hat \mu) - \phi(x) }{a} + \mathcal{O}(a) \ , \quad
	\partial_\mu \phi(x) = \frac{ \phi(x + \hat \mu) - \phi(x- \hat \mu) }{2a} + \mathcal{O}(a^2)  \quad
	\end{equation}
The central difference converges at a higher order, but it extends over three points: $x-\hat \mu$, $x$ and $x + \hat \mu$.

Lattice Field Theory constitutes a formulation of the theory \emph{before} the continuum limit is taken. 
Just like our trivial example above illustrates, a number of such formulations is possible, each one with its own strengths and weaknesses during numerical evaluation.

\section{Principle}

\begin{figure}[htb]
	\centering
	\includegraphics[width=0.6\textwidth]{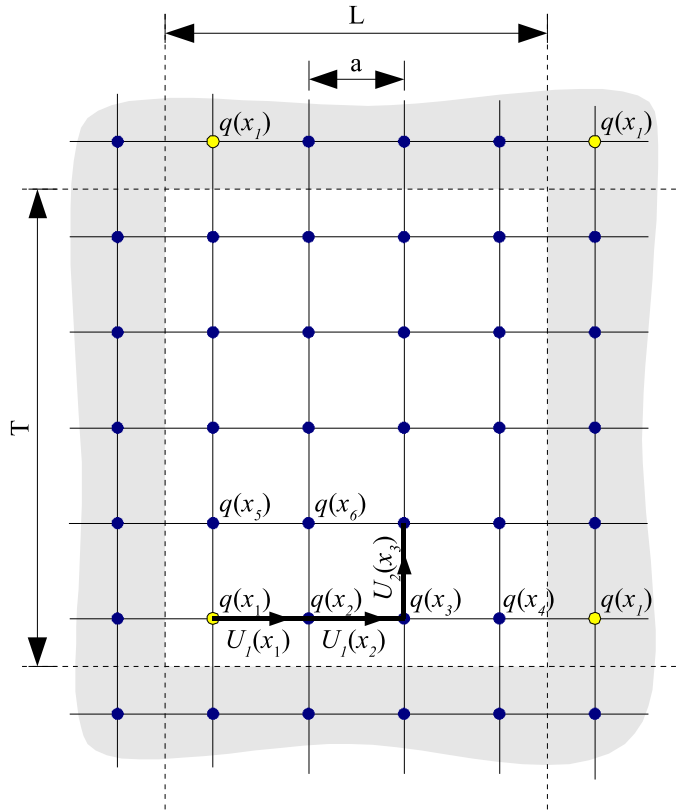}
	\caption{Lattice in $1+1$ dimensions.}
	\label{fig-thelattice}
	\end{figure}

In the path integral formalism of a continuous field theory, the expectation value of an observable $O[\phi,\partial \phi]$ is
\begin{equation}
	\braket{O} 
	= \frac{\int \mathcal{D}\phi\ O[\phi,\partial \phi] \exp( i S[\phi,\partial \phi] )}
	  {\int \mathcal{D}\phi\ \exp( i S[\phi,\partial \phi] )}
	\label{eq-pathintegral}
	\end{equation}
where the action is
\begin{equation}
	S[\phi,\partial \phi] \equiv \int d x^0 \int dx^1 \int dx^2 \int dx^3 \
		\mathcal{L}\big(\phi(x),\partial \phi(x)\big)
	\label{eq-action}
	\end{equation}
Eq. (\ref{eq-pathintegral}) almost has the form of a weighted average over field configurations $\phi$. The exponential $\exp( i S[\phi,\partial \phi] )$ yields a wildly oscillating phase factor except in the proximity of stationary configurations. For a numerical evaluation of the path integral this behavior is undesirable. Therefore, we perform a Wick-rotation of space-time in eq. \ref{eq-action}. The integration path $-\infty \ldots \infty$ of the time coordinate $x^0$ is rotated to $-i \infty \ldots i \infty$. This leaves the value of the integral invariant as long as the integrand has no poles in the area swept over by the rotating integration path. Equivalently, we can replace $x^0$ by ``imaginary time'' $x^4 \equiv i x^0$. In terms of the Euclidean action
\begin{equation}
	S_E[\phi,\partial \phi] \equiv  \int dx^1 \int dx^2 \int dx^3 \int d x^4\
	\mathcal{L}\big(\phi(x),\partial \phi(x)\big)
	\label{eq-euclaction}
	\end{equation}
the path integral reads 
\begin{equation}
	\braket{O} 
	= \frac{\int \mathcal{D}\phi\ O[\phi,\partial \phi] \exp( - S_E[\phi,\partial \phi] )}
	  {\int \mathcal{D}\phi\ \exp( -S_E[\phi,\partial \phi] )}
	\label{eq-eupathintegral}
	\end{equation}
Typically, modern lattice calculations work on a uniform lattice of lattice spacing $a$ :
\begin{equation}
	\big\{ x_1, x_2, x_3, ... \big\}
	 = \big\{\ (a n^1, a n^2, a n^3, a n^4)^T \ \big|\ n^1,n^2,n^3,n^4 \in \mathbb{Z}\ \big \}
	\end{equation}
(Here the lower index identifies a lattice site; it is not to be confused with a Lorentz index). The action is now approximated in terms of field values on the lattice sites $x_n$:
\begin{equation}
	\mathcal{S}_E\big[\phi,\partial \phi\big] \rightarrow 
	\mathcal{S}_\text{lat}\big( \phi(x_1), \phi(x_2), \phi(x_3), ...  \big)
	\end{equation}
How this is done explicitly is discussed in the following sections.
On the lattice, the path integral becomes
\begin{equation}
	\braket{O} 
	= \frac{\int d\phi(x_1) \int d\phi(x_2) \cdots\ O\big( \phi(x_1),\phi(x_2),... \big)\ 
	  \exp\left( - S_\text{lat}\big( \phi(x_1),\phi(x_2),... \big) \right)}
	  {\int d\phi(x_1) \int d\phi(x_2) \cdots\ 
	  \exp \left( - S_\text{lat}\big( \phi(x_1),\phi(x_2),... \big) \right)}
	\label{eq-latpathintegral}
	\end{equation}
So far, there are still (countably) infinitely many integrals and field variables $\phi(x_n)$ in the expression.
Since computers can only work with a finite number of values, one applies periodic boundary conditions\footnote{or, as the case may be, antiperiodic boundary conditions}:
\begin{equation}
	\phi(x_n+L_\text{lat}\hat\mu) = \phi(x_n) \text{ for } \mu=1..3 \qquad
	\phi(x_n+T_\text{lat}\hat\mu) = \phi(x_n) \text{ for } \mu=4
	\end{equation}
Typical modern lattices have $L_\text{lat}\approx30$, $T_\text{lat} \approx 60$. We abbreviate $L\equiv a L_\text{lat}$, $T \equiv a T_\text{lat}$. All in all, the calculation takes place in a box of the size $L \times L \times L \times T$ in discrete Euclidean space-time, compare fig. \ref{fig-thelattice}. 

Note that in Fourier space, only wave functions $\phi(x) \propto \exp(i k_\mu x^\mu)$ with discrete momenta $k$ can be represented on the lattice. Due to periodicity and finite lattice spacing  
\footnote{wave functions $\phi(x) \propto \exp(i k_\mu x^\mu)$ and 
$\phi(x) \propto \exp(i (k_\mu+2\pi/a) x^\mu)$ have the same representation on the lattice. We are interested in the low energy domain; large momenta should be cut off anyway. Therefore we interpret this ambiguity as a boundary on our momenta.}
\begin{equation}
	k \in \left( \frac{2\pi\mathbb{Z}}{L}, \frac{2\pi\mathbb{Z}}{L}, \frac{2\pi\mathbb{Z}}{L}, \frac{2\pi\mathbb{Z}}{T} \right)^T , \qquad
	-\frac{\pi}{a} < k_\mu \leq \frac{\pi}{a}
	\label{eq-latticemomenta}
	\end{equation}

As we have seen in section \ref{sec-renorm}, loop integrals of the form $\int d^4 k$ in a continuum theory can be divergent. Working with a finite set of momenta, the lattice imposes automatically both an \terminol{ultraviolet} and an \terminol{infrared} cutoff: Since integrals appear as discrete and finite sums over all allowed momenta, see eq. (\ref{eq-latticemomenta}), very small and very large (Euclidean) momenta do not contribute. Therefore, all quantities on the lattice are automatically finite. Nevertheless, they need to be interpreted within this \terminol{lattice regularization scheme}.
	
\section{Free Fermions}

\begin{figure}[htb]
	\centering
	a) \topalignbox{\includegraphics[width=0.35\textwidth]{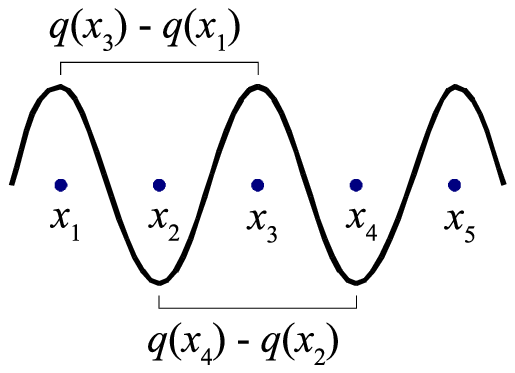}} \hfill
	b) \topalignbox{\includegraphics[width=0.35\textwidth]{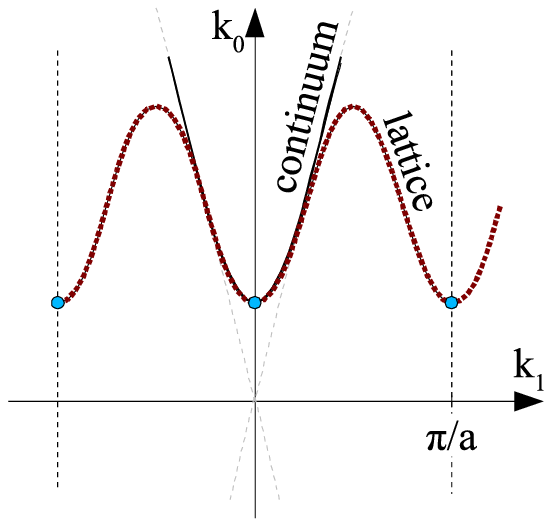}} \hfill{} 
	\caption{Fermion doubling problem: a) The central difference is blind to a wave with wavevector $\pi/a$.
	b) Dispersion relation in the continuum and on the lattice for simple Fermions.}
	\label{fig-fdoubling}
	\end{figure}

To illustrate how the discretization of the action is done, consider the action of free Fermions\footnote{We ignore here details of notation in Euclidean space}
\begin{equation}
	S_F = \int d^4  x\ \Big( \overline{q}(x)\; \gamma_\mu \partial_\mu\; q(x) + \overline{q}(x)\; m\; q(x) \Big)
	\end{equation}
Replacing the integral by a sum over all lattice sites $n$ and expressing the derivative as a central difference yields
\begin{equation}
	S^F_\text{lat} = a^4 \sum_n \left( \sum_\mu \left( \frac{1}{2 a}\; \overline{q}(x_n)\; \gamma_\mu\; \left( q(x_n+\mu) - q(x_n-\mu) \right) \right)
	+ \overline{q}(x_n)\; m\; q(x_n) \right) 
	\end{equation}
Writing Dirac indices as $\alpha,\beta,...$, the Fermionic action is of the general form
\begin{equation}
	S^F_\text{lat} = \sum_{n,m}\; \overline{q}_\alpha(x_n)\ K^{nm}_{\alpha \beta}\ q_\beta(x_m)
	\label{eq-freefermionact}
	\end{equation}
For this simple lattice action, the integrals in eq. (\ref{eq-latpathintegral}) can be carried out analytically, and one can check whether the dispersion relation approximates the physical one. For small momenta $k_\mu \ll \pi/a$, this is the case. Unfortunately, at this point, one has to deal with an unpleasant property of the central difference used in the construction: Consider the configuration $q(x) \propto \exp(i k_\mu x^\mu)$ for a momentum on the ``Brillouin zone'', for example with $\vec k = (\pi/a,0,0)$. Here, the central difference in the action vanishes for $\mu=1..3$: $q(x_n+\mu) - q(x_n-\mu) = 0$. This is due to the fact that the central difference works only with lattices sites $2a$ apart, and therefore cannot deal properly with momenta larger than $\pi/2a$, see fig. \ref{fig-fdoubling}a. As a result, there are spurious low energy modes of Fermions with non-zero momentum on the Brillouin zone, see fig. \ref{fig-fdoubling}b. In a lattice calculation, they would be excited along with the ``true'' Fermions. This is the \terminol{Fermion doubling problem}. The workarounds involve compromises:
	
\begin{figure}[htb]
	\centering
	a) \topalignbox{\includegraphics[width=0.35\textwidth]{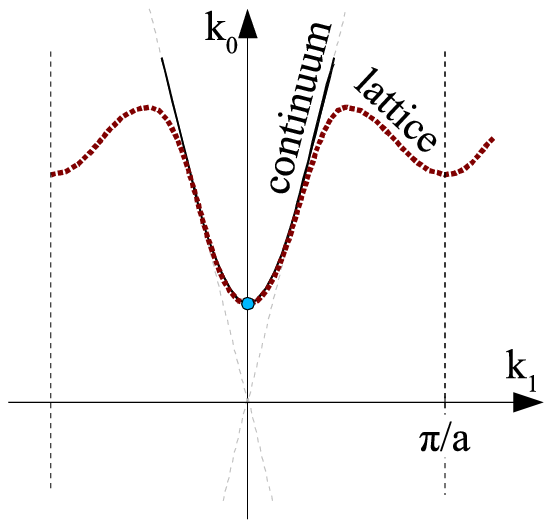}} \hfill
	b) \topalignbox{\includegraphics[width=0.30\textwidth]{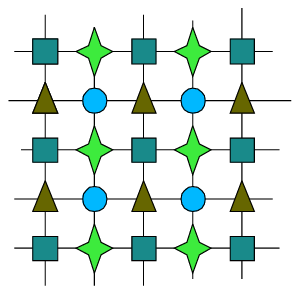}}
	\caption{Circumventing Fermion doubling: a) dispersion relation for Wilson Fermions b) staggered lattice made up of coarser lattices of lattice spacing $2a$, as indicated by the different types of nodes.}
	\label{fig-counterfdoubling}

\end{figure}	

\begin{itemize}
	\item
	One could use a non-central derivative, but that destroys Hermiticity of $K^{nm}_{\alpha \beta}$.
	\item
	For \terminol{Wilson Fermions}, an additional term 
	\begin{equation}
		\sum_\mu \frac{r}{2a}\; \overline{q}(x)\; \big( q(x+\hat \mu) - 2 q(x) + q(x-\hat \mu) \big)\
		\equiv a\, \frac{r}{2}\, \overline{q}(x)\, \square\, q(x)
		\end{equation}
	is added to the Lagrangian, its strength being controlled by the parameter $r$.
	The operator $\square$ is $\partial_\mu \partial_\mu$ in the continuum limit.
	The Wilson term treats Fermions like scalar particles, and vanishes in the continuum limit. It lifts the energy of the spurious modes on the Brillouin zone, see fig. \ref{fig-counterfdoubling}a. It also shifts the quark mass, which is now no longer $m$.
	Unfortunately, Wilson Fermions do not exhibit chiral symmetry for vanishing quark mass. A number of modifications have been found to restore chiral symmetry, though at considerable computational cost.
	\item
	\terminol{Staggered Fermions} by Kogut and Susskind address the Brillouin-zone problem by assembling the lattice from overlapping coarser lattices of lattice spacing $2a$, as illustrated in fig. \ref{fig-counterfdoubling}b. Fermionic degrees of freedom on each of the coarser lattices exhibit the correct dispersion relation. However, this approach requires the introduction of a new unphysical quantum number: there are now four different quark ``tastes''. Therefore, staggered Fermion calculations exhibit ``taste splittings'' for example in the mass spectrum. The splittings should vanish for $a \rightarrow 0$ and are therefore an indicator of systematic errors.
\end{itemize}

\section{Gluons -- The Gauge Field}

Introducing gauge symmetry of the QCD Lagrangian starts out with the wish that one be able to choose the ``coordinate system'' of color freely at every point in space-time. That is the meaning of the transformation law eq. (\ref{eq-gaugetrafo}).  It is possible to implement exact color gauge symmetry on the lattice, even at finite lattice spacing $a$. This justifies the name ``lattice \emph{gauge} theory''. The construction procedure on the lattice is even more straightforward than in the continuum. Note that the resulting gauge invariant lattice action differs from a lattice action derived by naive discretization of the continuum action.  

Let us neglect flavor here, and write the quark fields as color vectors $q = (q_\za)_{\za=1..3}^T$. We have seen that \emph{free} quarks on the lattice have an action of the form eq. (\ref{eq-freefermionact}).\footnote{Spelling out the color indices explicitly, it is a sum of color singlets of quark field bilinears $\overline{q}_{\alpha \za}(x_n) q_{\beta \za}(x_m)$, relating the field value on one lattice site to the field value on another.} We must simply ensure that Fermion bilinears remain invariant under gauge transformations 
$q(x_n) \rightarrow W(x_n)\; q(x_n)$. 
For two adjacent lattice sites $x_n$ and $x_n+\hat \mu$, this can be achieved by introduction of a transformation matrix $U_\mu(x_n) \in \mathrm{SU}(3)$, which specifies how the color coordinate systems of the two lattice sites are rotated with respect to each other. $U_\mu(x_n)$ is called a \terminol{link variable}, and is in general depicted as an arrow connecting $x_n$ and $x_n+\hat\mu$. Its transformation property 
\begin{equation}
	U_\mu(x_n) \rightarrow W(x_n)\; U_\mu(x_n)\; W^\dagger(x_n+\hat\mu) \equiv \tilde U_\mu(x_n) 
	\end{equation}
guarantees that an appropriately modified bilinear remains invariant:
\begin{equation}
	X \equiv \overline{q}_\alpha(x_n)\; U_\mu(x_n)\; q_\beta(x_n+\hat\mu) \rightarrow 
	\overline{q}_\alpha(x_n)\; W^\dagger(x_n)\; \tilde U_\mu(x_n) \; W(x_n+\hat\mu)\; q_\beta(x_n+\hat\mu) = X
	\end{equation}
Replacing the quark bilinears $\overline{q}_{\alpha}(x_n) q_{\beta}(x_m)$ in eq. (\ref{eq-freefermionact}) by their gauge invariant counterparts $\overline{q}_\alpha(x_n)\; U_\mu(x_n)\; q_\beta(x_n+\hat\mu)$ and embedding the link variables $U\equiv \big( U_\mu(x_n) \big)_{\mu,n}$ into the coefficient matrix $K^{nm}_{\alpha \beta}$, the gauge invariant Fermionic action attains the form
\begin{equation}
	S^F_\text{lat} = \sum_{n,m,\alpha,\beta}\; \overline{q}_\alpha(x_n)\ K^{nm}_{\alpha \beta}[U]\ q_\beta(x_m)
	\label{eq-ginvfermionact}
	\end{equation}
\parpic[r][r]{
	\includegraphics[]{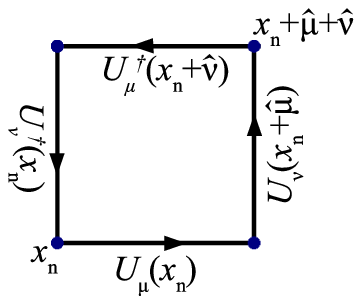} 
	}
In order to relate color coordinate systems of nonadjacent lattice sites, we can form products of the link variables. For example, both $U_\mu(x_n) U_\nu(x_n+\hat\mu)$ and $U_\nu(x_n) U_\mu(x_n+\hat\nu)$ describe a rotation of the color coordinate system when going from lattice site $x_n$ to $x_n+\hat\mu+\hat\nu$. Initially, one might require that both rotations be equal. Consider the product of link variables 
\begin{equation}
	U_{\mu \nu}(x_n) \equiv U_\mu(x_n) U_\nu(x_n+\hat\mu) U^\dagger_\mu(x_n+\hat\nu) U^\dagger_\nu(x_n) 
	\end{equation}
which describes a closed path (a ``Wilson loop'') around a square of lattice sites of side length $a$ (a ``plaquette''). With our requirement, we should have $U_{\mu \nu}(x_n)=\Eins$. Actually, demanding this would take us back to interaction-free Fermions. So let us allow $U_{\mu \nu}(x_n) \neq \Eins$. In analogy to parallel transport in general relativity, this means that color space is internally curved. The curvature causes tension, i.e., a restoring force, if we add the following term to the action
\begin{equation}
	S^G_\text{lat} 
	= \frac{\beta}{6} \sum_{\square} \Tr \left( 2 \cdot  \Eins - U_{\mu \nu}(x_n) - U_{\mu \nu}^\dagger(x_n) \right)
	\end{equation}
where $\beta$ is a constant, and $\sum_{\square}$ is a sum over all plaquettes on the entire lattice. It turns out that in the continuum theory a link variables $U_\mu(x_n)$ correspond to an object
\begin{equation}
	U(x_n,x_n+\mu) = \mathcal{P} \exp \left( i g \frac{\lambda_\zA}{2}\  \int_{x_n}^{x_n+\hat\mu} dz\; A_{\mu,\zA}(z) \right)
	\end{equation}
where $\mathcal{P}$ denotes a path ordering. Thus $U_\mu(x_n)$ actually encodes the gauge fields $A_{\mu,\zA}$. Furthermore, $S^G_\text{lat}$ reproduces the kinetic term of the gluon field:
\begin{equation}
	S^G_\text{lat} \xrightarrow{a \rightarrow 0} 
	\frac{1}{4} \mathcal G_{\mu \nu\, \zA}\ \mathcal G_{\mu \nu\,\zA} 
	\qquad \text{ for } \beta \equiv \frac{6}{g^2} 
	\end{equation}
We now have all the ingredients for a discretized gauge invariant QCD action 
$S_\text{lat} = S^F_\text{lat} + S^G_\text{lat}$.

As already mentioned, it is possible to find \terminol{improved actions} which converge more quickly to the continuum limit, e.g., with a discretization error $\mathcal{O}(a^2)$. 

\section{The Calculation Scheme}

With the lattice action, the path integral looks like
\begin{equation}
	\braket{O} \propto
	\int [dU]\int [dq]\int [d\overline{q}]\ O\big[U,\overline{q},q]\ 
	  \exp\left( - S^G_\text{lat} - \sum_{n,m}\; \overline{q}_\alpha(x_n)\ K^{nm}_{\alpha \beta}[U]\ q_\beta(x_m) \right)
	\label{eq-fermlatpathintegral}
	\end{equation}
Here, $[dU]$,$[dq]$,$[d\overline{q}]$ symbolize integration over all degrees of freedom at all lattice sites.
It is possible to directly integrate out Fermions, treating the Fermions as Grassmann variables. The result is of the form
\begin{equation}
	\braket{O} \propto
	\int [dU]\; \tilde O\big[U]\ 
	  \exp\left( - S^G_\text{lat} \right)\; \det(K[U]) 
	\label{eq-latpathintegral3}
	\end{equation}
The remaining integral still has a huge number of degrees of freedom of the order of $L_\text{lat}^3 T_\text{lat}$. 
Its numerical evaluation is achieved by clever sampling, most prominently by random ``Monte Carlo'' methods like the Metropolis algorithm. At each sampling step, the algorithm produces a set of numbers specifying the values of all the link variables $U$ on the lattice. Such a set of values is called a \terminol{gauge configuration}.
In the end, one obtains a whole \terminol{ensemble} $E_U$ of $N$ gauge configurations which contains a configuration $U$ with a probability
\begin{equation}
	P[ U \in E_U ] \propto \exp\left( - S^G_\text{lat} \right)\; \det(K[U])
	\end{equation}
The calculation of the ensemble is very costly. In particular the determinant $\det(K[U])$ is problematic. In the \terminol{quenched approximation} it is simply set to $1$. This is equivalent to neglecting spontaneous quark fluctuations -- sea quarks. Nowadays it is possible to take the Fermion determinant into account. This diploma thesis will make use of results from such \terminol{unquenched} (or \terminol{fully dynamical}) simulations exclusively. Luckily, once an ensemble is calculated, it can be used for the evaluation of many different observables, simply using the sample average
\begin{equation}
	\braket{O} \propto \frac{1}{N} \sum_{U \in E_U} \tilde O\big[U]
	\label{eq-lataverage}
	\end{equation}
The usage of samples and averages leads to statistical errors depending on the ensemble size as $1/\sqrt{N}$.

\section{Extracting Masses}

In order to determine the mass of a hadron, one takes an operator $X^\dagger$ which carries the quantum number of the desired particle. For the pion, one might take 
\begin{equation}
	X^\dagger(\vec p,\tau) = \sum_{\vec x_n} e^{i \vec p\cdot \vec x_n}\; 
	\overline{u}_\za(\vec x_n,\tau) \gamma_5 d_\za(\vec x_n,\tau)
	\end{equation}
which creates a pseudo-scalar color singlet of an anti-down up quark pair. Such an operator generates a whole spectrum of states, but only the pion is of interest, which is the mode of lowest energy.\footnote{In practise, it is advisable to use an operator which is not exactly point like. For such a \terminol{smeared} operator, the high energy contributions are somewhat dampened from the start.} Consider the propagation of the hadron for a certain Euclidean time $\tau$~:
\begin{equation}
	c(\vec p,\tau) \equiv \braket{X(\vec p,\tau)X^\dagger(\vec p,0)} =
	\sum_\alpha
	\bra{\Omega_0} X(\vec p,0)\ket{\alpha} e^{-\; E_{\alpha,\vec p}\; \tau} 
	\bra{\alpha} X^\dagger(\vec p,0)\ket{\Omega_0}
	\end{equation}
In the sum over all intermediate states $\ket{\alpha}$, high energy modes fall off very quickly with $\tau$. In the limit of large $\tau$, only the excitation $\ket{\pi}$ of lowest energy $E_{\pi,\vec p}$ contributes: 
\begin{equation}
	c(\vec p,\tau) \approx \text{const} \cdot e^{- E_{\pi,\vec p} \, \tau} \qquad \text{ for } \tau \rightarrow \infty
	\label{eq-propfalloff}
\end{equation}
The idea is to calculate $c(\vec p,\tau)$ on the lattice and to determine the rest mass $m_\pi = (E_{\pi,\vec p}^2 \nobreak-\nobreak \vec p^2)^{1/2}$ of $\ket{\alpha_\pi}$ from a fit to eq. (\ref{eq-propfalloff}) at large $\tau$. To that end, one substitutes an operator $O = X(\vec p,\tau)X^\dagger(\vec p,0)$ in eq. (\ref{eq-fermlatpathintegral}). Integrating out Fermions analytically, one obtains the operator $\tilde O$ in terms of the gauge fields $U$, which can then be evaluated according to (\ref{eq-lataverage}). Note that due to the periodicity of the lattice in the time direction, $\tau$ must be sufficiently small compared to half of the lattice extent $T$, otherwise the operator $X(\vec p,\tau)$ will ``feel'' the periodic image $X^\dagger(\vec p,T)$ of $X^\dagger(\vec p,0)$.

\section{Finding the Physical Length Scale}

The natural unit of length on the lattice is the lattice spacing $a$. For the calculation, it is practical to work with quantities like fields, masses, and couplings multiplied by an appropriate power of $a$, such that they become dimensionless in terms of natural units ($\hbar = c = 1$). For example, the quark mass is parametrized by $a m$. The fundamental parameters of QCD become dimensionless coupling strengths on the lattice. As an example, collaborations using Wilson Fermions (or improved versions thereof) typically specify $\beta$ and $\kappa \equiv (8r +\nobreak 2 a m)^{-1}$ as their simulation parameters. These two numbers encode the gluonic coupling strength and the quark mass. For the analysis of lattice results, observables must be converted back to units of the real world.
For instance, a pion mass calculation as described in the previous section yields $a m_\pi$. To translate this into $m_\pi$ in units $\mathrm{GeV}$, we need to know the lattice spacing $a$ in physical units.

The determination of $a$ requires the calculation of a suitable reference observable, which can then be compared to an experimental result. Early analyses used the $\rho$ mass as their reference observable. Nowadays, the static quark potential is compared to empirical data from heavy quarkonia, such as $\Upsilon$, a $\overline{b} b$ system. This method is advantageous, because it is independent of valence quark masses, and the lattice calculation is simple \cite{Be00}.\footnote{Basically, the operator involved is a Wilson loop.} From the spectrum of $\overline{b} b$ states one can calculate an effective potential. The same potential can be explored on the lattice by placing two infinitely heavy quarks on the lattice and varying their distance $r$. Typically, a reference length $r_0$ called the \terminol{Sommer scale} is defined in terms of the attractive force $F(r)$ between the quarks according to 
\begin{equation}
	F(r_0) r_0^2 = 1.65
	\label{eq-defsommerscale}
	\end{equation}
According to phenomenological models, $r_0 \cong 0.5 \units{fm}$ \cite{So93}. So if an analysis shows that the condition (\ref{eq-defsommerscale}) is met for heavy quarks that are $r_0/a$ lattice nodes apart, we know that their distance equals $0.5\units{fm}$ in nature.\footnote{Note that the LHPC group prefers a reference scale $r_1$, which is defined via $F(r_1) r_1^2 = 1$.}
Thus a mass $m_X$ is converted to units $\units{GeV}$ via
\begin{equation} 
	\boxed{
		m_X = 
		\frac{(a m_X)}{a} \hbar c
			= (a m_X)\frac{(r_0/a)}{r_0} \hbar c 
			= (a m_X)\frac{(r_0/a)}{0.5\units{fm}} 0.1973 \units{GeV\; fm}
	}
	\label{eq-latconversion}
	\end{equation}
The dimensionless quantities $a m_X$ and $r_0/a$ are those that are commonly tabularized in the literature.
	
\section{Uncertainties and Artefacts in Lattice Data}

In order to make calculations more realistic, larger and finer lattices would be required that demand a calculation power which modern supercomputers do not yet possess. At present, it is 
important to take the following uncertainties and artefacts into consideration when making a physical interpretation of lattice data:
\begin{itemize}
	\item
	Deviations from the physical value due to unphysically large quark masses
	(if not accounted for by matching to a quark mass dependent theory, such as $\chi$PT).
	\item
	Deviations from the physical value due to the finite size of the lattice
	(if not accounted for by a finite size correction within the theoretical framework, e.g., $\chi$PT).
	\item
	Statistical uncertainties from the ensemble average eq. $(\ref{eq-lataverage})$. Usually, lattice groups specify a one-standard-deviation error, and the fluctuations may be assumed to follow a Gaussian distribution.
	\item
	Discretization artefacts due to the finite lattice spacing $a$. Ideally, these are further reduced by continuum extrapolations, e.g. using specially adapted forms of $\chi$PT.
	\item
	Uncertainty about the lattice scale $a$. Lattice collaborations do their best to estimate this error, see e.g. \cite{Au04}. The determination of $a$ involves
	\begin{itemize}
		\item statistical errors from the lattice calculation of the static quark potential,
		\item systematic errors from a continuum extrapolation, 
		\item systematic errors from the use of the quarkonia model,
		\item experimental errors from the quarkonium spectrum,
		\item systematic uncertainties entering due to heavier sea quarks than in nature\footnote{The magnitude of this artifact can be studied to some extent by comparing to the quenched approximation \cite{Be00}.}
		\end{itemize}
\end{itemize}

\chapter{Methods of Statistical Error Analysis}
\label{sec-formalstatistics}

\newcommand{\true}[1]{{#1}^\text{true}}
\newcommand{\est}[1]{{#1}^\text{est}}
\newcommand{\CL}{\text{CL}}

This section explains the statistical concepts necessary for error analysis in a formal manner. It is based on concepts found for example in ref. \cite{EDJS71}. Important aspects of the $\chi^2$ method are given in ref. \cite{minuit}. In order to make the connection to the concrete task clear, the text describes the form of the problem and some basic techniques for the simple example of a fit of the pion mass expansion of the nucleon mass to two-flavor lattice data.

\section{How Statistics Relates Theory to Experiment}

An experiment produces a collection of numbers $\ul{y}$. Because of uncertainties, fluctuations, and other random behavior, $\ul{y}$ represents a sample of a random variable $\ul{Y}$. Together with knowledge about the experimental uncertainties, a theory provides the probability distribution of $\ul{Y}$. Usually, theories depend on parameters $\ul{p}$. Thus the object encoding the experimental predictions of the theory is a probability density $w_Y(\ul{y}|\ul{p})$ for the occurrence of $\ul{y}$ under the assumption of parameters $\ul{p}$. If the theory is true, there exists a ``\terminol{true}'' parameter set $\true{\ul{p}}$, such that $w_Y(\ul{y}|\true{\ul{p}})$ is the probability density of $\ul{Y}$ realized in nature. 

\begin{figure}[ht]
	\centering
	\includegraphics{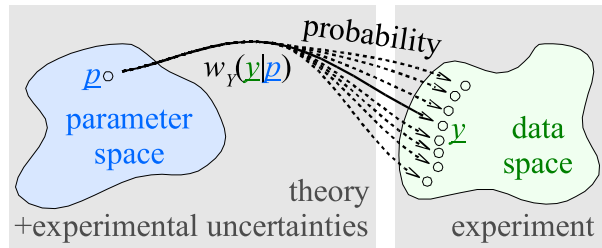}
	\caption{How statistics relates theory to experiment.}
	\label{fig-paramdataspace}
	\end{figure}

The task is to extract an estimate $\est{\ul{p}}(\ul{y})$ of the true parameters $\true{\ul{p}}$ given data samples $\ul{y}$. We would also like to know how precise our estimate is. Finally, we want to assess the plausibility of our original assumption that the theory is true.

\section{Definition of the Problem}
\label{sec-statproblem}

In our simplest case, we take from lattice calculations the pion mass and the nucleon mass associated with it. For $N$ different lattice simulation parameters, we have a set of points
\begin{equation}
	(\ m_\pi = \bar x_i \pm \Delta x_i,\ 
	m_N = \bar{y}_i \pm \Delta y_i\ ) \qquad \text{for\ } i=1..N
\end{equation}
The $\bar{x}_i$ and $\bar{y}_i$ are the numerically calculated values, and the $\Delta x_i$ and $\Delta y_i$ are one-standard-deviation errors. For the moment, let us assume the pion mass uncertainties play no significant role, and that the errors in the nucleon mass are uncorrelated. \par

For our purposes, the lattice data take on the role of experimental data, in the sense that they form a source of information independent of our theoretic calculations within $\chi$PT. \par

Chiral perturbation theory provides a function $m_N(x;\ul{p})$ which predicts the nucleon mass $m_N$ for a given pion mass $x$ and a vector of \terminol{parameters} $\ul{p}=(p_1,..,p_n)^T$, the low energy constants. The first task is to determine the low energy constants and their associated uncertainties from a fit to lattice data. A parameter set $\ul{p}$ that comes close to the truth should yield predictions $f_i(\ul{p}) \equiv m_N(\bar x_i;\ul{p})$ close to the calculated nucleon masses $\bar y_i$ from the lattice. In a symbolic notation, the conditions we want to fulfill optimally by a suitable choice of the parameters $\ul{p}$ are:
\begin{equation}
	\text{``\ } f_i(\ul{p}) \approx \bar{y}_i \pm \Delta y_i \text{\ ''} \qquad
	\text{for\ } i=1..N
	\label{eq-fitcond}
	\end{equation}
Later it will turn out that much more general fitting tasks can be brought into this simple form. Among other things, one can treat uncertainties in the pion mass, perform simultaneous fits and deal with some systematic uncertainties. 

Using the fit conditions (\ref{eq-fitcond}) one can calculate an estimate $\est{\ul{p}}$ and its statistical uncertainty. Plugging the estimate into our formula from $\chi$PT, one obtains the optimal interpolant $m_N(x;\est{\ul{p}})$. The statistical uncertainty of the estimate can be displayed as a band of interpolation functions at a given confidence level. \par

In the following, the statistical problem associated to (\ref{eq-fitcond}) is rendered more precise.

\section{$\chi^2$ and Maximum Likelihood Estimate}
\newcommand{\normq}[1]{\ensuremath{\left\Vert {#1} \right\Vert_2}} 
\newcommand{\norms}[1]{\ensuremath{\left\Vert {#1} \right\Vert_2^2}} 

Let us assume that there are no systematic errors in lattice calculations. Were the lattice calculations performed over and over again, the resulting $\bar{y}_i$ would be scattered around ``true'' values $\true{y}_i$. Thus the data points $\bar{y}_i$ are samples of random variables $Y_i$. For our analysis, we assume that the $Y_i$ are statistically independent and have normal distributions, with expectation values $\true{y}_i$ and standard deviations $\Delta y_i$. Let us also assume that the theory is correct, i.e.,
\begin{equation}
	f_i(\true{p}) = \true{y}_i 
\end{equation}
(Whether this assumption is plausible,i.e., whether the theory appears to be true, can be checked a posteriori). All in all we assume that the data follows a probability distribution \par

\begin{equation}
	w_Y(\ul{y}|\ul{p}) = \prod_{i=1}^N \frac{1}{\sqrt{2 \pi} \Delta y_i} \exp \left(-\frac{(f_i(\ul{p})-\bar{y}_i)^2}{2\ \Delta y_i}  \right)
	\end{equation}

Interpreted as a function of $\ul{p}$, $w_Y$ is called the \terminol{likelihood function}.
It is convenient to introduce normalized quantities $s_i \equiv \bar{y}_i / \Delta y_i$ and $t_i(\ul{p}) \equiv f_i(\ul{p}) / \Delta{y_i}$ as well as
\begin{equation}
	\chi^2(\ul{s}|\ul{p}) \equiv \chi^2(\ul{s},\ul{t}(\ul{p})) \equiv \sum_{i=1}^{N} (t_i(\ul{p}) - s_i)^2 = \norms{ \ul{t}(\ul{p}) - \ul{s} }
	\end{equation}
such that the probability density for the occurrence of lattice results $\ul{s}$ takes the form
\begin{equation}
	w_S(\ul{s}|\ul{p}) = (2 \pi)^{-N/2} \exp \left(-\frac{1}{2}\; \chi^2(\ul{s}|\ul{p}) \right)
	\end{equation}
For given input data $\ul{s}$, one can maximize the likelihood $w(\ul{s}|\ul{p})$ by tuning the parameters $\ul{p}$. The resulting parameter set $\est{\ul{p}}$ is called \terminol{maximum likelihood estimate}. Taking many different sets of input values $\ul{s}$ would reveal that $\est{\ul{p}}$ is scattered around $\true{\ul{p}}$. The numerical determination of  $\est{\ul{p}}$ is done by a minimization of $\chi^2(\ul{s}|\ul{p})$ with respect to $\ul{p}$. 

\section{Construction of Confidence Regions}
\label{sec-confreg}

Now we would like to construct a region $R$ of parameters which is known to contain the true parameters $\true{\ul{p}}$ with a given probability (\terminol{confidence level}) $\CL$. \footnote{In a sense, we are looking for something like a ``probability density $w_P(\ul{p}|\ul{y})$'' of the parameter set $\ul{p}$ given samples $\ul{y}$. Such a density cannot be defined, since $\ul{p}$ is \emph{not} a sample of a probability variable, and no probability measure exists in parameter space. However, the probability density of the estimator $\est{\ul{p}}(\ul{Y})$ is well defined. Therefore, the construction of confidence regions relies on the estimator.} For a given assumption on parameters $\ul{p}$ of the system, we can calculate the  probability density of the estimate $\est{\ul{p}}(\ul{S})$. The first step is to define a region of parameters $\tilde R(\ul{p})$ which contains the estimate $\est{\ul{p}}$ with probability $\CL$ :
\begin{equation}
	P[\est{\ul{p}}(\ul{S}) \in \tilde{R}(\ul{p})] = \CL
	\end{equation}
Actually, we are faced with the inverse problem: $\est{\ul{p}}$ is given, and we need a region $R(\est{\ul{p}})$ which contains the parameters of the system $\true{\ul{p}}$ with probability $\CL$. Defining
\begin{equation}
	R(\est{\ul{p}}) = \Big\lbrace \ul{p} \ \Big\vert \ \est{\ul{p}} \in \tilde{R}(\ul{p}) \Big\rbrace
	\end{equation}
provides a solution to this problem, because $R(\est{\ul{p}})$ fulfills
\begin{equation}
	P \left[ \true{\ul{p}} \in R(\est{\ul{p}}(\ul{S})) \right] = P \left[ \est{\ul{p}}(\ul{S}) \in \tilde{R}(\true{\ul{p}}) \right] = \CL
	\end{equation}
Now the steps above must be carried out explicitly, making use of $\chi^2$. One chooses 
\begin{equation}
	\tilde{R}(\ul{p})
	= \Big\{ \ul{p}' \ \Big\vert\ \chi^2\left(\ \ul{t}(\ul{p}'),\ \ul{t}(\ul{p})\ \right)\ \leq\ \chi^2_{\CL,n}(\ul{p}) \Big\}
	\label{eq-tildeRchi}
	\end{equation}
where $\chi^2_{\CL,n}(\ul{p})$ is a threshold value for $\chi^2$ that must be appropriately chosen
to enclose the desired probability content $\CL$.

\begin{figure}[h]
	\centering{
	\hfill \includegraphics[scale=0.7]{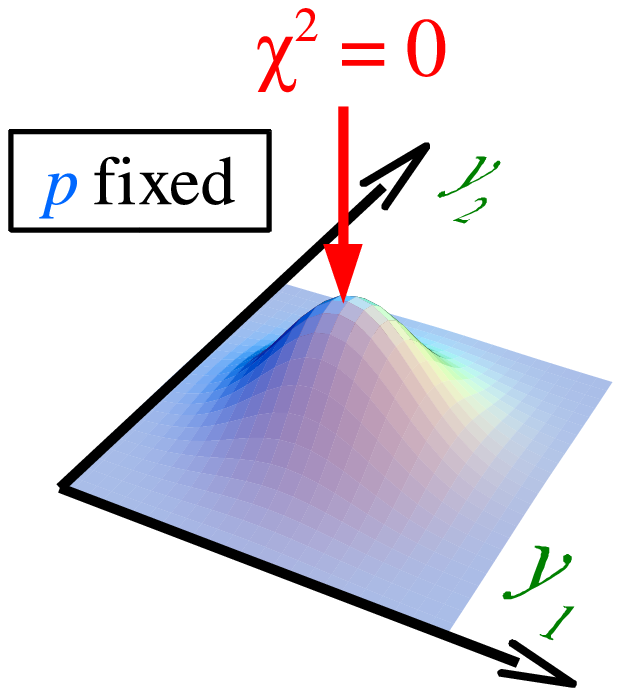} 
	\hfill \includegraphics[scale=0.7]{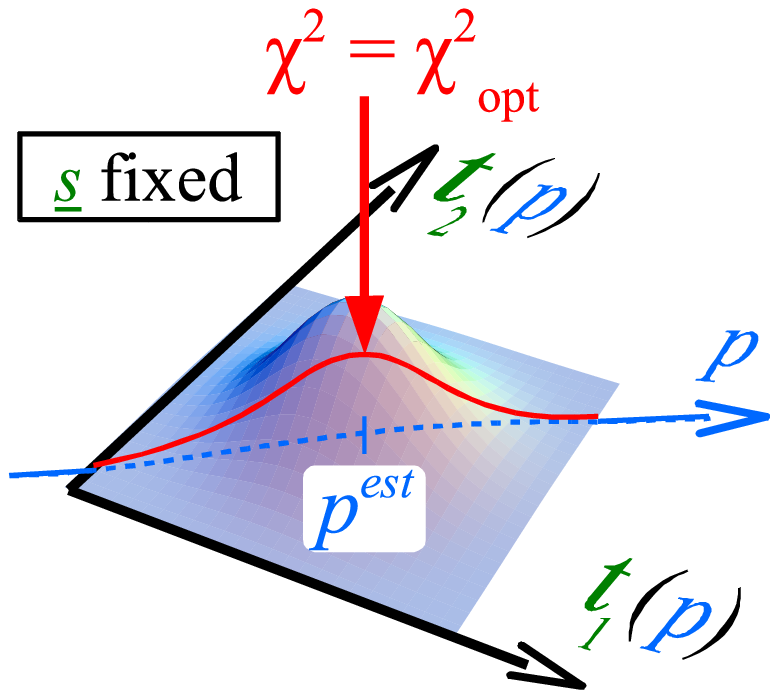} \hfill {} \par }
	\caption{The probability distribution $w_Y(\ul{s}|p)$ for a 2-dimensional data space and a 1-dimensional parameter space.}
	\label{fig-chisqrspace}
	\end{figure}

Then the error band is given by
\begin{equation}
	R(\est{\ul{p}})
	= \left\{ \ul{p} :\  \chi^2\left(\ \ul{t}(\est{\ul{p}}),\ \ul{t}(\ul{p})\ \right)\ \leq\ \chi^2_{CL,n}(\ul{p}) \right\}
	\label{eq-errbandnl}
	\end{equation}

\begin{figure}[h]
	\centering
	\includegraphics[scale=0.7]{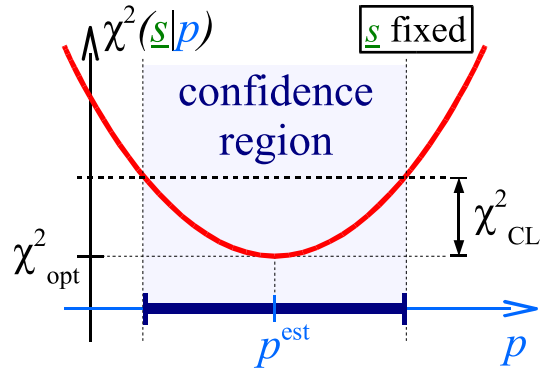} 
	\caption{Determination of the confidence region for a 1-dimensional parameter space.}
	\label{fig-chisqrthresh1}
	\end{figure}

\section{Calculating the $\chi^2$ Threshold}

\begin{figure}[h]
	\centering
	\includegraphics{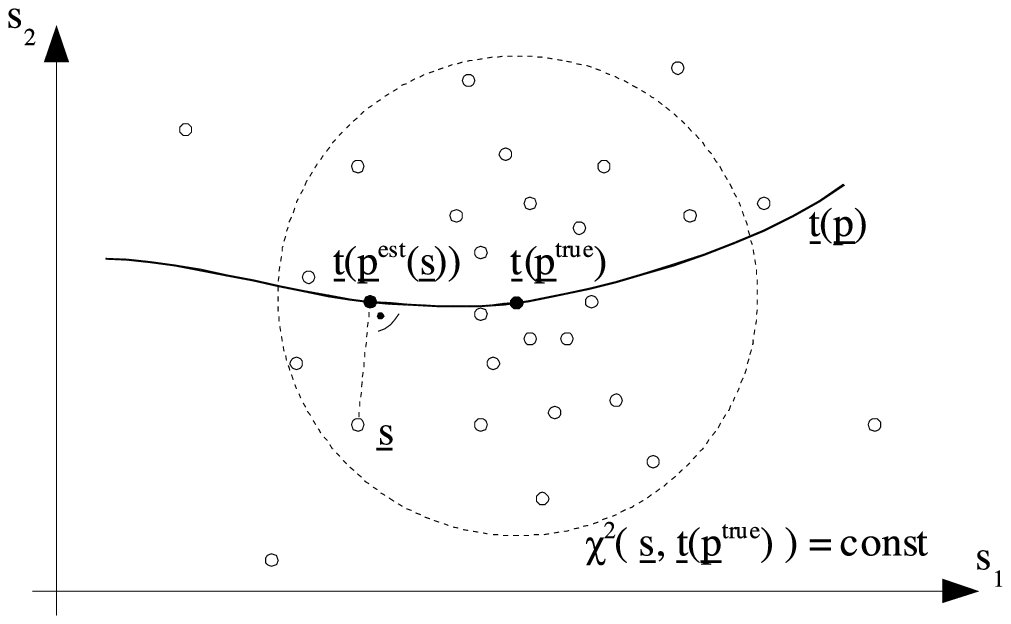}
	\caption{The manifold $\ul{t}(\ul{p})$.}
	\label{fig-chisqrmanifold}
	\end{figure}

We can visualize the way the $\chi^2$ method works by looking at the $N$ dimensional \terminol{data space} (fig. \ref{fig-chisqrmanifold}), where the measured results $\ul{s} \in \mathbb{R}^N$ appear as a point.  In data space, $\chi^2(\ul{a},\ul{b})=\Vert \ul{a}-\ul{b} \Vert_2^2$, i.e. $\chi^2$ simply measures the Euclidean distance between $\ul{a}, \ul{b} \in \mathbb{R}^N$. The function $\ul{t}(\ul{p})$ constitutes an $n$ dimensional manifold (a ``surface'') embedded in data space. 

\begin{figure}[h]
	\centering
	\includegraphics{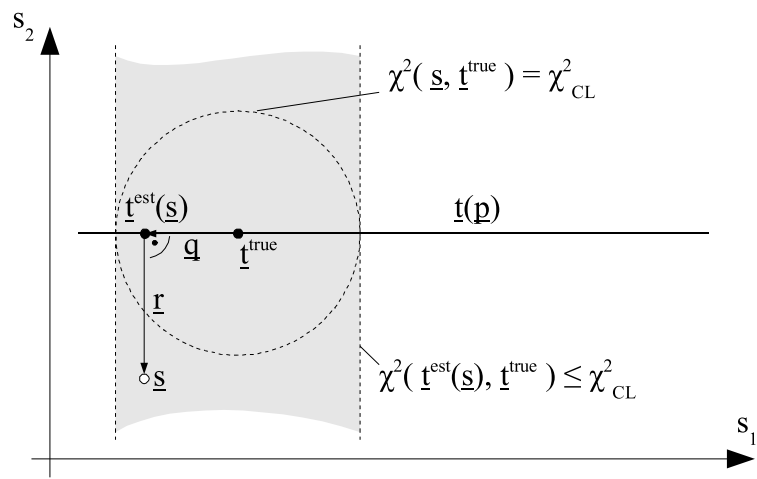}
	\caption{Flat manifold $\ul{t}(\ul{p})$ and $\chi^2$ threshold.}
	\label{fig-chisqrmaniflin}
	\end{figure}

The determination of the threshold $\chi^2_{\CL,n}$ becomes simple if we make the assumption that this manifold is \emph{planar/flat} in the vicinity of $\true{t}=\ul{t}(\true{\ul{p}})$. Note that a sufficient but not necessary criterion for flatness is linearity of $\ul{t}(\ul{p})$. In the applications presented in this thesis, the manifold $\ul{t}(\ul{p})$ is sufficiently flat. The assumption of flatness enables us to make the decomposition
\begin{equation}
	\ul{s} = \true{\ul{t}} + \ul{q} + \ul{r}
\end{equation}
where $\ul{q}$ is an element of the vector space $V_q$ parallel to the manifold and $\ul{r}$ is perpendicular to it (see fig. \ref{fig-chisqrmaniflin}):
\begin{equation}
	\ul{q} \in V_q \equiv \left\{ (\ul{t}(\ul{p}) - \true{\ul{t}})\ \text{for any}\ \ul{p} \right\}, \qquad
	V_q \oplus V_r = \mathbb{R}^N , \qquad
	\ul{r} \in V_r , \qquad
	\ul{r} \cdot \ul{q} = 0
	\end{equation}
The position of the maximum likelihood estimate $\ul{t}^\text{est}$ is at the point on the manifold closest to the data point $\ul{s}$
\begin{equation}
	\est{\ul{t}} = \true{\ul{t}} + \ul{q}
	\end{equation}
	
We can now determine the probability content of $\tilde{R}(\true{\ul{p}})$ as defined in eq. (\ref{eq-tildeRchi}).
Using the notation
\begin{equation}
	\Theta(  \langle\text{\it condition}\rangle )
	= \begin{cases} 1, & \langle\text{\it condition}\rangle\ \text{is true} \\ 0,&  \text{else} \end{cases}
	\end{equation}
we calculate
\begin{align}
	& P \left[ \est{\ul{p}}(\ul{S}) \in \tilde{R}(\true{\ul{p}}) \right] \nonumber\\
	&= \int_{\mathbb{R}^N}\ d\ul{s}\
		w_S(\ul{s} | \true{\ul{p}})\
		\Theta\left(\ \chi^2\left(\est{\ul{t}}(\ul{s}),\true{\ul{t}}\right)\ \leq \chi^2_{\CL,n}\ \right) \nonumber\\
	&= \int_{V_q}\ d\ul{q}\ \int_{V_r}\ d\ul{r}\
		w_S(\true{\ul{t}} + \ul{q}+\ul{r} | \true{\ul{p}})\
		\Theta\left(\ \chi^2\left(\true{\ul{t}} + \ul{q},\true{\ul{t}}\right)\ \leq \chi^2_{\CL,n}\ \right) \nonumber\\
	&= ( 2\pi )^{-N/2}
		\int_{V_q}\ d\ul{q}\ \int_{V_r}\ d\ul{r}\
		\exp\left( -\frac{1}{2} \chi^2( \true{\ul{t}} + \ul{q}+\ul{r} , \true{\ul{t}} )\ \right)\
		\Theta\left(\ \norms{\ul{q}} \leq \chi^2_{\CL,n} \right) \nonumber\\
	&= ( 2\pi )^{-N/2}
		\int_{V_q}\ d\ul{q}\ \int_{V_r}\ d\ul{r}\
		\exp\left( -\frac{1}{2} \norms{\ul{q}+\ul{r}} \right)\
		\Theta\left(\ \norms{\ul{q}} \leq \chi^2_{\CL,n}\ \right) \nonumber\\
	&= ( 2\pi )^{-N/2}
		\int_{V_q}\ d\ul{q}\ \int_{V_r}\ d\ul{r}\
		\exp\left( -\frac{1}{2} \norms{\ul{q}}\right) \exp\left(-\frac{1}{2} \norms{\ul{r}} \right)\
		\Theta\left(\ \norms{\ul{q}} \leq \chi^2_{\CL,n}\ \right)
	\label{eq-probcontent}
	\end{align}
Effectively, we integrate the probability content of the region shaded grey in fig. \ref{fig-chisqrmaniflin}.
The $\ul{r}$-integral is
\begin{equation}
	\int_{V_r}\ d\ul{r}\ \exp\left(-\frac{1}{2} \norms{\ul{r}} \right)\
	= \left[ \int_{-\infty}^\infty dx \exp\left(-\frac{1}{2} x^2 \right)\ \right]^{N-n}
	= ( 2\pi )^\frac{N-n}{2}
	\end{equation}
Thus
\begin{align}
	P \left[ \est{\ul{p}}(\ul{S}) \in \tilde{R}(\true{\ul{p}}) \right]
	&= ( 2\pi )^{-n/2}
		\int_{V_q}\ d\ul{q}\
		\exp\left( -\frac{1}{2} \norms{\ul{q}}\right)
		\Theta\left( \norms{\ul{q}} \leq \chi^2_{\CL,n} \right) \nonumber\\
	&= ( 2\pi )^{-n/2}
		\Omega_{n}
		\int_0^{\chi_\CL} dq\ q^{n-1} \exp\left( -\frac{1}{2} q^2\right) \nonumber\\
	&= ( 2\pi )^{-n/2}
		\Omega_{n} \frac{1}{2}
		\int_0^{\chi_\CL^2} dx\ x^\frac{n-2}{2} \exp\left( -\frac{x}{2} \right)
	\end{align}
where $\Omega_{d} = 2\pi^{d/2}/\Gamma(\frac{d}{2})$ is the area of the (d-1)-dimensional surface of the unit sphere.
Then finally
\begin{align}
	P \left[ \est{\ul{p}}(\ul{S}) \in \tilde{R}(\true{\ul{p}}) \right]
	&= \frac{1}{2^{n/2} \Gamma(\frac{n}{2})}
		\int_0^{\chi_\CL^2} dx\ x^{\frac{n}{2}-1} \exp\left( -\frac{x}{2} \right) \nonumber \\
	& \equiv \ W^{(n)}_{\chi^2} (\chi^2_{\CL,n})
	\ \mathop{=}^{!} \ \CL
	\end{align}
$W^{(n)}_{\chi^2} (\chi^2_{\CL,n})$ is the \terminol{cumulative $\chi^2$ distribution for $n$ degrees of freedom}.
Due to our flatness assumption, $\chi^2_{\CL,n}$ is independent of $\true{\ul{p}}$.

For the practical calculation of $R(\est{\ul{p}})$, we can use the Pythagorean theorem
\begin{equation}
	\norms{ \ul{t}(\ul{p}) - \est{\ul{t}}(\ul{s})} = \norms{\ul{t}(\ul{p}) - \ul{s}}
	- \underbrace{\norms{ \ul{s} - \est{\ul{t}}(\ul{s}) }}_{\displaystyle \equiv \chi^2_\text{opt}(\ul{s})}
	\end{equation}
to obtain from eq. (\ref{eq-errbandnl}) the prescription for the determination of the confidence region
\begin{equation} \boxed{
	R(\est{\ul{p}}(\ul{s})) \equiv R(\ul{s})
		= \Big\{ \ul{p}\ \Big\vert\  \chi^2\left(\ul{s}|\ul{p}\right)\ \leq\
		\chi^2_\text{opt}(\ul{s}) + \chi^2_{\CL,n} \Big\}
	} \label{eq-confregchi} \end{equation}
with $\chi^2_\text{opt}(\ul{s})$ being the smallest value $\chi^2\left(\ \ul{s},\ \ul{t}(\ul{p})\ \right)$ can take on for fixed $\ul{s}$:
\begin{equation} \boxed{
	\chi^2_\text{opt}(\ul{s}) \equiv \chi^2\left( \ul{s} | \est{\ul{p}}(\ul{s}) \right) = \min_{\ul{p}} \chi^2(\ul{s}|\ul{p})
	} \label{eq-chisqropt} \end{equation}

\section{Hypothesis Testing}

\begin{figure}[h]
	\centering
	\includegraphics{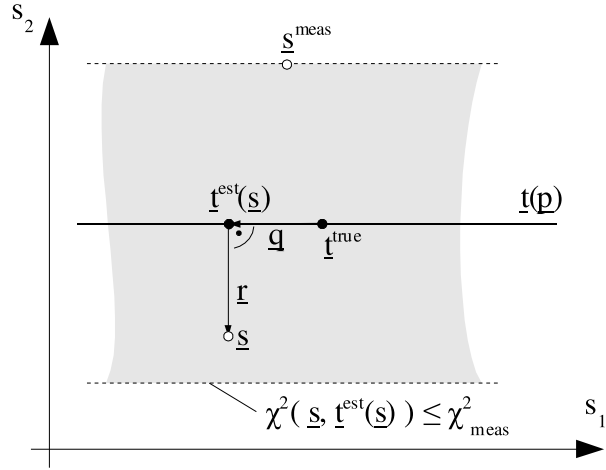}
	\caption{Hypothesis testing around a flat manifold $\ul{t}(\ul{p})$. }
	\label{fig-chisqrmanifhyp}
	\end{figure}
	
The $\chi^2$ measure also offers a simple way to judge how plausible it is that the data $\ul{s}$ follow the assumed distribution around the theory prediction  $\ul{t}(\ul{p})$. Again, a $\chi^2$ threshold obtained from the $\chi^2$ distribution is involved. It is important to note that here the relevant number of degrees of freedom (d.o.f.) is different than for the calculation of confidence regions.

For hypothesis testing, we calculate $\chi^2_\text{meas} \equiv \chi^2_\text{opt}(\ul{s}^\text{meas})$
for our measured sample $\ul{s}^\text{meas}$, and judge the hypothesis by the likelihood
of occurrence of a smaller or equal value of $\chi^2_\text{opt}(\ul{s})$.

\begin{align}
	P \left[ \chi^2_\text{opt}(\ul{S}) \leq \chi^2_\text{meas} \right]
	&= \int_{\mathbb{R}^N}\ d\ul{s}\
		w_S(\ul{s} | \true{\ul{p}})\
		\Theta\left( \chi^2_\text{opt}(\ul{s}) \leq \chi^2_\text{meas} \right) \nonumber \\
	&= ( 2\pi )^{-N/2}
		\int_{V_q}\ d\ul{q}\ \int_{V_r}\ d\ul{r}\
		\exp\left( -\frac{1}{2} \norms{\ul{q}+\ul{r}} \right)\
		\Theta\left(\ \norms{\ul{r}} \leq \chi^2_\text{meas}\ \right) \nonumber \\
	&= ( 2\pi )^{-\frac{N-n}{2}}
		\int_{V_r}\ d\ul{r}\
		\exp\left( -\frac{1}{2} \norms{\ul{r}} \right)
		\Theta\left(\ \norms{\ul{r}} \leq \chi^2_\text{meas}\ \right) \nonumber \\
	&= W^{(N-n)}_{\chi^2} (\chi^2_\text{meas})
	\end{align}

Obviously, for hypothesis testing we need to evaluate the $\chi^2$ distribution for $(N-n)$ degrees of freedom,
as opposed to the $n$ degrees of freedom appearing in the error band calculation.

The reason for this difference is the following:
For error band calculation, $\ul{q}$ plays the major role because we define our error band {\it on} the
manifold $\ul{t}(\ul{p})$.
Conversely, for hypothesis testing, the distance $\ul{r}$ to the manifold $\ul{t}(\ul{p})$ is
relevant, because it characterizes the ''distance to the theory''.

A value of $W^{(N-n)}_{\chi^2} (\chi^2_\text{meas})$ very close to $1$ means that the extremely unlikely accident has happened that the measured data are much farther away from theory than expected. That is -- if the theory is valid in the first place. Therefore a $W^{(N-n)}_{\chi^2} (\chi^2_\text{meas}) \approx 1$ casts doubt on the validity of the predictor $\ul{t}(\ul{p})$ or might indicate underestimated or correlated uncertainties.

On the other hand, a $W^{(N-n)}_{\chi^2} (\chi^2_\text{meas})$ close to $0$ represents a measurement that, compared to theoretic predictions, seems ``too good to be true''. Then one might want to check whether uncertainties are correlated or overestimated.

For a large number of degrees of freedom (d.o.f.), the expectation value of $\chi^2$ is approximately the number of d.o.f, namely $(N-n)$. Therefore, it is customary to specify ``\terminol{$\chi^2$ per d.o.f.}'' as an indicator of credibility. If $\chi^2_\text{meas}/(N-n)$ is much greater or well below $1$, one should be worried, for the same reasons as mentioned above. Note, however, that the distribution of $\chi^2_\text{meas}/(N-n)$ becomes more peaked with growing number of d.o.f.
	
\section{Confidence Regions for Single Parameters and Derived Quantities}
\label{sec-derivquant}

Suppose we ask for the value and confidence interval of one \terminol{single parameter} $p_i$, but do not care what values the other parameters might take on concurrently. This is a special case of a more general problem:
Consider some quantity $\ul{u}\in\mathbb{R}^m$, which can be calculated from the parameters using a function $\ul{h}\;:\;\mathbb{R}^n\;\rightarrow\;\mathbb{R}^m$. We would like to have an estimate of $\ul{u}$. How can the confidence region belonging to $\ul{u}$ be determined? \footnote{The special case $m=1$ and $\ul{h}(\ul{p})=p_i$ amounts to the single parameter analysis.}

In principle, we follow the method of construction presented in section \ref{sec-confreg}. Through the maximum likelihood estimator $\est{\ul{p}}$, we have an estimate of the value of $\ul{u}$, and $\est{\ul{u}}(S) \equiv \ul{h}(\est{\ul{p}}(\ul{S}))$ follows a known probability distribution. So under the assumption of $\ul{u}$ being the true value, we can select regions $\tilde{R}[\ul{h}](\ul{u}) \subset \mathbb{R}^m$ fulfilling
\begin{equation}
	P[\est{\ul{u}}(\ul{S}) \in \tilde{R}[\ul{h}](\ul{u})] = \CL
	\end{equation}
Then the desired confidence region is
\begin{equation}
	R[\ul{h}](\est{\ul{u}}) = \Big\lbrace u \ \Big\vert \ \est{\ul{u}} \in \tilde{R}[\ul{h}](\ul{u}) \Big\rbrace
	\end{equation}
$\tilde{R}[\ul{h}]$ is constructed with the help of $\chi^2$. We define level sets
\begin{equation}
	L(\ul{u}) \equiv \Big\lbrace \ul{p}\ \Big\vert\ \ul{h}(p)=\ul{u} \Big\rbrace
	\end{equation}
and an effective $\chi^2$ measure
\begin{align}
	\chi^2_\text{eff}(\ul{u},\ul{u}') 
	 & \equiv \min_{\ul{p} \in L(\ul{u}),\ \ul{p}' \in L(\ul{u}')}\ \chi^2(\ \ul{t}(\ul{p}),\ \ul{t}(\ul{p'})\ ) \\
	\chi^2_\text{eff}(\ul{s}|\ul{u}) 
	 & \equiv \min_{\ul{p} \in L(\ul{u})}\ \chi^2(\ \ul{s}\ ,\ \ul{t}(\ul{p})\ ) 
	\end{align}
Aside from pathologic cases, for each $\ul{u}$, $M(\ul{u}) \equiv \ul{t}(L(\ul{u}))$ is an $(n-m)$-dimensional connected manifold in data space, at least in the vicinity of $\true{\ul{t}}$. Furthermore, we assume here that all the manifolds $M(\ul{u})$ and the manifold swept out by $\ul{t}(\ul{p})$ are planar in the region of interest. For different $\ul{u}$, the $M(\ul{u})$ are then automatically parallel (otherwise they would intersect). The following definition provides confidence regions of the desired confidence level:
\begin{equation}
	\tilde{R}[\ul{h}](\ul{u})
	= \Big\{ \ul{u}'\ \Big\vert\ \chi^2_\text{eff} \left(\ul{u},\ul{u}' \right)\ \leq\ \chi^2_{\CL,m} \Big\}
	\end{equation}
To see this, we make the decomposition
\begin{equation}
	\ul{s} = \true{\ul{t}} + \ul{v} + \ul{q}  + \ul{r}
\end{equation}
where $\ul{v}$ is parallel to the $M(\ul{u})$, $\ul{q}$ is perpendicular to $\ul{v}$ but parallel to the manifold $\ul{t}(\ul{p})$, and $\ul{r}$ is perpendicular to both $\ul{v}$ and $\ul{q}$, where now $V_r$ is $m$-dimensional. The probability content is calculated in analogy to eq. $(\ref{eq-probcontent})$:
\begin{align}
	& P \left[ \est{\ul{u}}(\ul{S}) \in \tilde{R}[\ul{h}](\true{\ul{u}}) \right] \nonumber\\
	&= \int_{V_s}\ d\ul{s}\
		w_S(\ul{s} | \true{\ul{p}})\
		\Theta\left(\ \chi^2_\text{eff} \left(\est{\ul{u}}(\ul{s}),\true{\ul{u}}\right)\ \leq \chi^2_{\CL,m}\ \right) \nonumber\\
	&= ( 2\pi )^{-N/2}
		\int_{V_v}\! d\ul{v} \int_{V_q}\! d\ul{q} \int_{V_r}\! d\ul{r}\
		\exp\left( -\frac{1}{2} \chi^2( \true{\ul{t}} +\ul{v}+ \ul{q}+\ul{r} , \true{\ul{t}} )\ \right)\
		\Theta\left(\ \norms{\ul{q}} \leq \chi^2_{\CL,m} \right) \nonumber\\
	&= ( 2\pi )^{-N/2}
		\int_{V_q}\! d\ul{q}\ 
		\exp\left( -\frac{1}{2} \norms{\ul{q}}\right) \Theta\left(\ \norms{\ul{q}} \leq \chi^2_{\CL,m}\ \right)\ 
		\int_{V_v}\! d\ul{v}\ \int_{V_r}\! d\ul{r}\ \exp\left(-\frac{1}{2} \norms{\ul{r}+\ul{v}} \right)\ \nonumber \\
	&= ( 2\pi )^{-m/2}
		\int_{V_q}\ d\ul{q}\ 
		\exp\left( -\frac{1}{2} \norms{\ul{q}}\right) \Theta\left(\ \norms{\ul{q}} \leq \chi^2_{\CL,m}\ \right)\ 
		\nonumber \\
	&= W^{(m)}_{\chi^2} (\chi^2_{\CL,m}) = \ \CL
	\label{eq-uprobcontent}
	\end{align}
Making use of the Pythagorean theorem again, we find
\begin{align}
	R[\ul{h}](\est{\ul{u}}(\ul{s})) \equiv R[\ul{h}](\ul{s}) 
	& = \big\{ \ul{u} \ \big\vert\  \chi^2_\text{eff} \left(\ \est{\ul{u}},\ \ul{u}\ \right)\ \leq\ \chi^2_{CL,m} \big\}  \nonumber \\
	& = \big\{ \ul{u}\ \big\vert\ \chi^2_\text{eff} \left(\ \est{\ul{t}}(\ul{s}),\ \ul{u}\ \right)\ \leq\ \chi^2_{CL,m} \big\}  \nonumber \\
	& = \big\{ \ul{u}\ \big\vert\ \chi^2_\text{eff} \left(\ \ul{s}\ |\ \ul{u}\ \right)\ \leq\ \chi^2_\text{opt}(\ul{s}) + \chi^2_{CL,m} \big\}  
	\label{eq-uconfregeff} \\
	& =  \big\{ \ul{h}(\ul{p}) \ \big\vert\ \chi^2 \left(s|\ul{p} \right)\ \leq\ \chi^2_\text{opt}(\ul{s}) + \chi^2_{CL,m} \big\}
	\label{eq-uconfreg}
	\end{align}
where again $\chi^2_\text{opt}(\ul{s}) \equiv \chi^2 \left(\est{\ul{t}}(\ul{s}), \ul{s} \right)$.

Thus we deduce the following simple procedure to find the confidence region for $\ul{u}$: Determine the confidence region in parameter space using a $\chi^2$ threshold for $m$ parameters, and calculate the image under $\ul{h}$.

\section{Errors in the Pion Mass}

In section \ref{sec-statproblem}, the errors in the pion masses $m_\pi = \bar{x}_i \pm \Delta x_i$ were mentioned, but only now will it be shown how they can be taken into account. To get a feeling for the problem, let us look at an intuitive solution: In analogy to eq. (\ref{eq-fitcond}), the constraints take the form
\begin{equation}
	\text{`` } m_N(\bar{x}_i \pm \Delta x_i;\; \ul{p}) \approx \bar{y}_i \pm \Delta y_i \text{ ''}
	\end{equation}
The standard method of linear error propagation tells us how the error in $\bar{x}_i$ contributes to the error in $m_N$:
\begin{equation}
	\text{`` }m_N(\bar{x}_i; \ul{p}) \pm \left| \frac{\partial}{\partial \bar{x}_i} m_N(\bar{x}_i; \ul{p}) \right| \Delta x_i \approx \bar{y}_i \pm \Delta y_i \text{ ''}
	\end{equation}
If the errors in $\bar{x}_i$ and $\bar{y}_i$ are independent, the two contributions to the error may be added quadratically:
\begin{equation}
	\text{`` }m_N(\bar{x}_i; \ul{p})  \approx \bar{y}_i \pm 
	\sqrt{ (\Delta y_i)^2 + \left( \frac{\partial}{\partial \bar{x}_i} m_N(\bar{x}_i; \ul{p})  \Delta x_i \right)^2 }
	\text{ ''}
	\label{eq-efferrorintuition}
	\end{equation}
It will be shown that a more careful analysis of the problem gives the same result. 

It is possible to specify an optimization problem of the form (\ref{eq-fitcond}) which accounts for the errors in $\bar{x}_i$. We introduce a \emph{new parameter} $x_i$ for each of the $N$ data points. The complete set of optimization parameters is now $\ul{p} \equiv (u_1, ..., u_n, x_1, ..., x_N)^T$. Here the $u_i$ denote the parameters previously denoted $p_i$ and control the shape of the interpolation function $m_N(x;\ul{u})$. The fit constraints can then be specified as
\begin{equation} \begin{array}{lclcccccl}
	\text{`` } f_i(\ul{p}) & \equiv & m_N(x_i;\;\ul{u}) & \approx & \bar{y}_i & \pm & \Delta y_i & \qquad & i=1..n \text{ ''}\\
	\text{`` } f_{i+n}(\ul{p}) & \equiv & x_i & \approx & \bar{x}_i & \pm & \Delta x_i & \qquad & i=1..N \text{ ''}
	\label{eq-fitcondpi}
	\end{array} \end{equation}
Using this notation, the statistical toolkit presented in the previous sections can be applied. However, it is neither practical to have so many parameters, nor are we really interested in the values of the $x_i$. Luckily, they are tightly constrained by the linear conditions in the second line of eq. (\ref{eq-fitcondpi}). Making an approximation, the parameters $x_i$ can be eliminated from the problem, according to the following general procedure.

\section{Eliminating Linearly Constrained Parameters}
\label{sec-elimlinconst}

Suppose we have $n=m+k$ optimization parameters $\ul{p}=(u_1,...,u_m,x_1,...,x_k)^T$ and a problem of the form (\ref{eq-fitcond}). The $x_i$ are to be eliminated. The constraint functions $f_i$ must be approximately linear in the $x_i$:

\begin{equation}
	t_i(\ul{p}) \equiv \frac{f_i(\ul{p})}{\Delta y_i} \approx A(\ul{u}) \ul{x} + g_i(\ul{u})
	\end{equation}

Here $A(\ul{u}) \in \mathbb{R}^{N \times k}$ is a matrix. If the only quantities we want to analyze statistically are the $u_i$, we can apply the recipe of section \ref{sec-derivquant}. According to eq. (\ref{eq-uconfregeff}), the confidence region is given by the condition
$\chi^2_\text{eff} ( \ul{s}, \ul{u} )\ \leq\ \chi^2_\text{opt} + \chi^2_{CL,m}$ .
All that needs to be done is to find an approximation for $\chi^2_\text{eff}$, which is given by
\begin{equation}
	\chi^2_\text{eff} ( \ul{s}, \ul{u} ) = \min_{\ul{x}} \norms{ A(\ul{u}) \ul{x} + \ul{g}(\ul{u}) - \ul{s} }  
	\end{equation}
With the abbreviation $\ul{b}(\ul{u}) \equiv \ul{s} - \ul{g}(\ul{u})$, we are left with the famous problem to minimize $\norms{A \ul{x} - \ul{b}}$. It can be solved by a matrix decomposition 
\begin{equation}
	A = Q\;R
\end{equation}
where $Q \in \mathbb{R}^{N \times N}$ is an orthonormal matrix, and $R \in \mathbb{R}^{N \times k}$ is an upper triangular matrix. The norm remains invariant under application of $Q$ or $Q^T$, so
\begin{equation}
	\norms{A \ul{x} - \ul{b}} = \norms{Q^T\;(Q\;R\; \ul{x} - \ul{b})} = \norms{R\; \ul{x} - Q^T \;\ul{b}}
	\end{equation}
We assume here that $R$ has full rank $k$.\footnote{If it did not, the minimum of $\chi^2$ would not be unique. This would be a badly formulated problem.} The norm reaches its minimum when $\ul{x}$ is adjusted in such a way, that the $k$ first components of $R\; \ul{x} - Q^T \;\ul{b}$ vanish. Then only the last $m$ components contribute to the norm. We decompose the matrix $Q^T$ into an upper and a lower part
\begin{equation}
	Q^T = \begin{pmatrix} \tilde{Q}^T \\ \hline \hat{Q}^T \end{pmatrix} \qquad 
	\hat{Q}^T \in \mathbb{R}^{m \times N} 
	\end{equation}
and thus find\footnote{The following equation gives a prescription that is numerically both stable and fast.} 
\begin{equation} \boxed{
	\chi^2_\text{eff} ( \ul{s}, \ul{u} ) = \norms{ \hat{Q}^T \; \ul{b} } 
	= \norms{ \hat{Q}^T(\ul{u})\;( \ul{s}-\ul{g}(\ul{u}) ) } 
	} 
	\label{eq-chieffapprox}
	\end{equation}

Let us look again at the case of uncertainties in the pion mass. For simplicity, suppose there is only one lattice data point $(\bar{x} \pm \Delta x, \bar{y} \pm \Delta{y})$. The constraints are
\begin{equation} \begin{array}{lclcl}
	\text{`` } f_1(x,\ul{u}) 
	& \approx & \displaystyle \frac{\partial m_N(\bar x;\;\ul{u})}{\partial \bar{x}} \;(x-\bar{x}) 
	+ m_N(\bar x;\;\ul{u})
	& \approx & \bar{y} \pm \Delta y \text{ ''}\\
	\text{`` }f_2(x,\ul{u}) 
	& = & x
	& \approx & \bar{x} \pm \Delta x\text{ ''}
	\end{array} \end{equation}
with the abbreviations
$\displaystyle c \equiv \frac{\partial m_N(\bar x;\;\ul{u})}{\partial \bar{x}}$, 
$y \equiv m_N(\bar{x};\;\ul{u})$, we read off
\begin{equation}
	A = \begin{pmatrix} c/\Delta y \\ 1/\Delta x \end{pmatrix} \qquad
	\ul{b} = \begin{pmatrix} (c\bar{x} - y + \bar{y})/\Delta{y} \\ \bar{x}/\Delta{x} \end{pmatrix}
\end{equation}
$QR$-factorization of $A$ gives
\begin{equation}
	R = \begin{pmatrix} r \\ 0 \end{pmatrix}\ \text{with}\ r^2=\frac{c^2}{(\Delta y)^2} + \frac{1}{(\Delta x)^2} , \quad
	Q = \frac{1}{r}  \begin{pmatrix} c/\Delta{y} & -1/\Delta{x} \\ 1/\Delta{x} & c/\Delta{y} \end{pmatrix} 
	= \begin{pmatrix} \tilde Q \ \vline \ \hat Q \end{pmatrix} \quad \Rightarrow
	\end{equation}
\begin{align}
	\chi^2_\text{eff}(\ul{s},\ul{u}) & = \norms{\hat{Q}^T \ul{b}} 
	= \frac{1}{r^2} \norms{ \begin{pmatrix} \frac{-1}{\Delta x} & \frac{c}{\Delta y} \end{pmatrix}
	\begin{pmatrix} (c\bar{x} - y + \bar{y})/\Delta{y} \\ \bar{x}/\Delta{x} \end{pmatrix} }
	= \frac{\left(y - \bar{y}\right)^2}{c^2 (\Delta x)^2 + (\Delta y)^2} 
	\end{align}
The resulting expression for $\chi^2_\text{eff}$ is in agreement with the result of the intuitive approach eq. (\ref{eq-efferrorintuition}): The error in $\bar{x}$ effectively enlarges the error in $\bar{y}$ to 
$\sqrt{c^2 (\Delta x)^2 + (\Delta y)^2}$. It is easy to generalize the result to several lattice data points:
\begin{equation} \boxed{
	\chi^2_\text{eff} = \sum_i \frac{\left(m_N(\bar{x_i};\;\ul{u}) - \bar{y_i}\right)^2}{\left( \frac{\partial m_N(\bar x_i;\;\ul{u})}{\partial \bar{x}_i} \Delta x_i \right)^2 + (\Delta y_i)^2}
	\label{eq-chisqreffpionerr}
	} \end{equation}
We can directly use this simple formula in our computer program.

\section{Quadratic Approximation of $\chi^2$, Error Matrix}

The crucial quantity for the determination of confidence regions and intervals is $\chi^2(\ul{s}|\ul{p}) \equiv \chi^2(\ul{s},\ul{t}(\ul{p}))$, see eq. (\ref{eq-confregchi}). For many purposes, it is sufficient to perform a leading order expansion of $\chi^2(\ul{s}|\ul{p})$ in terms of $\ul{p}$ around its minimum $\chi^2_\text{opt}$ at $\est{\ul{p}}$, resulting in a quadratic form:

\begin{equation}
	B_{kl} \equiv \frac{\partial^2 \chi^2(\ul{s} | \ul{p}^\text{est})}{\partial p_k\, \partial p_l} 
	\qquad \Rightarrow \qquad
	\chi^2(\ul{s}|\ul{p}^\text{est}+\ul{\delta p}) 
	\approx  \chi^2_\text{opt} +  \frac{1}{2} {\delta p}_k\, B_{kl}\, {\delta p}_l
	\label{eq-quadraticapproxchi}
	\end{equation}

We take a look at the hyper-ellipsoid 
\begin{equation}
	H(T) \equiv \Big\lbrace \est{\ul{p}}+\delta \ul{p} \ \Big\vert \ 
	\frac{1}{2} \delta \ul{p}^T\, B\, \delta \ul{p} \leq T \Big\rbrace
	\end{equation}
For a threshold $T=\chi^2_{CL,n}$, this hyper-ellipsoid is the joint confidence region of all parameters.
Consider now a parameter dependent quantity $u \in \mathbb{R}$ which can be calculated from the parameters using the function $h(\ul{p})$. In section \ref{sec-derivquant} we saw that its confidence interval is $h(\;H(\chi^2_{\CL,1})\;)$. So methods to determine the image of $H(T)$ under $h$ are needed. For example, we could directly determine the upper bound $u_\mmax$ of the interval by maximizing $h$ with the constraint $\frac{1}{2} {\delta p}_k\, B_{k,l}\, {\delta p}_l \leq T$. Numerically, this is an involved task. It is more convenient to employ the reparametrization
\begin{equation}
	\delta p_k(\ul{q}) = \sqrt{2\; T}\; B^{\scriptstyle -1/2}\, \frac{\ul{q}}{\normq{\ul{q}}}\; \sin \normq{\ul{q}}
	\label{eq-reparamp}
	\end{equation}
where $B^{\scriptstyle -1/2}$ is the symmetric matrix square root of the inverse of $B$. Then $\est{p}+\delta \ul{p}(\ul{q})$ lies within the hyper-ellipsoid automatically, and an unconstrained maximization
\begin{equation}
	u_\mmax = \max_{\ul{q}}\ h(\est{p}+\delta \ul{p}(\ul{q}))
	\label{eq-umax}
	\end{equation}
does the job. If $h(\ul{p})$ is approximately linear within the confidence region, 
\begin{equation}
	h(\ul{p}) \approx h(\est{\ul{p}}) + \ul{v}^T\; \delta \ul{p}
	\label{eq-ulinearexpan}
	\end{equation}
we have an even faster method. $h$ will take on its maximum value on some point of the boundary of the hyper-ellipsoid, and there its gradient with respect to $\ul{p}$ must be parallel to that of $\chi^2(\ul{s}|\ul{p})$, so for some constant $\alpha$
\begin{equation}
	\frac{\partial \chi^2(\ul{s}|\ul{p})}{\partial \ul{p}} \approx B\; \delta \ul{p} \mathop{=}^! \alpha\; \ul{v} \qquad \Rightarrow \qquad 
	\delta \ul{p} = \alpha\; B^{-1}\; \ul{v}
	\end{equation}
$\alpha$ is determined so as to place $\delta \ul{p}$ on the boundary of the ellipsoid:
\begin{equation}
	T \mathop{=}^{!} \frac{1}{2}\; \delta \ul{p}^T\, B\, \delta \ul{p}
	= \frac{\alpha^2}{2}\; \ul{v}^T\; B^{-1}\; \ul{v} 
	\qquad \Rightarrow \qquad
	\alpha = \sqrt{\frac{2 \; T}{\ul{v}^T B^{-1}\; \ul{v}}}
	\end{equation}
Substituting this into the expansion eq. (\ref{eq-ulinearexpan}), we get
\begin{equation}
	u_\mmax = h(\est{\ul{p}})+ \alpha \; \ul{v}^T\; B^{-1}\; \ul{v}
	= h(\est{\ul{p}}) + \sqrt{2 \; T\; \ul{v}^T B^{-1}\; \ul{v}}
	\end{equation}
Because it is such a useful quantity, one introduces the \terminol{error matrix} $E \equiv 2 \; T\; B^{-1}$. Then we may write
\begin{equation} \boxed{
	u = h(\est{\ul{p}}) \pm \sqrt{\ul{v}^T E\; \ul{v}}
	} \label{eq-errmatusage} \end{equation}
The error matrix encodes the uncertainty of any parameter dependent quantity. For example, the uncertainties of the individual parameters $p_i$ are found directly on the diagonal of the error matrix:
\begin{equation}
	p_i = \est{p}_i \pm \sqrt{E_{ii}}
	\end{equation}
	
In order to determine the uncertainty of a single scalar quantity, irrespective of other quantities, the threshold $T$ is set to $\chi^2_{\CL,1}$. For a confidence level $\CL=68\%$, corresponding to one standard deviation, one finds $\chi^2_{\CL,1}=1$. 
	
\section{Connection to Error Propagation}

The representation (\ref{eq-errmatusage}) with $T=\chi^2_{\CL,1}=1$ allows us to make the connection to the standard method of error propagation, according to which
\begin{equation}
	(\Delta u)^2 = \sum_k \left( \frac{\partial u(\est{\ul{p}}(\ul{s}))}{\partial s_k} \Delta s_k \right)^2
	= \sum_k \left( \frac{\partial u(\est{\ul{p}})}{\partial p_i} \frac{\partial \est{p}_i(\ul{s})}{\partial s_k} \right)^2
	= v_i \frac{\partial \est{p}_i(\ul{s})}{\partial s_k} \frac{\partial \est{p}_j(\ul{s})}{\partial s_k} v_j
	\end{equation}
Here we have made use of $\Delta s_k = 1$. Einstein's summation convention is implied for all indices except for those explicitly summed over. $\est{\ul{p}}$ always minimizes $\chi^2(\ul{s}|\ul{p})$, so we determine its derivative by shifting the data $\ul{s}$ a little and analyzing the necessary shift $\delta \ul{p}$ in $\ul{p}$ to stay in the minimum:
\begin{equation}
	0
	= \frac{\partial \chi^2(\ul{s}+\delta\ul{s}|\ul{p}^\text{est}+\delta\ul{p})}{\partial p_l}
	= \frac{\partial^2 \chi^2(\ul{s}|\ul{p}^\text{est})}{\partial p_l \partial p_i} \delta p_i
	+ \frac{\partial^2 \chi^2(\ul{s}|\ul{p}^\text{est})}{\partial p_l \partial s_k} \delta s_k
	+ ...
	\label{eq-varychi}
	\end{equation}
\begin{equation}
	\Rightarrow \qquad
	\frac{\partial p_i^\text{est}(\ul{s})}{\partial s_k}
	= - (B^{-1})_{il} \frac{\partial^2 \chi^2(\ul{s}|\ul{p}^\text{est})}{\partial p_l \partial s_k} 
	= - (B^{-1})_{il} \frac{\partial t_k(\est{p})}{\partial p_l}
	\end{equation}
Calculating $B$ explicitly gives
\begin{equation}
	B_{lm} = 2 \sum_k \left( \frac{\partial t_k(\est{p})}{\partial p_l} \frac{\partial t_k(\est{p})}{\partial p_m}
	+ (s_k - t_k(\est{\ul{p}})) \frac{\partial^2 t_k(\est{p})}{\partial p_l \;\partial p_m} \right)
	\approx 2 \frac{\partial t_k(\est{p})}{\partial p_l} \frac{\partial t_k(\est{p})}{\partial p_m}
	\label{eq-propapprox}
	\end{equation}
if we can neglect the contribution proportional to the second derivatives of $\ul{t}(\ul{p})$. Obviously this is always the case when the fit is very good. Then 
\begin{equation}
	(\Delta u)^2 = v_i\; (B^{-1})_{il} \frac{\partial t_k(\est{p})}{\partial p_l}  \frac{\partial t_k(\est{p})}{\partial p_m} (B^{-1})_{jm}\; v_j \approx 
	v_i\; (B^{-1})_{il}\; \frac{1}{2} B_{lm}\; (B^{-1})_{jm}\; v_j
	\end{equation}
and finally, with the definition of the error matrix for $T=1$
\begin{equation}
	(\Delta u)^2 \approx v_i\; \frac{1}{2} (B^{-1})_{ji}\; v_j = v^T\; E \; v
	\end{equation}
Thus we have shown that error propagation applied to the maximum likelihood estimate is identical to error analysis using a $\chi^2$ threshold, as long as $\chi^2(\ul{s}|\ul{p})$ is approximately quadratic in $\ul{p}$ and the approximation (\ref{eq-propapprox}) is valid.
	
\section{Error Bands and Interpretation}

A central task of this thesis is the visualization of the uncertainties in our fit curve. Instead of just plotting the optimal fit curve, we display a band. There are two different formal concepts of constructing such a band, leading to bands of different width and requiring different interpretation:
\begin{itemize}
\item
One concept is to determine a confidence region $R$ at a given confidence level and then define the band as the set of curves
\begin{equation}
	\Big\{ g : \mathbb{R} \rightarrow \mathbb{R}\ \Big\vert\  g(x)=m_N(x;\ul{p})\ ,\ \ul{p}\in R\ \Big\} 
\end{equation}
Plotting all these curves, we sweep over an area consisting of the set of points
\begin{equation}
	\mathcal{B}_\text{global} \equiv \Big\lbrace (x,m_N(x;\ul{p})) \ \Big\vert \ \ul{p} \in R \ \Big\rbrace
\end{equation}
(The appropriate range for $x$ will be subject of later discussions). This error band covers all curves that belong to a parameter set in the multi-parameter confidence region. All curves contributing to this \terminol{global error band} are likely candidates for the true function. 

\item
The other concept is to go first to a fixed pion mass $x$. Here, the nucleon mass is a function of the parameters, $h_x(\ul{p}) \equiv m_N(x;\ul{p})$. One may now calculate the corresponding confidence interval $R[h_x]$, using the formalism discussed in section \ref{sec-derivquant}. Assembling the confidence intervals at various values of $x$, we obtain a band 
\begin{equation}
	\mathcal{B}_\text{local} \equiv \Big\lbrace (x,y) \ \Big\vert \ y \in R[h_x] \Big\rbrace
\end{equation}
In contrast to the global error band, the width of this \terminol{local error band} $\mathcal{B}_\text{local}$ at a given $x$ directly corresponds to the statistical uncertainty about $m_N$ at that point. However, the local error band should not be interpreted as the graph of a family of functions, because it does \emph{not} cover all curves within the confidence region.
\end{itemize}

In practise, $\mathcal{B}_\text{global}$ and $\mathcal{B}_\text{local}$ can be calculated using the same algorithm. There is but one difference: The $\chi^2$-threshold $\chi^2_{\CL,n}$ of the $\mathcal{B}_\text{global}$ is larger than the than the $\chi^2$-threshold $\chi^2_{\CL,1}$ of $\mathcal{B}_\text{local}$. Consequently, the local error band is more narrow.
 
Throughout this diploma thesis, the \emph{global} error band $\mathcal{B}_\text{global}$ is shown. It provides an answer to the question ``Where might the true interpolation curve lie?''.

\section{Uncertainties of Fixed Parameters}

It is not possible to extrapolate the pion mass dependence of the nucleon mass using a fit to the corresponding lattice data only. More information is needed. The literature offers a number of estimates for the LECs, obtained from the analysis of empirical data or other lattice calculations. At the present stage of research, it is sensible to fix parameters $p_i$ in our fit formula $m_N(x;\ul{p})$ to these ``empirical'' values wherever reliable data is available. The interpolation function then takes the form $m_N(x;\ul{p},\ul{b})$, where $\ul{b}$ are the $m$ fixed parameters and $\ul{p}$ are the $n$ parameters that remain to be fitted.

The external estimates for the LECs $b_i$ have uncertainties, given in the form of intervals $U_{b_i}=[b_{i,\mmin},b_{i,\mmax}]$. We have no information about the correlation of these uncertainties. We cannot assume that two different fixed parameter values $b_i$ and $b_j$ are samples of independent statistical processes. A conservative treatment of error propagation in this case is to consider all possible choices of the $b_i$ within their error intervals, and then to take the union of the resulting confidence regions. Thus for each choice 
$\ul{b} \in U_{\ul{b}} \equiv U_{b_1} \times U_{b_2} \times \cdots \times U_{b_m}$
we calculate the confidence region $R_{\ul{b}}$ of the parameters $\ul{p}$ using the fit function $m_N(x;\ul{p},\ul{b})$. From these regions, we form the envelope
\begin{equation}
	R_\text{env} = \bigcup_{\ul{b} \in U_{\ul{b}}} R_{\ul{b}} 
	\end{equation}
This envelope is our conservative estimate of the uncertainties of $\ul{p}$. Analogously, the error band is now an envelope of error bands at specific choices of $\ul{b}$.

\section{Implementation of Statistical Error Analysis}

Two different algorithms have been implemented to produce parameter uncertainties and error bands:
\begin{itemize}
\item a Monte Carlo exploration of the parameter space
\item a parabolic approximation in Mathematica
\end{itemize}
The Monte Carlo method has the ability to handle deformed confidence regions, while the parabolic approximation always produces hyperellipsoids. However, in the final form of the fit problem, this approximation turns out to be sufficient.

\subsection{Hybrid MINUIT and Monte Carlo Algorithm}
\label{sec-algoMC}

The idea of the Monte Carlo search is simple. First, a bounding box is determined for the confidence region, i.e., minimal and maximal values for each parameter $p_i$ are calculated. Then, the parameters are chosen randomly inside the bounding box. If the resulting $\chi^2$ is larger than an appropriately chosen threshold, this choice of parameters is discarded. Otherwise, the interpolation function is calculated, and the upper and lower bounds of the error band are modified wherever necessary to accommodate the new representant. 

The implemented program makes use of the CERN ROOT library, which includes a version of the MINUIT minimization routines. The program has an option to perform the band calculation using the error matrix provided by MINUIT instead of a Monte Carlo search.
To determine upper and lower limits of the band, the function $m_N(x;\ul{p},\ul{b})$ is evaluated on a grid of pion masses $\xi_1,...,\xi_{n_\text{res}}$. The choice of this grid only affects the extent and the resolution of the plot. Errors in the pion mass are neglected, so that $\chi^2$ is of the form
\begin{equation}
	\chi^2(\ul{\bar y}|\ul{p}) = \sum_{i=1}^N \left( \frac{\bar{y}_i - m_N(\bar x_i;\ul{p},\ul{b})}{\Delta y_i} \right)^2
	\end{equation}
Now the algorithm looks as follows
\begin{itemize}
	\item
	Select initial values for the fixed parameters: $\ul{b} \leftarrow \ul{b}_\text{start}$.
	\item
	Minimize $\chi^2$ using MINUIT's {\tt MIGRAD} command. \\
	This returns a set of optimal parameters $\est{\ul{p}}$.
	\item
	Initialize the boundary variables of the band: \\
	$\mathcal{B}_\mmin(\xi_k) \leftarrow m_N(\xi_k;\est{\ul{p}},\ul{b})$ \quad $\forall k=0..n_\text{res}$ \\
	$\mathcal{B}_\mmax(\xi_k) \leftarrow m_N(\xi_k;\est{\ul{p}},\ul{b})$ \quad $\forall k=0..n_\text{res}$
	\item
	loop until user aborts or desired number of iterations is reached
	\begin{itemize}
		\item Scan the fixed parameter range $R_{\ul{b}}$ on a uniform grid. For each $\ul{b}$
		\begin{itemize}
			\item perform a Monte Carlo search of 200000 steps \hspace{1cm} {\it or}
			\item minimize $\chi^2$ and perform a band calculation using the error matrix provided by MINUIT, applying eq. (\ref{eq-errmatusage})
			\end{itemize}
		(In all steps, the variables for $\mathcal{B}_\mmin(\xi_k)$, $\mathcal{B}_\mmax(\xi_k)$ are only overwritten when
		the extent of the band becomes larger.)
		\end{itemize}
	\end{itemize}

On each call to a Monte Carlo search
\begin{itemize}
	\item
	Find a bounding box $U_{\ul{p}}$ for the parameters $\ul{p}$ fulfilling \\
	$\chi^2(\ul{\bar y}|\ul{p}) < \chi^2_\text{opt}(\ul{\bar y}) + \chi^2_{\CL,n} \quad \forall \ul{p} \in U_{\ul{p}}$
	\item
	for each Monte Carlo step, do
	\begin{itemize}
		\item choose $\ul{p}$ randomly from within the bounding box $U_{\ul{p}}$
		\item if $\chi^2(\ul{\bar y}|\ul{p}) > \chi^2_\text{opt}(\ul{\bar y}) + \chi^2_{\CL,n}$ , skip the rest of the loop
		\item for $k=1,..,n_\text{res}$ do
		\begin{itemize}
			\item if $\mathcal{B}_\mmin(\xi_k) > m_N(\xi_k;\ul{p},\ul{b})$,
				store $\mathcal{B}_\mmin(\xi_k) \leftarrow m_N(\xi_k;\ul{p},\ul{b})$
			\item if $\mathcal{B}_\mmax(\xi_k) < m_N(\xi_k;\ul{p},\ul{b})$,
				store $\mathcal{B}_\mmax(\xi_k) \leftarrow m_N(\xi_k;\ul{p},\ul{b})$
			\end{itemize}
		\end{itemize}
	\end{itemize}

The bounding box can be optimally determined using MINUIT's {\tt mnerrs}-command, applied once for each parameter $p_i$. Staying always in the minimum of $\chi^2(\ul{\bar y}|\ul{p})$ with respect to all other parameters $p_1,...,p_{i-1},p_{i+1},...,p_n$, this procedure searches the largest interval $[p_{i,\mmin},p_{i,\mmax}]$ fulfilling $\chi^2(\ul{\bar y}|\ul{p}) < \chi^2_\text{opt}(\ul{\bar y}) + \chi^2_{\CL,n}$. The bounding box in question is then
\begin{equation}
	U_{\ul{p}} = [p_{1,\mmin},p_{1,\mmax}] \times \cdots \times [p_{n,\mmin},p_{n,\mmax}] \nonumber
	\end{equation}

\begin{figure}[h]
	\centering
	\includegraphics[width=0.6\textwidth]{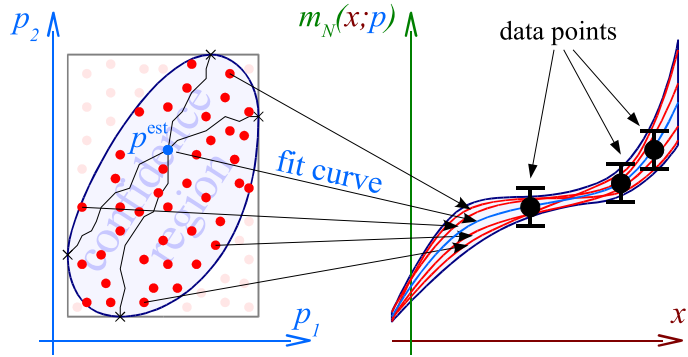}
	\caption{Hybrid MINUIT and Monte Carlo Method.}
	\label{fig-algorithm}
	\end{figure}

The principle of the algorithm is illustrated in fig. \ref{fig-algorithm}. A simple $\chi^2$ minimization yields $\est{\ul{p}}$, with which one can plot the best fit curve (blue). The MINUIT {\tt mnerrs}-command explores the parameter space, finding the bounding box that encloses the confidence region (zigzag paths terminated by crosses on the box). Then a random search selects points (red dots) within the bounding box. They are discarded if they exceed the $\chi^2$-limit (light red dots), otherwise they lie within the confidence region (shaded blue). The corresponding interpolant is calculated and contributes to the band (red curve).

\note{Note on performance:}{
	In the ideal case, with no correlations and $\chi^2$ depending quadratically on the parameters, the relative volume taken up by the confidence region in the bounding box corresponds to that of a hyper-sphere in a hyper-cube. In $n$ dimensions, the volume of a unit sphere divided by the volume of its bounding box is
	\begin{equation}
		\frac{2 \pi^{n/2}}{n\; \Gamma(\frac{n}{2})} \Big/ 2^n \quad \mathop{=}^{n\text{ even}} \quad
		\frac{1}{\left(\frac{n}{2}\right)!}\left(\frac{\sqrt{\pi}}{2}\right)^n
	\end{equation}
	which decreases rapidly with growing $n$. This means that the number of hits inside the confidence region decreases rapidly with a growing number of parameters. However, a certain number of hits within the confidence region is required to obtain a sufficiently accurate representation of the confidence region. Thus the algorithm becomes very inefficient for large $n$. Experiments carried out with $n=6$ still performed well enough. Yet, in order to apply the algorithm in a more challenging situation, the uniform distribution of trial points within the box would have to be replaced by something smarter. Metropolis-like exploration of the parameter space could be fruitful. Additionally, this would work without the help of a traditional minimizer like MINUIT, and one might be able to deal with non-Gaussian probability distributions this way. 
	}

\subsection{Quadratic Approximation in Mathematica}
\label{sec-algomathematica}

Mathematica offers greater flexibility and a friendlier interface for rapid development. Therefore, the statistical analysis has also been implemented in Mathematica, in quadratic approximation. Instead of MINUIT, Mathematica's native minimization routines {\tt FindMinimum} and {\tt NMinimize} are used. The matrix $B$ of eq. (\ref{eq-quadraticapproxchi}) is determined by differentiating $\chi^2$ twice, using Mathematica's numerical differentiation. Upper and lower limits of the error band and the parameter confidence intervals are determined using the reparametrization eq. (\ref{eq-reparamp}) and (\ref{eq-umax}), or, as a cross check, using the error matrix as in eq. (\ref{eq-errmatusage}). Evaluating
\begin{equation}
	\sqrt{\frac{\left| \chi^2(\ul{\bar y}|\ul{p}) - \chi^2_\text{opt}(\ul{\bar y}) \right|}{\chi^2_{\CL,n}}} - 1
	\label{eq-algomatherr}
	\end{equation}
at random points $\ul{p}$ on the boundary of the confidence region gives an estimate of the relative error in the uncertainty caused by non-quadratic behavior of $\chi^2$. 
Uncertainties in fixed parameters $\ul{b}$ are treated in the same way as in the Monte Carlo approach: The box $U_{\ul{b}}$ is scanned on a uniform grid (Typically, it the constraints $f_i(\ul{p},\ul{b})$ behave monotonously in $\ul{b}$, and it is sufficient to include the corners). 

In our nucleon mass fit, errors in the pion mass are treated using an effective $\chi^2$ as in eq. (\ref{eq-chisqreffpionerr}). The requirement for setting up this effective $\chi^2$ is the approximate linearity of $m_N(x;\ul{p},\ul{b})$ for $x \approx \bar{x_i} + \mathcal{O}(\Delta x_i)$. If this holds, we may replace the derivative by a finite difference with a gap size of the order of $\Delta x_i$. This is numerically advantageous. Altogether, the $\chi^2$ employed here is
\begin{equation}
	\chi^2(\ul{\bar y}|\ul{p}) = \sum_{i=1}^N 
	\frac{ \left[ \bar{y}_i - m_N(\bar x_i;\ul{p},\ul{b})\right]^2}
	{\left[ m_N(\bar x_i+\frac{1}{2}\Delta x_i;\;\ul{p},\ul{b}) 
	- m_N(\bar x_i-\frac{1}{2}\Delta x_i;\;\ul{p},\ul{b})  \right]^2 
	+ \left[\Delta y_i\right]^2} 
	\end{equation}
The multi-parameter $\chi^2$-increment employed is $\chi^2_{\CL,n}$, corresponding to the number $n$ of fitted parameters $\ul{p}$. 

\section{Note on Conventions}

Throughout this work, we stick to a confidence level $\CL = 68 \%$, unless specified otherwise.
Errors of fit parameters are specified in terms of single parameter error bounds, using $\chi^2_{\CL,1} = 1$. This complies with the standard conventions in the literature, and corresponds to the errors given by \texttt{MINUIT} in the default mode. Note that the \emph{multi parameter} confidence region is not covered entirely by the bounding box from single parameter errors \cite{minuit}. 

\chapter{Theoretical Errors}
\label{sec-theoryerr}

\section{General Remarks}

Previous sections have discussed at length how uncertainties in the input data propagate into uncertainties about our fit parameters and the interpolation curve. Yet, this statistical error analysis always forced us to make a central assumption: that the theory describes the data correctly and accurately. We know, however, that this is not the case. For one thing, the data may be subject to systematic distortions. Furthermore, the fit function from theory is just a series expansion truncated at some order of small momenta $p$. Even if the expansion coefficients (LECs) were known precisely, our fit function deviates from truth significantly once the expansion variables reach a certain magnitude. These deviations are often referred to as \terminol{higher order effects}. Our error analysis must therefore work in two steps:

\begin{enumerate}
\item
\label{item-stepstat}
\terminol{Statistical error analysis:}
Make an assumption on the range of applicability of the theory. Assume that the theory is exact in this range, i.e. ignore theoretical uncertainties\footnote{Including theoretical uncertainties in the statistical analysis is difficult. It has been attempted in the course of this diploma thesis, but the results were not convincing, perhaps in part due to a lack of enough redundance in the input data.}.
Pick data from the range of applicability. Fit theory parameters to data in that range. Using the formalism presented in chapter \ref{sec-formalstatistics}, determine the uncertainty of theoretic parameters (here the LECs) resulting from the uncertainty in \terminol{input data} (here e.g. the lattice data points).
\item
\label{item-steptheory}
\terminol{Theory error analysis:} Verify that the assumption of a negligible/tolerable theory error in step \ref{item-stepstat} is justified. If it is, the uncertainties determined in step \ref{item-stepstat} can be accepted as the final result of our analysis. If not, try to estimate an improved range of applicability. For an effective theory this means: Make an intelligent guess about the magnitude of the deviation of the fit function from truth, which results from the fact that the fit function is a truncated series expansion. To make this guess, the approximate size of the LECs is needed. There is no other choice but to make use of the results of step \ref{item-stepstat}. Find the upper limits of the expansion variables where theoretic errors are expected to be acceptable, and go back to task \ref{item-stepstat}.
\end{enumerate}

Despite the interplay of the two steps, theoretical and statistical error estimation require entirely different methods. It is dangerous to confuse the two different tasks, see section \ref{sec-gafpi}. Ways of estimating higher order effects will be illustrated in situ. In general, it is a ``dirty'' task, lacking mathematical rigor. 

\section{Convergence Properties}

Heuristic arguments show that perturbative QCD yields \terminol{asymptotic series} \cite{West99}. To understand what an asymptotic series is, consider a Taylor series expansion $S_n(z) \equiv \sum_{i=0}^n a_i z^i$ of some function $f(z)$. $S_n(z)$ is said to form an asymptotic series if 
\begin{align}
	\text{for fixed }z\ &: \qquad \lim_{n \rightarrow \infty} S_n(z) \neq f(z) , \\
	\text{but for fixed n}\ &: \qquad \exists C_n \ : \quad \left| S_n(z) - f(z) \right| \leq C_n z^{n+1} \quad \forall z
	\end{align}
In most cases, the series is divergent, i.e. $\lim_{n \rightarrow \infty} S_n(z)$ does not even exist.
Even though $S_n(z)$ does not converge towards the correct result for $n \rightarrow \infty$, it does yield good approximations for $f(z)$ at some order $n$ of the series, if only $z$ is sufficiently small. 
At fixed $z$, there is some order $n=n_\text{best}(z)$ where the series gives its best approximation to $f(z)$. When calculating $S_n(z)$ to higher and higher orders, results start becoming worse again beyond $n_\text{best}(z)$. Typically, the lower $z$, the higher is $n_\text{best}(z)$. For QCD matrix elements, the expansion parameter $z$ is related to the (running) coupling constant $\alpha_s$.

It is reasonable to suspect that $\chi$PT yields asymptotic series as well. For our nucleon mass formula, the expansion parameter $z$ is related to the pion mass $m_\pi/\Lambda_\chi$. In particular at larger pion masses, we must take into consideration the possibility that efforts to go to the next order are in vain, and turn out to \emph{worsen} our result.

\chapter{Chiral Extrapolation of Nucleon Mass Lattice Data}

We turn to the numerical application of the machinery introduced in the previous chapters, fitting infrared regularized SU(2) B$\chi$PT nucleon mass expressions to unquenched two-flavor lattice data. 

\section{Survey of Available Lattice Data}

Our primary collection of two-flavor lattice data is that of ref. \cite{K04}, listing fully dynamical simulation results from the QCDSF, UKQCD, CP-PACS and JLQCD collaboration. Secondly, we make use of data presented in ref. \cite{OLS05}, comprising recent results from SESAM and related projects. The selection of input data has to match the capabilities of the model function. In a later section we will explore the ability of $\chi$PT to correct for artifacts due to the finite lattice simulation volume. Therefore, we already show here data with noticeable shifts due to smaller lattice volumes. Table \ref{tab-latticedat} lists the two-flavor lattice data used in plots and fits, up to a pion mass $m_\pi<1.0\units{GeV}$. The horizontal lines separate ''volume groups'' of equal simulation parameters ($\beta$, $\kappa$, collaboration). For each group, the simulation on the largest lattice is marked ''large $L$''. The following selection criteria are supposed to keep spurious lattice artifacts at a tolerable level:
	
\begin{itemize}
	\item lattice spacing 
		$a < \begin{cases} 0.15\units{fm} & \text{for }\mathcal{O}(a)\text{-improved Fermions; i.e. data from \cite{K04}} \\ 
		                   0.10\units{fm} & \text{for } \text{unimproved Fermions; i.e. data from \cite{OLS05}} \end{cases}$
	\item simulation volume side length $ L > 1.0 \units{fm} $ , because, according to \cite{OLS05}, 
		finite volume corrections based on $\chi$PT seem to be applicable in this regime.
	\item For each volume group, the simulation in the largest volume fulfills $m_\pi L > 5$.
		These points are marked ''large $L$'' and exhibit negligible finite volume effects. 
		(The pion cloud, whose size is of the order of the Compton wave length $1/m_\pi$,
		must fit well into the simulation volume.)
	\end{itemize} \par
	
The criteria for the ''large $L$'' points for improved Fermions are those of ref. \cite{K04}, also employed in ref. \cite{PHW04}.

For points 53 and 54, no calculation has been performed for the Sommer radius $r_0/a$, which is needed to determine the lattice spacing $a$. In accordance with ref. \cite{OLS05}, $r_0/a$ for these two points is copied from point 56, which features the same simulation parameters $\beta$, $\kappa$.

\begin{figure}[h]
	\centering
	\includegraphics{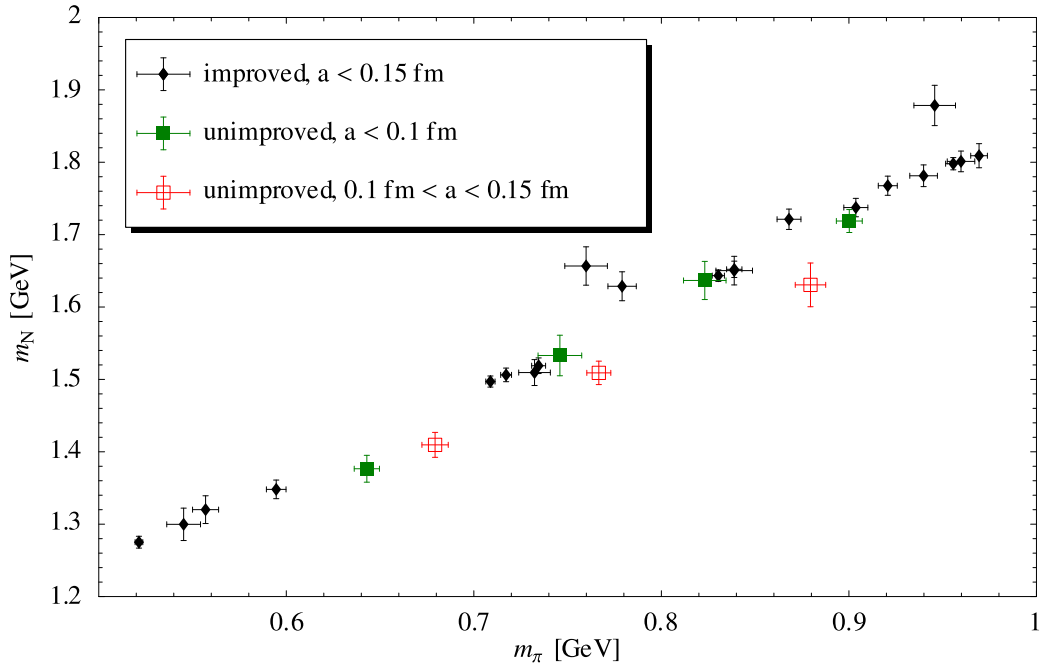}
	\caption{''Large $L$'' ($m_\pi L > 5$) lattice data from simulations with improved Fermions \cite{K04} and and unimproved Wilson Fermions \cite{OLS05}.}
	\label{fig-discretization}
	\end{figure}

For the unimproved Fermions of ref. \cite{OLS05}, we expect larger discretization errors at comparable lattice spacings $a$. Looking at fig. \ref{fig-discretization}, the decision to place the cut for unimproved Fermions at $0.10\units{fm}$ seems reasonable: Unimproved Fermion calculations for coarser lattices visibly deviate from the general trend of high quality data points. We have marked them ``coarse'' in table \ref{tab-latticedat}.

The conversion of lattice data to physical units has been performed according to eq. \ref{eq-latconversion}. Statistical errors in masses $am$ and in the length scale $r_0/a$ have been propagated quadratically.
	

\section{Application to Large Volume Lattice Data}
\label{sec-infvolband}

A fit of the expression $m_N^{(\leq 4\cdot)}$ as in eq. (\ref{eq-massp4}) to lattice data has already been shown in ref. \cite{PHW04}.  This is to be supplemented with a thorough discussion of uncertainties. 

\subsection{Higher Order Effects in the Pion Mass}
\label{sec-highordpi}

\begin{figure}[h]
	\centering
	a) \topalignbox{\includegraphics[width=0.4\textwidth,clip=1,trim=-22 0 0 -20]{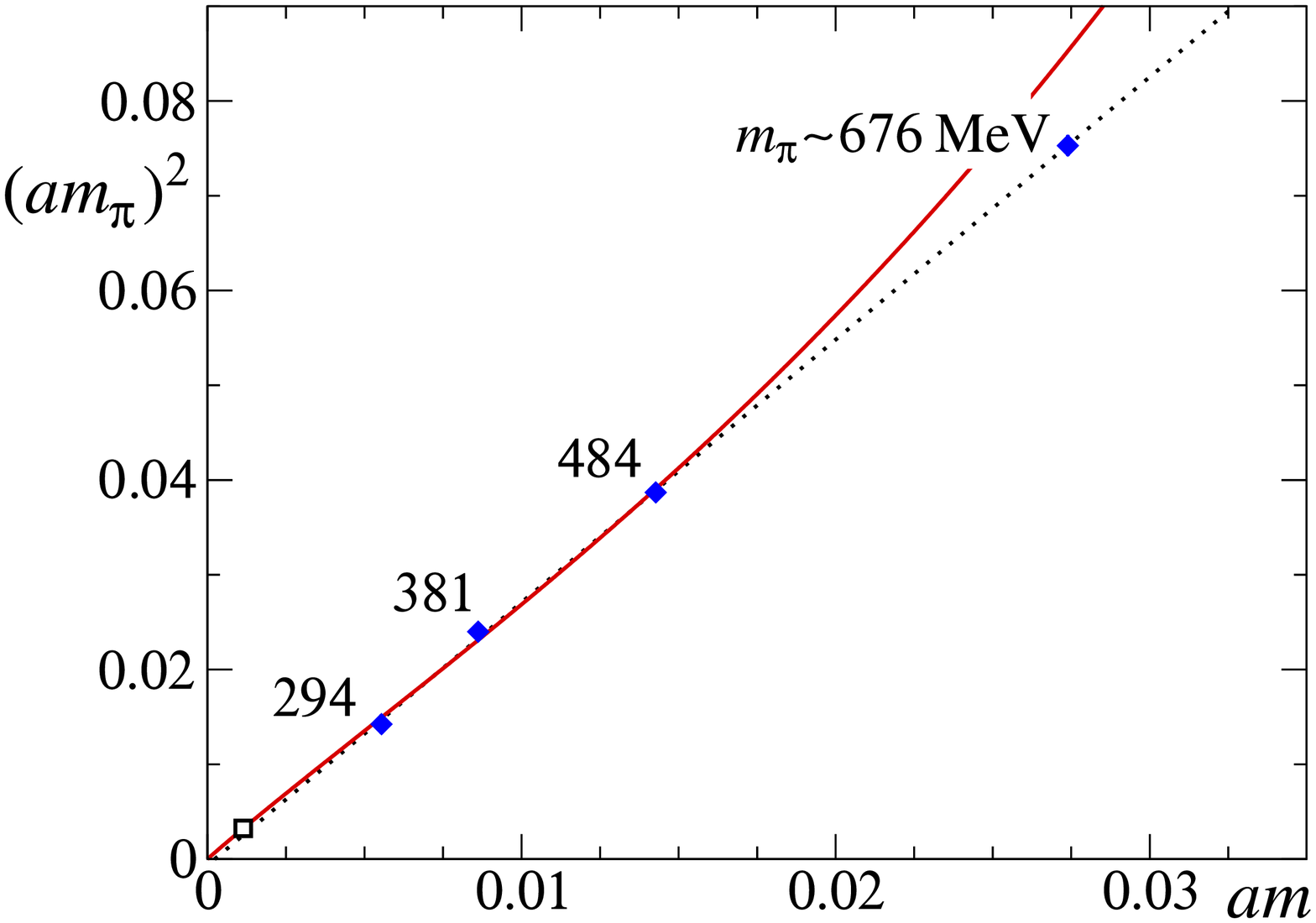}} \hfill
	b) \topalignbox{\includegraphics[width=0.5\textwidth]{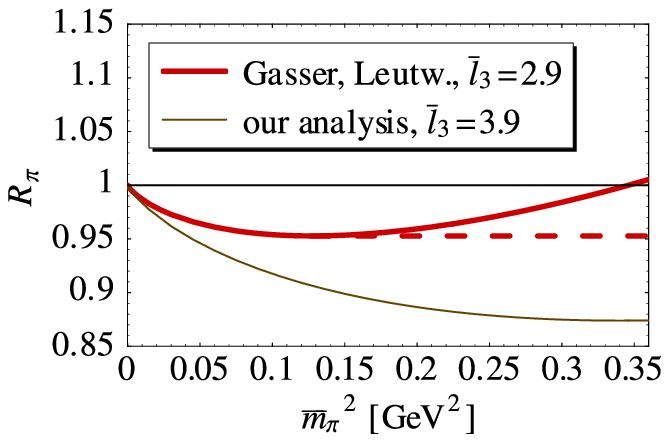}} \hfill {}\\
	\caption{a) Taken over from ref. \cite{Lue05}: SU(2) lattice results for the pion mass plotted versus the current quark mass $m$ in lattice units. The dotted line is a linear fit of all four data points. Below $m_\pi=500\units{MeV}$, one-loop chiral perturbation theory at order $p^4$ fits the data equally well (full line). The point represented by an open square is obtained by extrapolation to the physical pion mass. \quad
	b)~$p^4$~correction factor $R_\pi$ to GMOR. The thick solid line depicts the central value of the result of ref. \cite{GL84} $\overline{l}_3=2.9 \pm 2.4$. The low lying brown line corresponds to our analysis in chapter \ref{sec-mesonfit} finding $\overline{\elll}_3=3.9$, see also fig. \ref{fig-mesonfitrpi}. The dashed line figuratively suggests that effects beyond the $p^4$ approximation might level out the correction factor $R_\pi$, which is what lattice data indicates.}
	\label{fig-lueschpl}
	\end{figure}

Actually, 
$m_N^{(\leq 4\cdot)} = m_N^{(\leq 4\cdot)}(\overline{m}_\pi; m_0,g_A,f_\pi,c_1,c_2,c_3,e_1^{(4)}(\lambda))$ 
is a function of the leading order pion mass $\overline{m}_\pi = \sqrt{2 B \hat{m}}$ -- not the physical pion mass $m_\pi$ provided by lattice calculations. Our assumption here is that the difference is negligible. This assumption seems to be justified, because the formula
\begin{equation}
	m_\pi^2 \cong 2 B_{1} \hat m 
	\label{eq-latGMOR}
	\end{equation}
describes lattice calculations of the pion mass as a function of the quark mass very well, if $B_{1}$ is fitted to the data, see fig. \ref{fig-lueschpl}a. It appears self-evident that this means that that the GMOR relation is very accurate without further corrections at higher orders:
\begin{equation}
	m_\pi^2 \cong \overline{m}_\pi^2 = 2 B \hat m 
	\end{equation}
In other words, the slope $B_1$ observed on the lattice ``naturally'' is identical to the parameter $B$. We are going to accept this common presumption for our SU(2) fits. Yet, I would like to mention that this is still an open issue. Let us take a look at the pion mass expansion to order $p^4$ \cite{GL84}:
\begin{equation}
	m_\pi^2 = \overline{m}_\pi^2 + \frac{\overline{m}_\pi^4}{(4 \pi f_\pi^0)^2} \left(2 (4\pi)^2 \elll_3^r(\lambda) + \frac{1}{2} \ln  \frac{\overline{m}_\pi^2}{\lambda^2} \right) + \mathcal{O}(p^6) = \overline{m}_\pi^2 R_\pi
	\label{eq-pionmasssu2}
	\end{equation}
where $R_\pi$ can be expressed as 
\begin{equation}
	R_\pi = 1 + \frac{\overline{m}_\pi^2}{2(4\pi f_\pi^0)^2} 
	\ln \frac{\overline{m}_\pi^2}{\Lambda_\pi^2}  + \mathcal{O}(p^6)\ ,\qquad
	\Lambda_\pi^2 \equiv \lambda^2 \exp\left[ -4 (4\pi)^2 \elll_3^r(\lambda)\right]
	\end{equation}
The correction term vanishes in the chiral limit and at $\overline{m}_\pi^2 = \Lambda_\pi^2$. In between lies a minimum whose magnitude is proportional to $\Lambda_\pi^2$. No matter how we choose $\Lambda_\pi$, we get sizeable corrections for pion masses below $0.6\units{GeV}$. Note that lattice calculations have not reached pion masses much lower than $0.3\units{GeV}$. It is possible that deviations from eq. (\ref{eq-latGMOR}) set in there. Let us take, for example, the central value of the estimate $\overline{\elll}_3 = -4 (4\pi)^2 \elll_3^r(\overline{m}_\pi) = 2.9 \pm 2.4$ obtained from the study of meson and quark mass ratios in ref. \cite{GL84}. As in ref. \cite{Lue05}, we plot $R_\pi$ over $\overline{m}_\pi^2$, using $f_\pi^0 \approx 0.0924\units{GeV}$, see fig \ref{fig-lueschpl}b. Since the value of $B$ is determined from the fit, any functional behavior of $R_\pi(\overline{m}_\pi)$ which is approximately constant for $m_\pi>0.3\units{GeV}$ is compliant with lattice data. In other words, we can still fulfill eq. \ref{eq-latGMOR} in the range of available lattice data, if we make higher order corrections starting at order $p^6$ responsible for the formation of a plateau (figuratively displayed as the dashed line). Note however, that now $B \neq B_1$. Since we surely work in the plateau region of $R_\pi$, we could now utilize
\begin{equation}
	\overline{m}_\pi = \frac{B}{B_1} m_\pi
	\end{equation}
as input to our nucleon formula -- if we only knew $B/B_1$. Just to get an impression of a possible magnitude of this ratio, let us make the assumption that the plateau lies at the level of the minimum of $R_\pi$, as plotted in fig. \ref{fig-lueschpl}. Then for $\overline{l}_3 = 2.9$ we obtain $B_1/B=0.95$.
An attempt to extract information from MILC SU(3) data in section \ref{sec-mesonfit} even suggests $B_1/B \approx 0.9$ in the relevant range, see the low lying brown line in fig. \ref{fig-lueschpl}b.

For the practical purposes in the following SU(2) nucleon mass fits, we have ignored the concerns about higher order effects in the pion mass, and have made no difference between $m_\pi$ and $\overline{m}_\pi$.

\subsection{Setting up the Fit}

Without loss of generality, we set the regularization scale to $\lambda=1\units{GeV}$.
Restricting ourselves to ''large $L$'' lattice data with the smallest available pion masses
we pick points no. 19, 41, 8 and 23 from table \ref{tab-latticedat}.\footnote{Our choice to select data with $m_\pi < 0.6 \units{GeV}$ was initially motivated by a recommendation in ref. \cite{BHM04} not to go beyond pion masses larger than $0.6\;\ldots\;0.7\units{GeV}$.}
In order to avoid an under-determined fit, we must fix some parameters.
We eliminate $c_1$ by substituting the experimental values of pion and nucleon mass
\begin{equation*}
	m_\pi^\text{phys} = 0.138\units{GeV}\ , \qquad m_N^\text{phys} = 0.938\units{GeV} 
	\end{equation*}
into formula (\ref{eq-massp4}):
\begin{equation}
	m_N^{(\leq 4\cdot)}(m_\pi^\text{phys}; m_0,g_A^0,f_\pi^0,c_1,c_2,c_3,e_1^{(4)}(\lambda)) \mathop{=}^{!} m_N^\text{phys}
	\quad \Rightarrow \quad c_1(m_0,g_A^0,f_\pi^0,c_2,c_3,e_1^{(4)})
	\nonumber
	\end{equation}
The LEC $c_2$ is set to $3.2\units{GeV^{-1}}$ as determined from pion-nucleon scattering in ref. \cite{FMS},
and $c_3$ is fixed at $-3.4\units{GeV^{-1}}$, according to the $NN$ phase shift analysis in ref. \cite{EM02}.
Particularly the uncertainty about $c_3$ is substantial, but difficult to quantify.
For the moment, we ignore uncertainties in $c_2$ and $c_3$, and defer the discussion to section \ref{sec-empconstr}.
As a first approximation, we set $g_A^0$ and $f_\pi^0$ equal to their values
at the physical point 
\begin{equation}
	g_A^0 \approx g_A^\text{phys} = 1.267, \qquad f_\pi^0 \approx f_\pi^\text{phys} = 92.4\units{MeV}
	\end{equation}
Actually, the chiral limit values $g_A^0$ and $f_\pi^0$ are expected to differ slightly from the values at the physical point. We can safely assume to find $g_A^0$ and $f_\pi^0$ in the intervals
\begin{equation}
	g_A^0 =  1.1\ ..\ 1.3 , \qquad
	f_\pi^0 =  86.2\units{MeV}\ ..\ (92.4\units{MeV}\!=\!f_\pi^\text{phys})
	\label{eq-gafpiinterv}
	\end{equation}
The range for $g_A^0$ is taken from \cite{HPW03}. For the lower boundary of $f_\pi^0$, we
take the lowest estimate we found in the literature: a numerical analysis of the
pion mass dependence of $f_\pi$ in ref. \cite{Goc04}.
Eventually, only two parameters remain free:  $m_0$ and $e_1^{(4)}(\lambda=1\units{GeV})$. 

In summary, we have 
fixed parameters $\ul{c}=(g_A^0,f_\pi^0,c_2,c_3)$ and 
fit parameters  $\ul{p}=(m_0,e_1^{(4)})$.

\subsection{The Calculations}

\begin{figure}[h!]
	\vspace{18pt}
	\topalignbox{a) }\topalignbox{\includegraphics[width=0.9\textwidth, trim = -18 0 30 40]{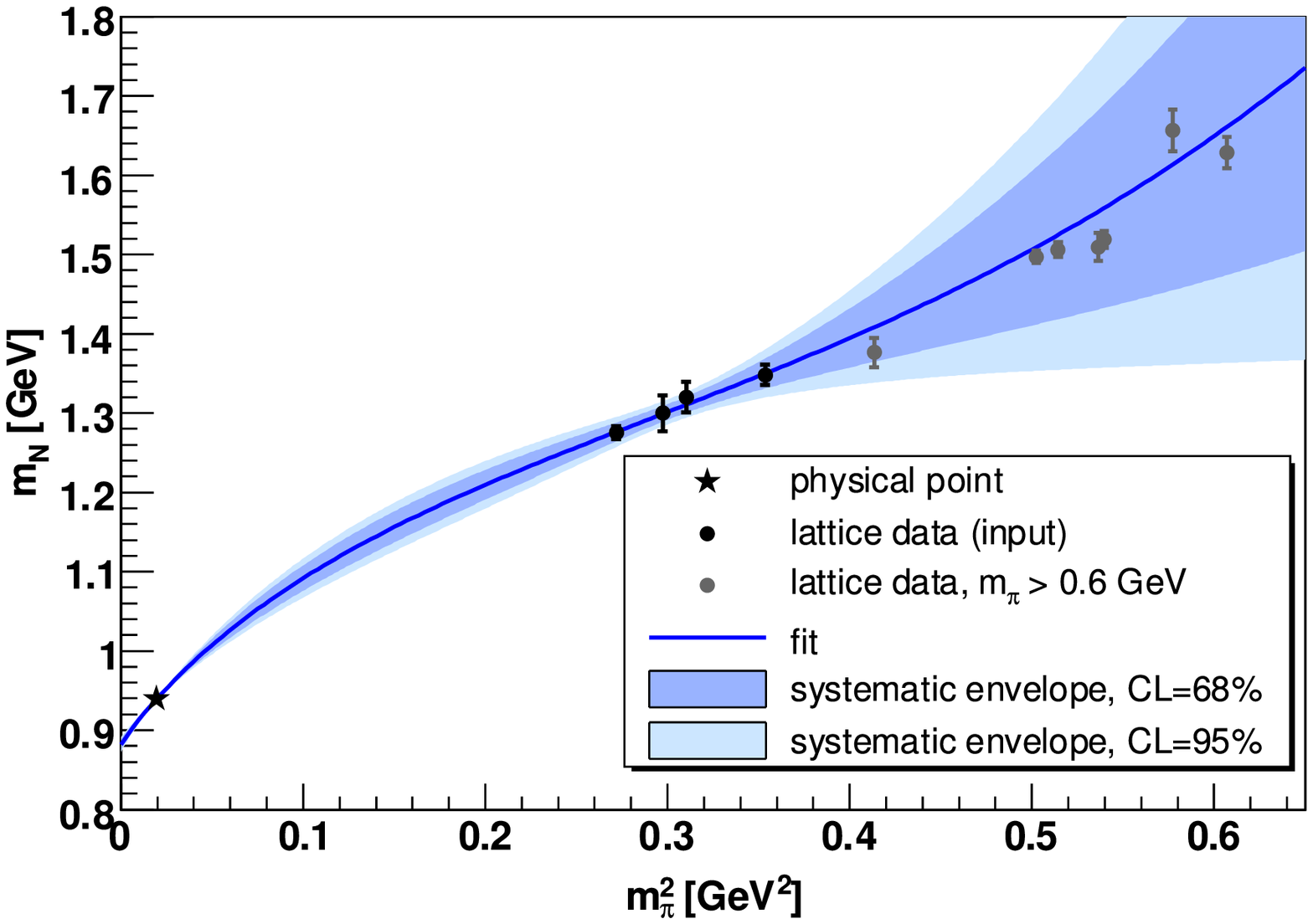}} \\
	\topalignbox{b) }\topalignbox{\includegraphics[width=0.9\textwidth]{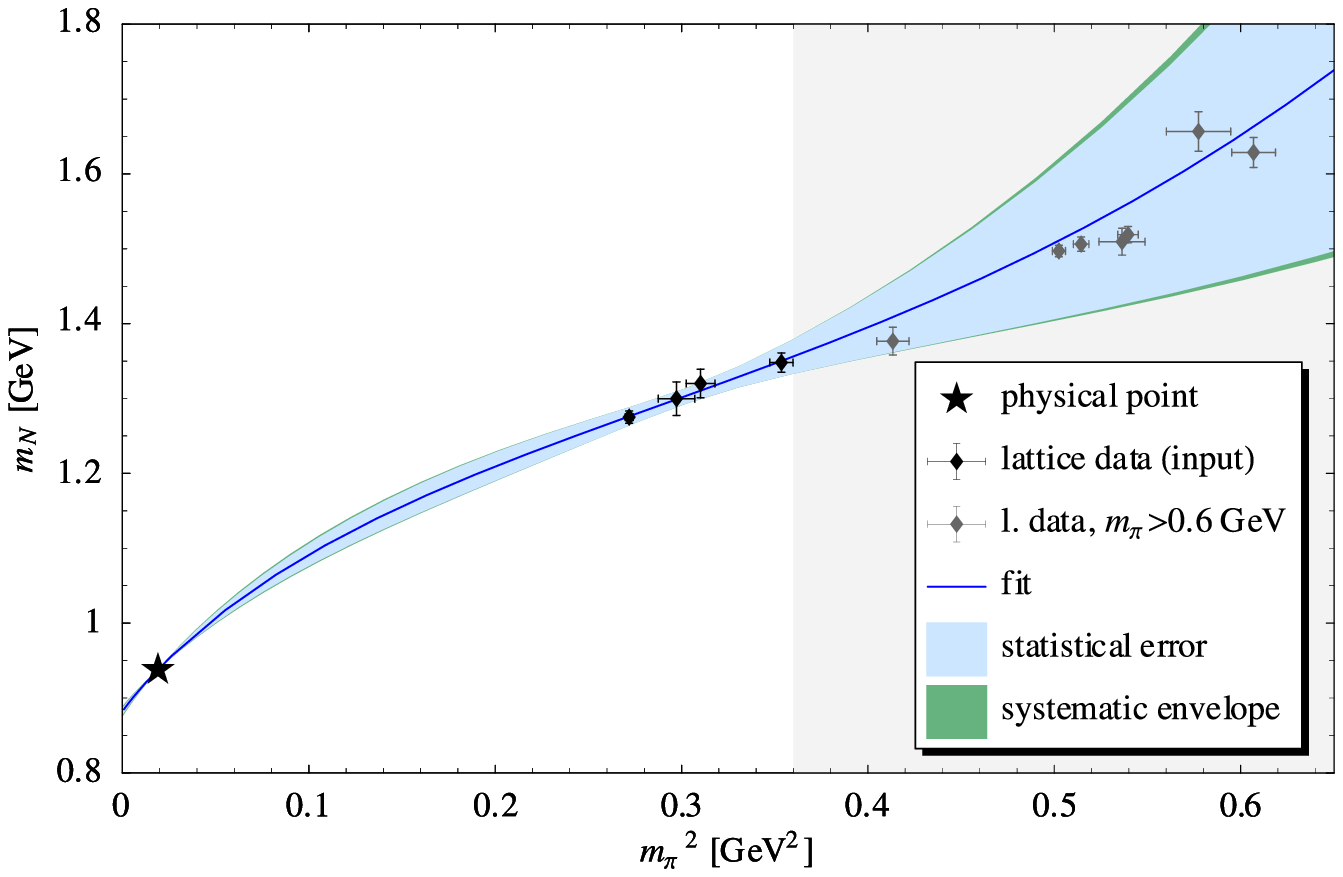}} \\
	\caption{Global statistical error band for a fit to large volume lattice data with $m_\pi<\nobreak 0.6\units{GeV}$. The physical point is included, and we fixed $c_2=3.2\units{GeV^{-1}}$, $c_3=-3.4\units{GeV^{-1}}$. The grey lattice data points have not been used as input. They are only shown for comparison. a) Fit II: systematic envelope from the Monte Carlo algorithm at confidence levels $\CL=68\%$ and $\CL=95\%$, pion mass errors ignored b) Fit IIb: error analysis in quadratic approximation at $\CL=68\%$, pion mass errors included.}
	\label{fig-band}
	\end{figure}

Originally, calculations were performed using the Monte Carlo algorithm described in section \ref{sec-algoMC}. Errors in the pion mass were neglected. $g_A^0$ and $f_\pi^0$ were varied on a uniform $11\times11$ grid spanning the range (\ref{eq-gafpiinterv}). The resulting error bands and confidence regions were combined to what we termed a ``\terminol{systematic envelope}''. As an example, for the band calculation of the $\CL=68\%$ systematic envelope, the number of parameter trials was in total $2 \times 10^8$. Thereof $2\times10^7$ were in the region of confidence. The band grew in some direction 1023 times.
For $g_A^0$ and $f_\pi^0$ fixed at the physical values, the fit results of this setup are identical to those of Fit II in ref. \cite{PHW04}. They are listed again for convenience in table \ref{tab-params}.\footnote{The tables also list resulting values of other observables $T^+$, $P_1^+$ and $\sigma_N$ which we will discuss later.} Systematic envelopes of error bands are shown for $\CL=68\%$ and $\CL=95\%$ in fig. \ref{fig-band}a. In the plots, we choose $m_\pi^2$ as our abscissa, because it is approximately proportional to the quark mass $\hat m$. \par

Later, the quadratic approximation described in section \ref{sec-algomathematica} was employed, because of Mathematica's greater flexibility. Random inspections using expression (\ref{eq-algomatherr}) show that the extent of the confidence region does not deviate from the hyperellipsoidal approximation by much more than $5\%$ in any direction. The lattice error in the pion mass is now taken into account, but the effect is hardly noticeable. Fixing $g_A^0$ and $f_\pi^0$ at their physical values yields the ``statistical error band'' of fig. \ref{fig-band}b, at a confidence level $\CL=68\%$. The systematic envelope was calculated at the same confidence level, letting $g_A^0$ and $f_\pi^0$ assume values on a uniform $5\times5$ grid. The confidence intervals of the parameters are listed in table \ref{tab-paramsb}. 

Fig. \ref{fig-band} shows the resulting error bands calculated with the two different algorithms.

\renewcommand{\arraystretch}{1.15}
\begin{table}[th]
	\caption{Fit II as in \cite{PHW04}: parameter values and observables}
	\label{tab-params}
	\centering
	\vspace{10pt}
\begin{tabular}{ll||cc}
	$e_1^{(4)}(1 \units{GeV})$ & $(\mathrm{GeV^{-3}})$  & 2.9 $\pm$ 0.7 & fitted  \\                
	$m_0$ & $(\mathrm{GeV})$                            & 0.8820 $\pm$ 0.0028 & fitted \\             
	$c_1$ & $(\mathrm{GeV^{-1}})$                       & -0.927 $\pm$ 0.037 & elim. \\          
	$g_A^0$ &                                           & 1.267 & fixed \\                    
	$f_\pi^0$ & $(\mathrm{GeV})$                        & 0.0924 & fixed \\                 
	$c_2$ & $(\mathrm{GeV^{-1}})$                       & 3.2 & fixed \\                        
	$c_3$ & $(\mathrm{GeV^{-1}})$                       & -3.4 & fixed  \\  
	\hline                    
	$\chi^2/\text{d.o.f.}$ &                            & 0.15 & \\     
	\hline        
	$\sigma_N$ & $(\mathrm{GeV})$                       & 0.0491 $\pm$ 0.0028 &          
	\end{tabular} \par    
\end{table}
\renewcommand{\arraystretch}{1}

\renewcommand{\arraystretch}{1.15}
\begin{table}[th]
	\caption{Fit IIb: parameter values and observables. Pion mass errors have been taken into account.}
	\label{tab-paramsb}
	\centering
	\vspace{10pt}
\begin{tabular}{ll||cc|cc}
	 & & \multicolumn{2}{l|}{(a) statistical error} & \multicolumn{2}{l}{(b) systematic envelope}  \\
	\hline
	$e_1^{(4)}(1 \units{GeV})$ & $(\mathrm{GeV^{-3}})$  &  3.0 $\pm$ 0.7 & fitted & 1.0 .. 4.6 & fitted \\   
	$m_0$ & $(\mathrm{GeV})$                            &  0.8820 $\pm$ 0.0030 & fitted & 0.876 .. 0.888 & fitted \\
	$c_1$ & $(\mathrm{GeV^{-1}})$                       & -0.927 $\pm$ 0.039 & elim. & -1.04 .. -0.82 & elim. \\     
	$g_A^0$ &                                           & 1.267 & fixed & 1.10 .. 1.30 & scanned \\         
	$f_\pi^0$ & $(\mathrm{GeV})$                        & 0.0924 & fixed  & 0.0862 .. 0.0924 & scanned \\    
	$c_2$ & $(\mathrm{GeV^{-1}})$                       & 3.2 & fixed & 3.2 & fixed \\                            
	$c_3$ & $(\mathrm{GeV^{-1}})$                       & -3.4 & fixed & -3.4 & fixed \\ 
	\hline                         
	$\chi^2/\text{d.o.f.}$ &                            & 0.13 & & 0.1268 .. 0.1346 & \\         
	$T^{+(\leq 3)}$ & $(\mathrm{GeV^{-1}})$             & 7.22 $\pm$ 0.35 & & 6.3 .. 9.6 & \\            
	$P_1^{+(\leq 3)}$ & $(\mathrm{GeV^{-3}})$           & 598.69 $\pm$ 0.45 & & 570. .. 770. & \\        
	$\sigma_N$ & $(\mathrm{GeV})$                       & 0.0490 $\pm$ 0.0029 & & 0.0446 .. 0.0537 & \\  
	\hline
	$\tilde T^{+(\leq 3)}$ & $(\mathrm{GeV^{-1}})$      & 2.67 $\pm$ 0.35 & & 1.8 .. 4.3 & \\
	$\tilde P_1^{+(\leq 3)}$ & $(\mathrm{GeV^{-3}})$    & 837.63 $\pm$ 0.45 & & 810. .. 1050. & \\
	\hline
	$e_1^{(3)}(1 \units{GeV})$ & $(\mathrm{GeV^{-3}})$  & 1.83 & & & fitted \\
	\end{tabular} \par    
\end{table}
\renewcommand{\arraystretch}{1}

\begin{figure}[h!]
	\topalignbox{a) }\topalignbox{\includegraphics[width=0.45\textwidth]{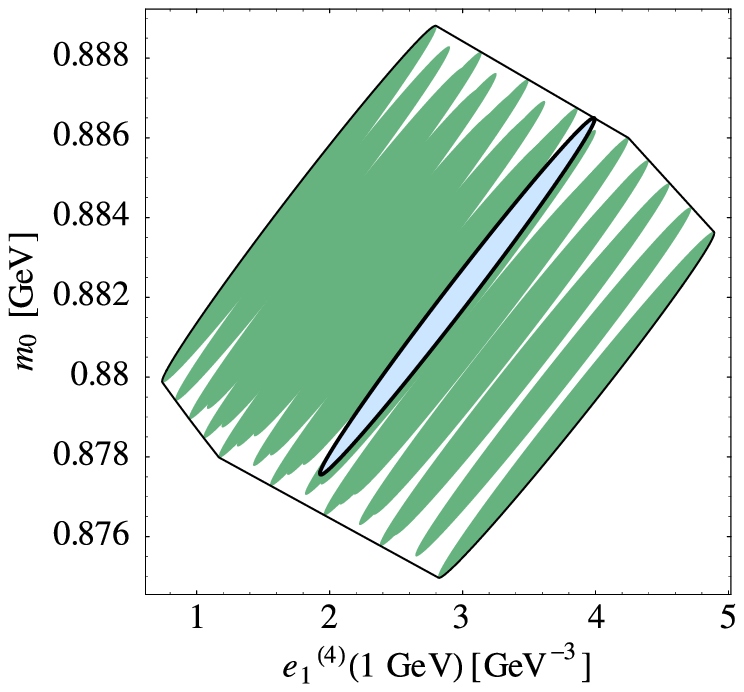}} 
	\hfill {}
	\topalignbox{b) }\topalignbox{\includegraphics[width=0.45\textwidth]{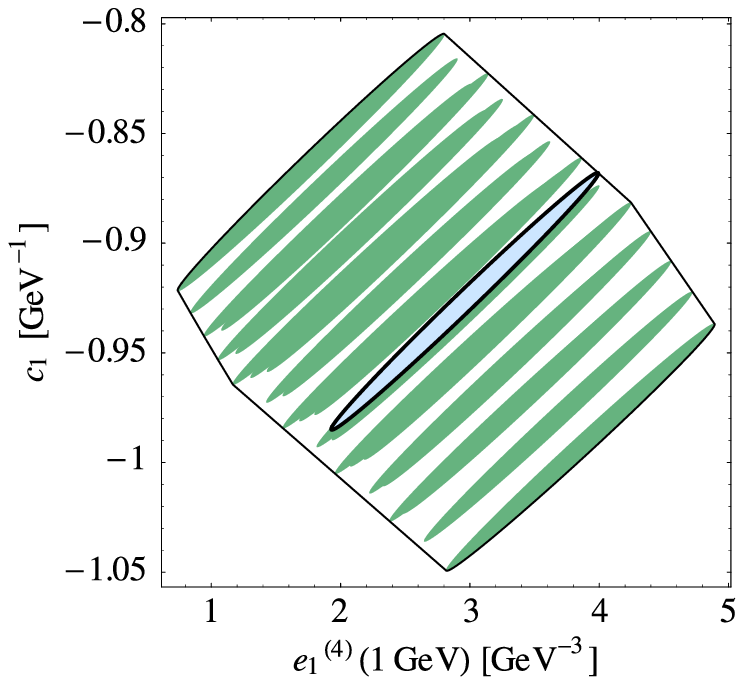}}
	\hfill {} \\
	\topalignbox{c) }\topalignbox{\includegraphics[width=0.45\textwidth]{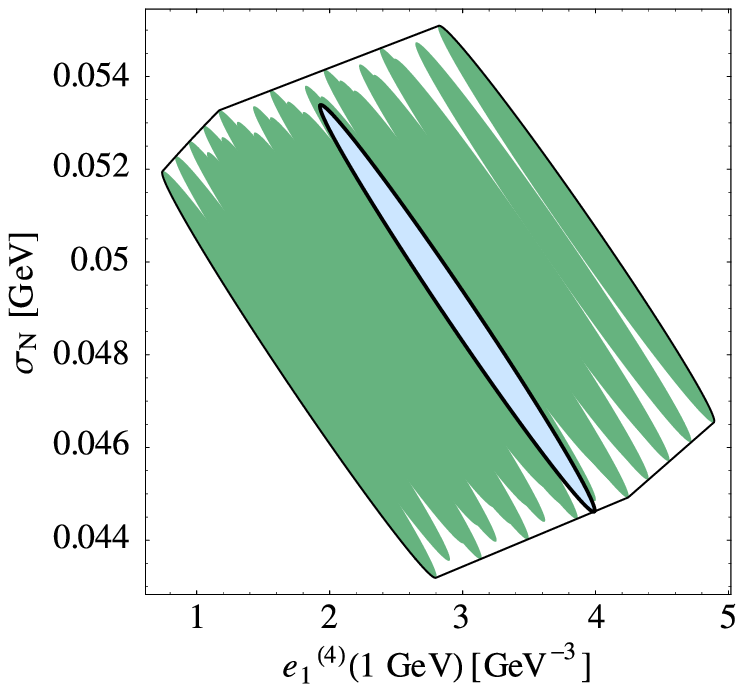}}
	\hfill {} 
	\topalignbox{d) }\topalignbox{\includegraphics[width=0.45\textwidth]{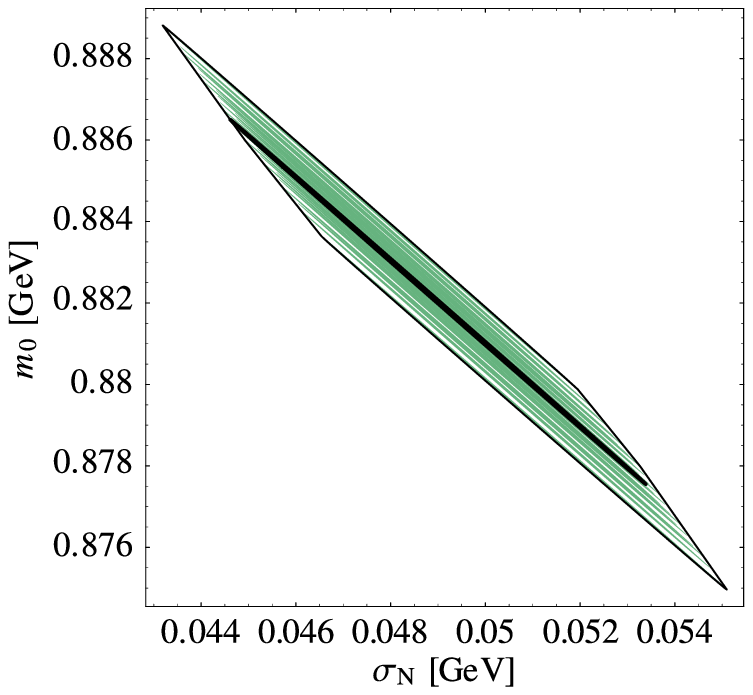}} 
	\hfill {}\\
	\caption{$\CL=68\%$ 2-parameter confidence region for Fit IIb. Light blue oval with thick line: confidence region for $g_A^0=g_A^\text{phys}$, $f_\pi^0=f_\pi^\text{phys}$. Green regions: confidence regions for a scan of $g_A^0$ and $f_\pi^0$. Together they form the systematic envelope, bordered by the outer line. The confidence region for the fit parameters a) has been translated in b), c), d) to show the statistical correlation between other quantities of interest.}
	\label{fig-confreg}
	\end{figure}
	
\subsection{Discussion of the Result}

In the limit $m_\pi^2 \rightarrow 0$, the error band remains narrow,
reflecting our ability to determine $m_0$ accurately.
(Of course, this prediction relies heavily on the inclusion of the physical point,
which is rather close to the chiral limit.)
For pion masses below $0.6\units{GeV}$, i.e. left of the four selected lattice points,
the band does not bulge much. As long as information about the physical point
and the LECs is included, chiral perturbation theory
provides a stable interpolant which is conditioned well and can make predictions within its
range of validity.
The agreement between best-fit curve and lattice data up to $m_\pi \approx 750\,{\rm{MeV}}$, 
close to the typical chiral symmetry breaking scale $\Lambda_\chi$, 
has to be considered accidental with present information in input. 
The diverging statistical band indicates that the predictive power of our analysis about the shape of the curve in the region $m_\pi > 600\,{\rm MeV}$ is low: the information we have in input (in the form of lattice data and parameters), constrains the shape of the interpolating functions weakly at pion masses larger than $600\,{\rm MeV}$. This coincides with our expectation that higher order effects become intolerable at larger pion masses.

Table \ref{tab-params} gives a summary of parameters and single parameter error bounds. 
Note that the individual fit parameters depend heavily on the choice of $g_A^0$ and $f_\pi^0$,
resulting in large ''systematic envelopes'' in table \ref{tab-params}, column (c).
The same phenomenon can be observed visually in plots of the confidence region, see fig. \ref{fig-confreg}.
Note that the statistical correlation between $e_1^{(4)}$ and $m_0$ is considerable, but still allows us to perform a stable fit. In contrast, a parametrization in terms of $m_0$ and the pion nucleon sigma term $\sigma_N$ would not have been possible -- here the confidence ellipse degenerates to a line, see fig. \ref{fig-confreg}d). This is due to the fact that $e_1^{(4)}$ is essential for the description of the curve in the range where we have data.


\subsection{Higher Order Effects in $g_A$ and $f_\pi$}
\label{sec-gafpi}

Like the nucleon mass, the axial-vector coupling constant $g_A$ and the pion decay constant $f_\pi$ can be expanded in terms of quark masses, or more conveniently $\overline{m}_\pi$. They differ from their chiral limit values by correction terms starting at order $\overline{m}_\pi^2$, see e.g. ref \cite{BKM95}:
\begin{equation}
	g_A^0 = g_A(\overline{m}_\pi) ( 1 + \mathcal{O}(\overline{m}_\pi^2) )\ , \quad
	f_\pi^0 = f_\pi(\overline{m}_\pi) ( 1 + \mathcal{O}(\overline{m}_\pi^2) )
	\end{equation}
When substituting these expressions into $m_N^{(\leq 4\cdot)}$, the corrections appear at order $\overline{m}_\pi^5$.

\begin{figure}[h!]
	\centering

	\unitlength=1mm
	\begin{fmffile}{axialcoupl}
	\fmfset{arrow_len}{3mm}
	\fmfset{arrow_ang}{15}

	\begin{tabular}{cccccccc}

		\begin{fmfgraph*}(27,13)
		\fmfleft{l}
		\fmfright{r}
		\fmfforce{(0.0w,0.0h)}{l}
		\fmfforce{(1.0w,0.0h)}{r}
		\fmfforce{(0.25w,0.0h)}{ol}
		\fmfforce{(0.75w,0.0h)}{or}
		\fmf{fermion}{l,ol}
		\fmf{fermion,label=$N$,l.side=right}{ol,or}
		\fmf{fermion}{or,r}
		\fmf{dashes,left=1,tension=0.5,label=$\pi$,l.side=right}{ol,or}
		\fmfblob{5mm}{ol,or}
		\end{fmfgraph*}
		&

		$\longrightarrow$ &
		
		\begin{fmfgraph*}(18,13)
		\fmfleft{l}
		\fmfright{r}
		\fmftop{t}
		\fmfforce{(0.0w,0.0h)}{l}
		\fmfforce{(1.0w,0.0h)}{r}
		\fmfforce{(0.5w,0.0h)}{oc}
		\fmfforce{(0.5w,1.0h)}{t}
		\fmf{fermion}{l,oc}
		\fmf{fermion}{oc,r}
		\fmf{wiggly}{oc,t}
		\fmfblob{5mm}{oc}
		\end{fmfgraph*}
		&
		
		= &

		\begin{fmfgraph*}(18,13)
		\fmfleft{l}
		\fmfright{r}
		\fmftop{t}
		\fmfforce{(0.0w,0.0h)}{l}
		\fmfforce{(1.0w,0.0h)}{r}
		\fmfforce{(0.5w,0.0h)}{oc}
		\fmfforce{(0.5w,1.0h)}{t}
		\fmf{fermion}{l,oc}
		\fmf{fermion}{oc,r}
		\fmf{wiggly}{oc,t}
		\fmfdot{oc}
		\end{fmfgraph*}
		&
		
		+ &
		
		\begin{fmfgraph*}(27,13)
		\fmfleft{l}
		\fmfright{r}
		\fmftop{t}
		\fmfforce{(0.0w,0.0h)}{l}
		\fmfforce{(1.0w,0.0h)}{r}
		\fmfforce{(0.25w,0.0h)}{ol}
		\fmfforce{(0.75w,0.0h)}{or}
		\fmfforce{(0.5w,0.0h)}{oc}
		\fmfforce{(0.5w,1.0h)}{t}
		\fmf{fermion}{l,ol}
		\fmf{fermion}{ol,oc}
		\fmf{fermion}{oc,or}
		\fmf{fermion}{or,r}
		\fmf{dashes,right=1,tension=0.5,label=$\pi$,l.side=right}{ol,or}
		\fmf{wiggly}{oc,t}
		\fmfdot{ol,or,oc}
		\end{fmfgraph*}
		&
		
		$\ldots$
				
		\vspace{33pt} \\
		(1) & & (2) & & (3) & & (4) &
		\end{tabular}
	\end{fmffile}

	\caption{Corrections entering the pion loop contribution to the nucleon self energy. The pion in diagram (1) couples to the axial-vector current of the nucleon. (2)-(4): Some corrections to the interaction vertex can be explored by probing the nucleon with an external axial-vector field (wiggly line). These corrections are subsumed in $g_A$.}
	\label{fig-garundiag}
	\end{figure}
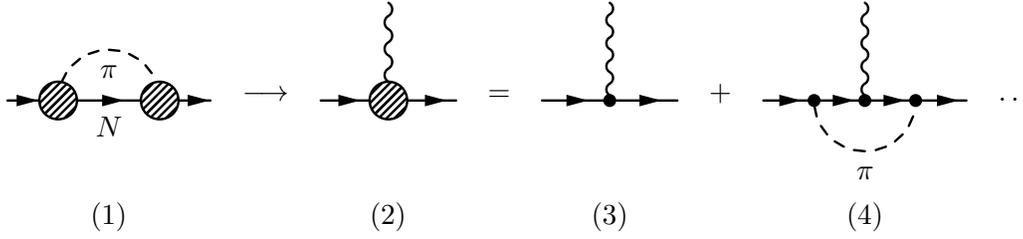
 
Using the \emph{full} axial-vector coupling $g_A(\overline{m}_\pi)$ and/or the \emph{full} pion decay constant $f_\pi(\overline{m}_\pi)$ as input to the nucleon mass formula
\begin{equation*}
	m_N^{(\leq 4\cdot)}(m_\pi,g_A(m_\pi),f_\pi(m_\pi),c1,c2,c3,e_1^{(4)})
	\end{equation*}
actually amounts to the inclusion of some graphs of higher order. In particular, it modifies the description of the coupling of the pion to the nucleon in diagram \ref{fig-massdiag} (a) which occurs at order $p^3$. The pion couples to the axial vector current of the nucleon, whose strength is parametrized by $g_A$. Thus corrections to $g_A$ will improve the description of the vertex, as illustrated in fig. \ref{fig-garundiag}. 

In this context, the findings in ref. \cite{GB99} should be mentioned: A two-loop heavy baryon calculation shows that the contributions to the nucleon mass at order $p^3$ and $p^5$ are
\begin{eqnarray}
	m_N^{(3)} + m_N^{(5)} 
	&=& -\frac{3\; g_{\pi N N}}{32 \;\pi\; m_N^2 } m_\pi^3 
	+ \frac{3\; g_A^2}{256\; \pi\; f_\pi^2\; m_0^2} m_\pi^5 + {\cal{O}}(m_\pi^6)
	\label{eq-massp5h}
	\end{eqnarray}
The term of order $m_\pi^5$ is already present in our expression $m_N^{(\leq 4 \cdot)}$, because it is a recoil correction automatically obtained from a relativistic calculation. All other contributions at order $p^5$ arise naturally through the use of the full pion-nucleon-nucleon coupling constant $g_{\pi N N}(m_\pi)$, the full pion mass $m_\pi$ and the full nucleon mass $m_N$ in the term proportional to $m_\pi^3$. There is a close relationship of $g_{\pi N N}$ to $g_A$ and $f_\pi$, the difference being called the Goldberger-Treiman discrepancy
\begin{equation}
	\Delta_{\pi N N} \equiv 1 - \frac{m_N\; g_A}{f_\pi\; g_{\pi N N}} = \mathcal{O}(m_\pi^2)
	\label{eq-goldbergertreiman}
	\end{equation}
$\Delta_{\pi N N}$ is known to be tiny at the physical point ($\sim 4 \%$ \cite{BKM95}), but we do not have information about its evolution with the pion mass yet. It is the Goldberger-Treiman discrepancy which is responsable for the fact that use of $g_A$ and $f_\pi$ instead of their chiral limit values does not suffice to promote our nucleon mass formula to order $p^5$.

All in all, the use $g_A(m_\pi)$ and/or $f_\pi(m_\pi)$ at least amounts to a partial evaluation of higher order diagrams, giving contributions starting at order $\overline{m}_\pi^5$.
Thus, as mentioned in ref. \cite{Beane}, evaluating
\begin{align}
	\delta m_N^{(5a)} & \equiv 
	m_N^{(\leq4\cdot)}(m_\pi,g_A(m_\pi),f_\pi(m_\pi),c_1,c_2,c_3,e_1^{(4)}) \nonumber \\
	& - m_N^{(\leq4\cdot)}(m_\pi,g_A^0,f_\pi^0,c_1,c_2,c_3,e_1^{(4)})
	\label{eq-deltagafpi}
	\end{align}
can serve as a measure of the size of some higher order contributions starting at order $\overline{m}_\pi^5$. Wherever these contributions are large, we must conclude that the next order correction is likely to be significant, and that the theoretical uncertainty should be estimated as high.

In an attempt to perform such a study on higher order contributions, ref. \cite{Beane} performs the following steps: Unknown parameters $\ul{p}$ are fitted to the data using $g_A^\text{phys}$ and $f_\pi^\text{phys}$. Next, confidence intervals for $g_A^0$ and $f_\pi^0$ similar to ours in eq. \ref{eq-gafpiinterv} are estimated. Now $m_N(\overline{m}_\pi,g_A^0,f_\pi^0,\ul{p})$ is plotted, keeping the fit parameters $\ul{p}$ constant and varying $g_A^0$ and $f_\pi^0$ within their confidence intervals. The result is a \emph{huge} band, giving the impression that higher order effects are intolerably large. However, to my present understanding, the author of ref. \cite{Beane} has confused statistical and theoretical error analysis. The produced band does neither possess the claimed meaning as an estimate of higher order effects, nor does it correctly reflect our uncertainties in the fitting procedure.
 
\begin{figure}[h!]
	\centering
	\includegraphics{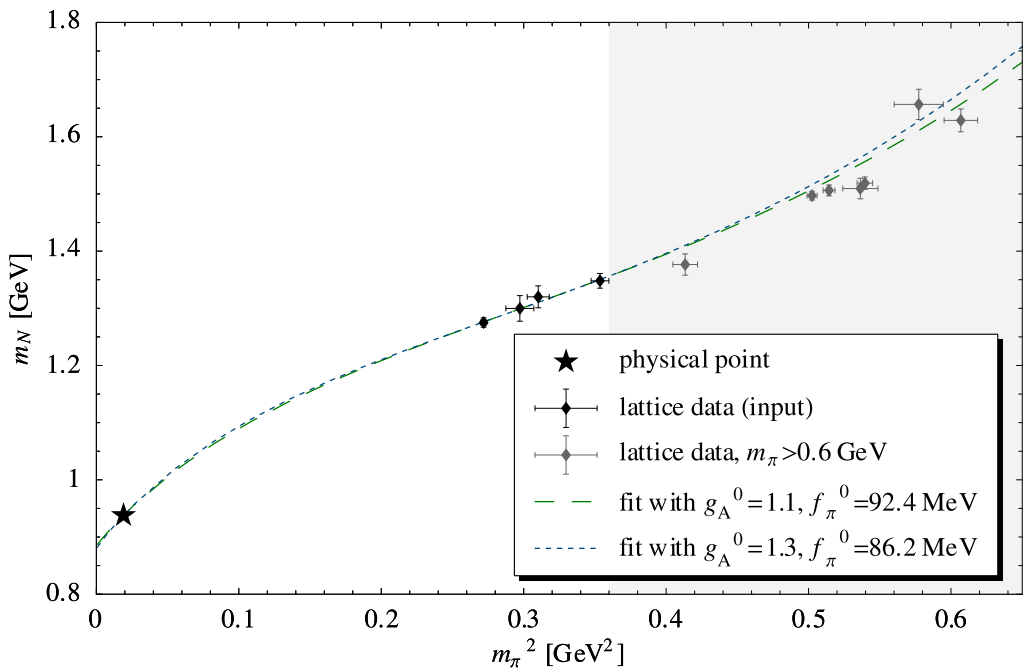}
	\caption{Fits at two different extreme choices of $g_A^0$ and $f_\pi^0$.}
	\label{fig-gafpivar}
	\end{figure}

Varying $g_A^0$ and $f_\pi^0$ within their confidence intervals is part of the \emph{statistical} analysis. Here, the impact of incomplete knowledge of \emph{input parameters} on our output function is assessed. In order to treat correlations correctly, it is necessary to \emph{refit} the parameters $\ul{p}$ for each choice of $g_A^0$ and $f_\pi^0$. Fig. \ref{fig-gafpivar} shows fits for two different extreme choices of $g_A^0$ and $f_\pi^0$. Obviously, the shape of the interpolant remains largely unaffected. The impact of uncertainties about $g_A^0$ and $f_\pi^0$ also manifests itself as the difference between our ``systematic envelope'' and our ``statistical error band'' in fig. \ref{fig-band}, which is quite small. Obviously, the effect of uncertainties in $g_A^0$ and $f_\pi^0$ is far less dramatic than pointed out in \cite{Beane}. Note that the magnitude of \emph{higher order corrections} does in no way depend on our \emph{uncertainties} in the input!

On the other hand, discussing higher order effects requires the knowledge of the \emph{functions} $g_A(m_\pi)$ and $f_\pi(m_\pi)$ over the whole range of $m_\pi$. Given these functions one may apply eq. \ref{eq-deltagafpi}. Of course, the procedure is consistent only if $g_A^0 = g_A(m_\pi=0)$ and $f_\pi^0 = f_\pi(m_\pi=0)$. The quark mass dependence of $f_\pi$ is still not very well known. Recent attempts to fit to lattice data do not perform well in the region where lattice data is available, see e.g. ref. \cite{Goc04}. For $g_A$, however, we may take the best fit curve with eq. (12) in ref. \cite{HPW03}. The formula therein has been calculated in non-relativistic SSE to order $\epsilon^3$. As usual, we fit $e_1^{(4)}$ and $m_0$ using our formula $m_N^{(\leq4\cdot)}(m_\pi,g_A(0),...)$, fixing $f_\pi^0=0.0924\units{GeV}$, $c_2=3.2\units{GeV^{-1}}$, $c_3=-3.4\units{GeV^{-3}}$. Next to this best fit, we plot $m_N^{(\leq4\cdot)}(m_\pi,g_A(m_\pi),...)$. The difference between the two plots is not severe, meaning that higher order effects from diagrams related to $g_A$ should not play much of a role. Again, we contradict the statement made in \cite{Beane}.

\begin{figure}[h!]
	\centering
	\includegraphics{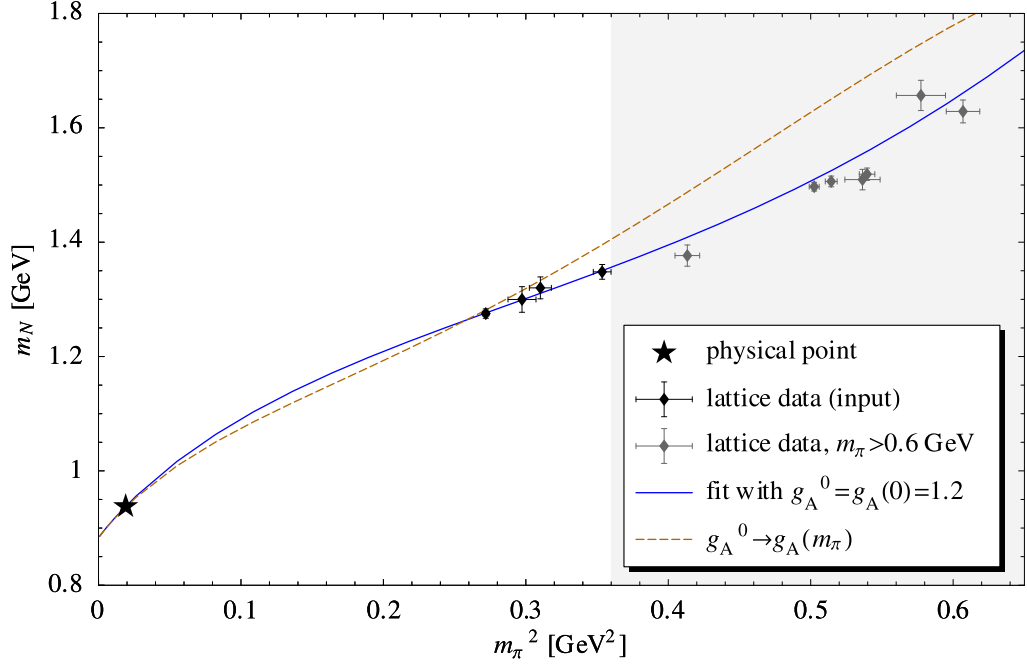}
	\caption{``Higher order effects'' made visible by replacing $g_A^0 \rightarrow g_A(m_\pi)$.}
	\label{fig-garun}
	\end{figure}

\subsection{Convergence}
\label{sec-conv}

The straightforward way of assessing convergence properties are plots at different orders in the perturbation expansion. For this purpose, we take parameters from our fit with the best fit function available to us: $m_N^{(\leq 4 \cdot)}$ from formula (\ref{eq-massp4}). Then we plot curves using these same parameters at lower orders, i.e., $m_N^{(\leq 3)}$ from formula (\ref{eq-massp3}) and $m_N^{(\leq 2)}$ from formula (\ref{eq-massp2}).
As mentioned in section \ref{sec-nucleonmass}, matters are complicated by the role of $\mathcal{L}_{\pi N}^{(4)}$ at different orders. $e_1^{(4)}(\lambda)$ and $e_1^{(3)}(\lambda)$ are actually not the same coupling constants. However, the functional form of $m_N^{(\leq 4\cdot)}$ does not permit us to discriminate between the contribution from $e_1^{(3)}(\lambda)$ and the additional contributions entering at order $p^4$. A naive approach to set $e_1^{(3)}(\lambda)=e_1^{(4)}(\lambda)$ is $\lambda$-dependent, because the running of the two coupling constants differs: Requiring scale independence of the individual formulae $m_N^{(\leq 3)}$ and $m_N^{(\leq 4 \cdot)}$, we obtain
\begin{align}
	e_1^{(3)}(\lambda) - e_1^{(3)}(\lambda_0) & = 
	- \frac{3}{32 \pi^2 (f_\pi^0)^2 } \frac{(g_A^0)^2}{m_0} \ln{\frac{\lambda}{\lambda_0}} \\
	e_1^{(4)}(\lambda) - e_1^{(4)}(\lambda_0) & = 
	- \frac{3}{32 \pi^2 (f_\pi^0)^2 }\left(\frac{(g_A^0)^2}{m_0} - 8 c_1+c_2+4c_3\right)\ln{\frac{\lambda}{\lambda_0}}
	\end{align}
The reason for the different running behavior are two more loop diagrams entering only in a genuine $p^4$ calculation, see fig. \ref{fig-massdiag} (b,c). 

At present, there is no way to figure out numerically which part of $e_1^{(4)}(\lambda)$ belongs to $e_1^{(3)}(\lambda)$. This thwarts our intention to import all parameters from the $p^4$-fit to $p^3$. The best we can do is to determine $e_1^{(3)}(\lambda)$ from a secondary one-parameter $p^3$-fit to the data. Of course, this compromises the stringency of our convergence test.

\begin{figure}[h!tb]
	\centering
	\includegraphics[width=\textwidth]{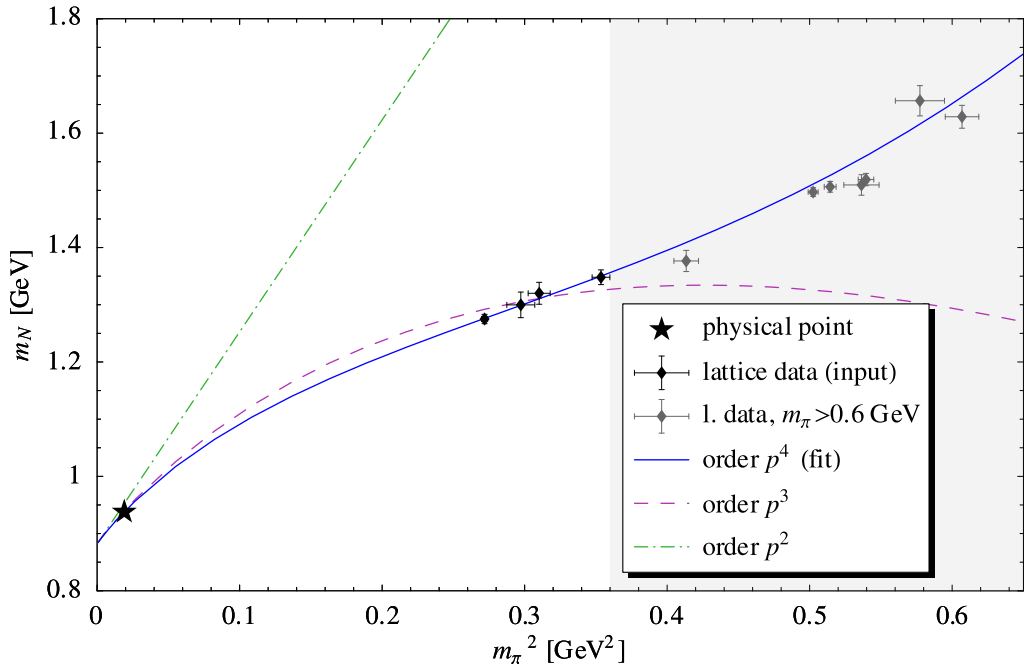}
	\caption{Study of convergence: nucleon mass predictions from formulae at order $p^2$, $p^3$ and $p^4$, using the parameters from the $p^4$ fit. At order $p^3$, $e_1^{(3)}(\lambda)$ cannot be imported from $p^4$ and needs to be determined from a secondary fit to the data. }
	\label{fig-conv}
	\end{figure}

For the plot in figure \ref{fig-conv}, the central values from the fit at order $p^4$ have been substituted into the formulae at order $p^4$, $p^3$ and $p^2$, leaving $e_1^{(3)}(\lambda)$ open. From a one-parameter fit of the $p^3$ function to lattice data we obtain $e_1^{(3)}(1\units{GeV}) = 1.38\units{GeV^{-3}}$. Figure \ref{fig-conv} conveys the impression that convergence is formidable in the whole area $m_\pi < 0.6\units{GeV}$. Here, the $p^3$ and $p^4$ curve run very closely together. In the shaded area $m_\pi > 0.6\units{GeV}$ the two curves separate quickly. At such large pion masses, the $p^4$ curve can no longer be treated as a small correction to the $p^3$ curve. It should be admitted, that this outcome of our convergence analysis is strongly influenced by the fit of $e_1^{(3)}(\lambda)$ to the data. Certainly, the intersection of the $p^3$ and $p^4$ curves in the midst of our data points is not an accident.

\begin{figure}[h!tb]
	\centering
	\includegraphics[width=\textwidth]{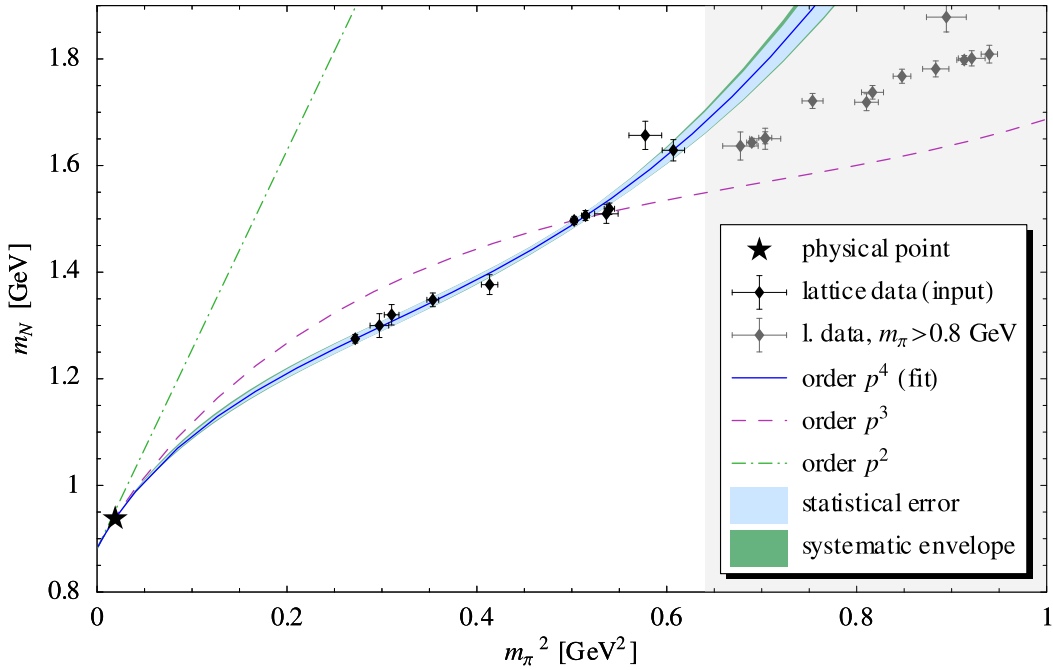}
	\caption{Study of convergence for a fit to data of pion  masses up to $0.8\units{GeV}$. }
	\label{fig-convhigh}
	\end{figure}

Therefore it is interesting to study a modification of our assumption on the range of applicability of the theory. Taking input from lattice data of pion masses up to $0.8\units{GeV}$, we obtain the results shown in fig. \ref{fig-convhigh}. The error band at order $p^4$ becomes more narrow due to the increased statistics. Again, the $p^3$ curve intersects the $p^4$ curve in the center of the lattice data range -- this time at a larger pion mass.
Thus our convergence analysis strongly depends on the placement of our input data. This is undesirable, yet it does not prevent us from drawing conclusions: The overall convergence inside our assumed range of applicability $m_\pi < 0.8\units{GeV}$ is visibly worse than it was for the  $m_\pi<0.6\units{GeV}$ fit, and that even though $e_1^{(3)}(\lambda)$ was chosen in the most favorable way.

\subsection{Ab-Initio Extrapolations?}

So far we needed to fix some LECs to values from previous empirical studies. This is not entirely satisfactory. What about our initial goal to make physical predictions from pure extrapolations of lattice data? Surely, such an ``ab-initio'' calculation would require lattice data points at lower pion masses. But how low? In order to address this question, we extend our input lattice data by placing ``fake'' data points on the best fit curve we already have. Then, band calculations are performed -- this time with less fixed parameters and without inclusion of the physical point. This way we can explore how future lattice calculations at lower pion masses will improve our ability to make predictions.

\begin{figure}[h!tb]
	\centering
	\includegraphics[width=\textwidth]{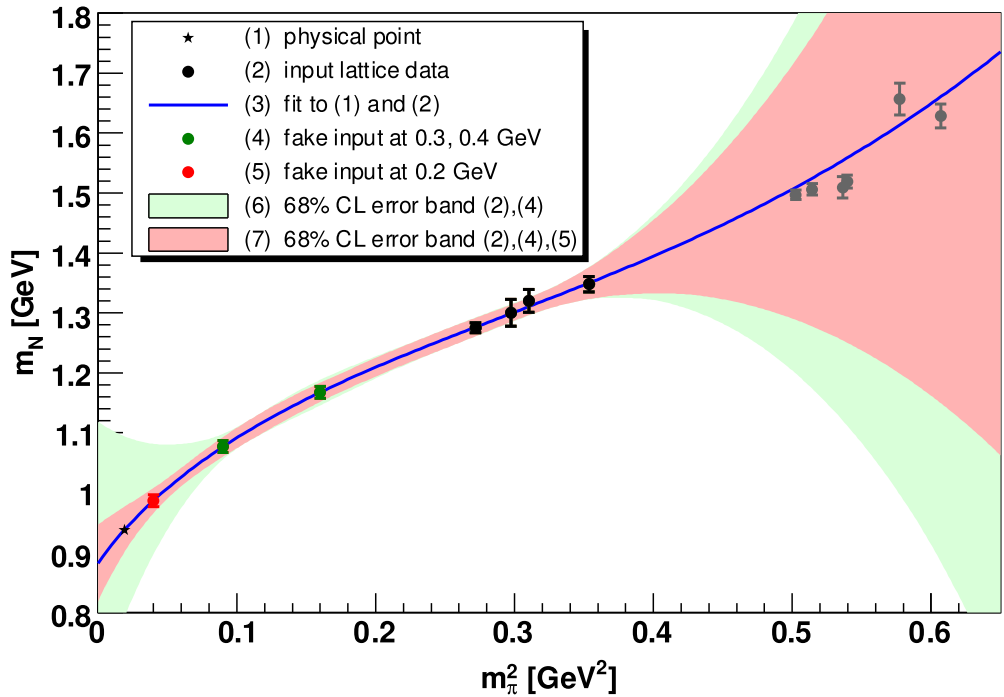}
	\caption{Ab-initio band calculations for fits to fake lattice data input at low $m_\pi$. }
	\label{fig-fakebands}
	\end{figure}

We begin our analysis by plotting Fit II from table \ref{tab-params}. On the fit curve, we calculate the points at $m_\pi = 0.3\units{GeV}$ and $m_\pi = 0.4\units{GeV}$. We append them to our set of input lattice data points, giving them uncertainties of $\pm 0.01\units{GeV}$. Next, a band calculation is performed on the extended input data set. This time, the physical point is not included. In our study, we keep $g_A^0 \approx g_A^\text{phys}$ and $f_\pi^0 \approx f_\pi^\text{phys}$ fixed, but we release all other LECs. It should be noted that the LECs parametrize the interpolation curve with some redundancy. As in ref. \cite{PHW04}, we eliminate this redundancy by introducing linear combinations 
\begin{equation}
	A \equiv e_1^{(4)}(1 \units{GeV}) + \frac{3 c_2}{128 \pi^2 (f_\pi^0)^2}\ , \qquad
	B \equiv c_2 + 4 c_3
	\end{equation}
Thus the variables $\ul{p}$ of the error analysis are $m_0$, $c_1$, $A$ and $B$. The result is the green outer band of fig. \ref{fig-fakebands}. Introducing one more ``fake'' data point at $m_\pi = 0.2\units{GeV}$ narrows down the band to the area shaded red.

The result of this and similar studies is clear: Within the forseeable future, an ab-initio extrapolation of lattice data down to the physical point using our fit function $m_N^{(4)}$ will not be possible. The problem is \emph{badly conditioned}. Actually, this is not so surprising. The fit function is similar to a polynomial. Polynomial extrapolations are known to have bad extrapolation properties.

Is this a reason to give up chiral extrapolations of lattice data altogether? No, it is not. First of all, we can study and fit \emph{multiple} observables with the same set of parameters simultaneously, thereby greatly improving the condition of the extrapolation problem. An example of such a simultaneous fit is given in section \ref{sec-mesonmatching}. Secondly, chiral perturbation theory can describe the effects of other lattice simulation parameters, like the size of the simulation volume. 
In section \ref{sec-voleff}, we demonstrate that this way valuable additional input becomes accessible.
Last but not least, we can incorporate and check compatibility with empirical information. We give an example of this possibility in the next section.

\section{Consistency with Empirical Constraints}
\label{sec-empconstr}

If results from lattice extrapolations are physically meaningful, they should be compatible with
empirical results for observables whose chiral expansions involve the same parameters. Here we study a collection of such quantities:
\begin{itemize}
	\item $T^+(m_\pi)$, the isospin-even $\pi N$ S-wave scattering amplitude at threshold 
	\item $P_1^+(m_\pi)$, the isospin-even non-spin-flip $\pi N$ P-wave scattering amplitude at threshold 
	\item $\sigma_N$, the pion nucleon sigma-term  
	\end{itemize}
The empirical results can be compared to the $\chi$PT predictions evaluated \emph{at physical conditions}, i.e. at $m_\pi = m_\pi^\text{phys}$.

\subsection{Monitored Quantities and their Chiral Expansions}

$T^+(m_\pi)$ is related to the isospin-even $\pi N$ scattering length $a^+$ via
\begin{equation}
	T^+(m_\pi) =  4\pi \left( 1 + \frac{m_\pi}{m_N} \right) a^+ 
	\end{equation}
From the precision study of pionic atoms, the authors of ref. \cite{ELT02} extract \\ 
\hbox{$a^+ = (-12 \pm 2 \pm 8) \cdot 10^{-4} m_{\pi^\pm}^{-1}$}, which provides us with an empirical value 
\begin{equation}
	\boxed{T^+ = (-0.12 \pm 0.11) \units{GeV^{-1}}}
	\end{equation}
Note that this value is anomalously small. For comparison, the isovector scattering length is $a^- = (895 \pm 3 \pm 13)\cdot 10^{-4} m_{\pi^\pm}^{-1}$ \cite{ELT02}. 
In HB$\chi$PT, contributions at order $p^2$ and $p^3$ 
read
\cite{BKMpiN}

\begin{align}
	T^{+(\leq 2)}
	&= \frac{2 m_\pi^2}{(f_\pi^0)^2}\left(c_2+c_3-2c_1-\frac{(g_A^0)^2}{8 m_0}\right) &
	T^{+(3)} &= \frac{3 (g_A^0)^2 m_\pi^3}{64 \pi (f_\pi^0)^4} 
	\end{align}


$P_1^+(m_\pi)$ can be expressed as a linear combination of scattering volumes
\begin{equation}
	P_1^+(m_\pi) = 4\pi \left( 1 + \frac{m_\pi}{m_N} \right) (4 a_{33} + 2 a_{31} + 2 a_{13} + a_{11} )
\end{equation}
One finds values for the empirical scattering volumes collected in \cite{TE03} with data from \cite{Koch86}, \cite{Koch80} and \cite{FM97}, totalling to
\begin{equation}
	\boxed{P_1^+ = (1044 \pm 38) \units{GeV^{-3}}}
\end{equation}
In HB$\chi$PT, contributions at order $p^2$ and $p^3$ read \cite{BKMpiN}
\begin{align}
	P_1^{+(\leq 2)}
	& = \frac{2}{(f_\pi^0)^2} \left( c_2 \frac{m_\pi}{m_0} - c_3 \right) + \frac{(g_A^0)^2 m_\pi}{4 (f_\pi^0)^2 m_0^2}
	& 
	P_1^{+(3)}
	& = - \frac{(g_A^0)^2 m_\pi}{12 \pi (f_\pi^0)^4} \left( (g_A^0)^2 + \frac{77}{32} \right)
	\label{eq-p1plus}
\end{align}
%

Last but not least, we employ information about the pion-nucleon sigma-term, see eq. (\ref{eq-sigmaterm}).
Ref. \cite{GLS91} provides us with the empirical value $\sigma_N = (0.045 \pm 0.008) \units{GeV}$,
and argues that a significantly larger value for $\sigma_N$ would lead to 
implausible consequences: Either the strangeness content of the nucleon would be substantial 
($\bra{N} m_s \bar{s} s \ket{N} \sim 0.4\units{GeV}$) or the GMOR relations would be accidental and the quark condensate $\langle \bar{q} q \rangle$ would have to be suppressed. Therefore, we choose here to stick to the value above, rather than to the more recent TRIUMF results \cite{PSWA02}, still under debate. The latter found $\Sigma_d = (0.067 \pm 0.006)\units{GeV}$, which according to ref.\cite{GLS91} relates to the sigma term via a correction $\Delta = (-0.003 \pm 0.001)\units{GeV}$, such that $\sigma_N = \Sigma_d + \Delta = (0.064 \pm 0.007) \units{GeV}$ comes out considerably larger. A short overview of present information on $\sigma_N$ can be found in ref. \cite{Sainio05}.

\subsection{First Numerical Experiments}

The empirical values for $T^+$, $P_1^+$ and $\sigma_N$ are to be compared with
the chiral formulae evaluated {\em at the physical pion mass},
i.e. at $m_\pi = m_\pi^\text{phys}$.

In a first step, let us assume that for $T^+$ and $P^+_1$ it is sufficient to look at 
the chiral expansions to order $p^3$. No new parameters are involved at this order.
(One might argue that contributions of higher orders should be negligible,
since we evaluate at the physical pion mass, comparatively close to the chiral limit.)
Therefore, we can readily evaluate the three observables with the parameters of the fits.
Table \ref{tab-paramsb} shows that $T^{+(\leq 3)}$ and $P_1^{+(\leq 3)}$ are completely incompatible
with the empirical values. The empirical values lie far away from the systematic envelope.\footnote{Note that these two quantities are very sensitive to $g_A^0$ and $f_\pi^0$, leading to a large systematic envelope.} First of all, let us assume that corrections at order $p^4$ and higher can be made responsible.
Evaluating the contributions of order $\leq 3$ using the central parameter values of Fit II, we get
\setlength{\savearraycolsep}{\arraycolsep} \setlength{\arraycolsep}{2pt}
\begin{equation*}
	\begin{array}{cccccccccccc}
	T^+ & = & & T^{+(\leq 2)} & + & T^{+(3)} & + & T^{+(\geq 4)} & & \stackrel{\displaystyle !}{=} & -0.22 \units{GeV^{-1}} \\  
	& = & ( & 6.4 & + & 0.86 & + & -7.4 & )\ \units{GeV^{-1}} & & \vspace{5pt} \\
	P_1^+ & = & & P^{+(\leq 2)} & + & P^{+(3)} & + & P^{+(\geq 4)} & & \stackrel{\displaystyle !}{=} & 1044 \units{GeV^{-3}} \\  
	& = & ( & 922 & + & -323 & + & 446 & )\ \units{GeV^{-3}} & & 
	\end{array}
	\end{equation*}
\setlength{\arraycolsep}{\savearraycolsep}
where $T^{+(\geq 4)}$ and $P_1^{+(\geq 4)}$ have been chosen in such a way that
the equations remain balanced. The magnitudes of $T^{+(\geq 4)}$ and $P_1^{+(\geq 4)}$ obviously come out so large that we cannot speak of a perturbative series any longer.\footnote{We have checked whether fourth order coupling of ``natural size'' can account for such large corrections. Formulae of $T^+$ and $P_1^+$ at order $p^4$ can be found in \cite{FM00}, making use of 
$T^{+(\leq 4)} = 4 \pi (1+m_\pi/m_0) a_{0+}^+ + \mathcal{O}({m_\pi}^5)$ and $P^{+(\leq 4)} = 4 \pi (1+m_\pi/m_0) (2a_{1+}^+ + a_{1-}^+) + \mathcal{O}({m_\pi}^3)$. For some of the LECs involved we found estimates from the literature. Assuming the 9 remaining LECs are scattered around zero following a Gaussian distribution, the standard deviation (i.e., the typical size) of these unknown LECs is $\approx 5\units{GeV}^{-3}$. I leave it to the judgement of the reader whether this result can be counted as evidence for the breakdown of the theory or not. }

To localize the origin of the discrepancy, we first determine the parameter ranges permitted by the empirical constraints without making use of lattice data. We do not apply the $\chi^2$ method here, because the empirical uncertainties may be correlated.
We demand that each one of the quantities $T^{+(\leq 3)}$, $P^{+(\leq 3)}_1$ and $\sigma_N$ evaluates to a value within its  empirical error bound. Again, we eliminate $c_1$ by requiring the nucleon mass to hit the physical point, i.e. $M_N^{(\leq 4)}(m_\pi^\text{phys}) = M_N^\text{phys}$. For $e_1^r$, we permit a generous range between $-10$ and $10\units{GeV^{-3}}$, and  
we allow values $g_A^0$ and $f_\pi^0$ in the range of eq. (\ref{eq-gafpiinterv}). Table \ref{tab-paramsemp} shows that the permissible values of $m_0$, $c_1$, $c_2$ comply well with our results from the lattice fit. Note however, that \emph{the constraints from pion nucleon scattering force $c_3$ to take on a value large in magnitude}, around $-5 \units{GeV^{-3}}$. Other analyses fitting to data from pion nucleon scattering find similarly large values \cite{FM00}. 

\begin{table}[h]
	\caption{Parameter values allowed by empirical constraints}
	\label{tab-paramsemp}
	\centering
	\vspace{10pt}
\begin{tabular}{ll||cc}
	$m_0$ & $(\mathrm{GeV})$ &                            0.868 .. 0.895 & \\
	$c_1$ & $(\mathrm{GeV^{-1}})$ &                       -1.27 .. -0.75 & \\
	$c_2$ & $(\mathrm{GeV^{-1}})$ &                        2.1 .. 3.8 & \\
	$c_3$ & $(\mathrm{GeV^{-1}})$ &                       -5.6 .. -4.3 & \\
	$e_1^{(4)}(1 \units{GeV})$ & $(\mathrm{GeV^{-3}})$ &  -10 .. 10 & input   \\
	$g_A^0$ & &                                            1.1 .. 1.3 & input  \\ 
	$f_\pi^0$ & $(\mathrm{GeV})$ &                         0.0862 .. 0.0924  & input  \\
	$T^{+(\leq 3)}$ & $(\mathrm{GeV^{-1}})$ &             -0.23 .. -0.01 & input   \\
	$P_1^{+(\leq 3)}$ & $(\mathrm{GeV^{-3}})$  &          1006 .. 1082 & input  \\
	$\sigma_N$ & $(\mathrm{GeV})$ &                        0.037 .. 0.053 & input  \\
	\end{tabular} \par 
\end{table}

We have checked that a tolerably good fit to lattice data is still possible with a value of $c_3$ around $-5 \units{GeV^{-3}}$. However, we will see in section \ref{sec-voleff}, that an improved fit to a wider range of data using a finite volume correction favours a value $c_3$ lower in magnitude, around our original input from $NN$ scattering $c_3 \approx -3 \units{GeV}$. What makes pion nucleon scattering different? It turns out that the influence of the delta resonance can explain our observations.

\subsection{The Role of the Delta Resonance}

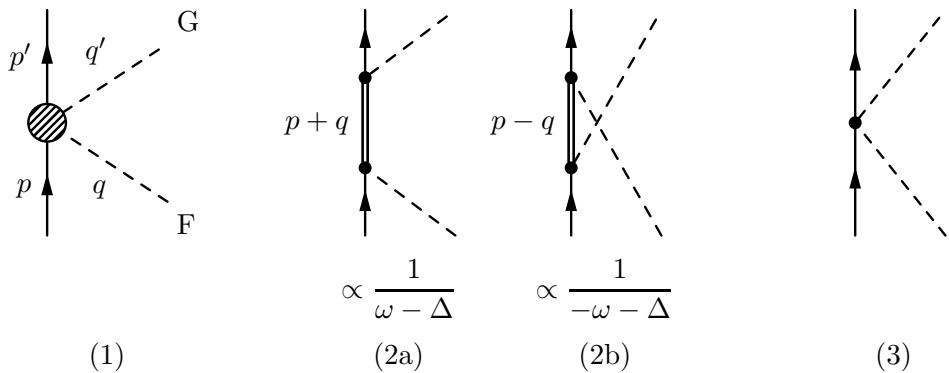
\begin{figure}[tbh]
	\centering

	\unitlength=1mm
	\begin{fmffile}{pindelta}
	\fmfset{arrow_len}{3mm}
	\fmfset{arrow_ang}{15}

	\setlength{\savearraycolsep}{\tabcolsep}
	\setlength{\tabcolsep}{15pt}
	\begin{tabular}{cccccc}

		\begin{fmfgraph*}(20,30)
		\fmftop{nout,piout}
		\fmfbottom{nin,piin}
		\fmfforce{(0.1w,0.0h)}{nin}
		\fmfforce{(0.1w,1.0h)}{nout}
		\fmfforce{(0.9w,0.15h)}{piin}
		\fmfforce{(0.9w,0.85h)}{piout}
		\fmfforce{(0.1w,0.5h)}{int}
		\fmf{fermion,label=$p$,label.side=left}{nin,int}
		\fmf{fermion,label=$p'$,label.side=left}{int,nout}
		\fmf{dashes,label=$q$,label.side=left}{piin,int}
		\fmf{dashes,label=$q'$,label.side=left}{int,piout}
		\fmflabel{$\zF$}{piin}
		\fmflabel{$\zG$}{piout}
		\fmfblob{5mm}{int}
		\end{fmfgraph*}
		&
		&
		\begin{fmfgraph*}(15,30)
		\fmftop{nout,piout}
		\fmfbottom{nin,piin}
		\fmfforce{(0.2w,0.0h)}{nin}
		\fmfforce{(0.2w,1.0h)}{nout}
		\fmfforce{(1.0w,0.0h)}{piin}
		\fmfforce{(1.0w,1.0h)}{piout}
		\fmfforce{(0.2w,0.3h)}{int1}
		\fmfforce{(0.2w,0.7h)}{int2}
		\fmf{fermion}{nin,int1}
		\fmf{dbl_plain,label=$p+q$,label.side=left}{int1,int2}
		\fmf{fermion}{int2,nout}
		\fmf{dashes}{piin,int1}
		\fmf{dashes}{int2,piout}
		\fmfdot{int1,int2}
		\end{fmfgraph*}
		&
		\begin{fmfgraph*}(15,30)
		\fmftop{nout,piout}
		\fmfbottom{nin,piin}
		\fmfforce{(0.2w,0.0h)}{nin}
		\fmfforce{(0.2w,1.0h)}{nout}
		\fmfforce{(1.0w,0.0h)}{piin}
		\fmfforce{(1.0w,1.0h)}{piout}
		\fmfforce{(0.2w,0.3h)}{int1}
		\fmfforce{(0.2w,0.7h)}{int2}
		\fmf{fermion}{nin,int1}
		\fmf{dbl_plain,label=$p-q$,label.side=left}{int1,int2}
		\fmf{fermion}{int2,nout}
		\fmf{dashes}{piin,int2}
		\fmf{dashes}{int1,piout}
		\fmfdot{int1,int2}
		\end{fmfgraph*}
		&
		&
		\begin{fmfgraph*}(15,30)
		\fmftop{nout,piout}
		\fmfbottom{nin,piin}
		\fmfforce{(0.2w,0.0h)}{nin}
		\fmfforce{(0.2w,1.0h)}{nout}
		\fmfforce{(1.0w,0.0h)}{piin}
		\fmfforce{(1.0w,1.0h)}{piout}
		\fmfforce{(0.2w,0.5h)}{int}
		\fmf{fermion}{nin,int}
		\fmf{fermion}{int,nout}
		\fmf{dashes}{piin,int}
		\fmf{dashes}{int,piout}
		\fmfdot{int}
		\end{fmfgraph*}
		\vspace{5pt} \\
		& & $\displaystyle \propto \frac{1}{\omega - \Delta}$ & $\displaystyle \propto \frac{1}{- \omega - \Delta}$ & & \vspace {7pt} \\
		(1) & & (2a) & (2b) & & (3) \\
		\end{tabular}
		\setlength{\tabcolsep}{\savearraycolsep}
		
	\end{fmffile}
	\caption{(1)~elastic $\pi N$ scattering, (2a),~(2b)~contribution from the delta resonance, (3)~contact interaction vertex.}
	\label{fig-piNpiNprocess}
	\end{figure}

An effective theory must encode the effects of excited states in its low energy constants.  The excitation energy of the $\Delta(1232)$ resonance lies only $\Delta = 0.294 \units{GeV} \approx 2 m_\pi^\text{phys}$ above the nucleon mass.
It is a spin-$3/2$ isospin-$3/2$ state. A pion-nucleon system can take on the same quantum numbers, so it can make the transition into a delta excitation. It has been known for a long time that this intermediate state plays an important role in the $\pi N$ scattering process \cite{EW88}. Already within a simple non-relativistic treatment of the delta resonance, we can shed light on our issue concerning $c_3$.\footnote{I am in debt to Prof. Wolfram Weise, who pointed this out.} 

Fig. \ref{fig-piNpiNprocess} (1) depicts a general elastic pion nucleon scattering process. Here $\zF$ and $\zG$ are the pion isospin indices. Consider the graphs fig \ref{fig-piNpiNprocess} (2a) and (2b). The double line symbolizes a delta resonance.
We work in the static limit, discarding all corrections suppressed with $1/m_0$. In the center of mass system the nucleon momentum is $p\approx(m_0,-\vec{q})$. The in- and out-going pion carries the energy $\omega \equiv (m_\pi^2+{\vec{q}\,}^2)^{1/2}=(m_\pi^2+{\vec{q}\,}'^2)^{1/2}$. The non-relativistic $\pi N \Delta$-vertex and $\Delta$-propagator can be written \cite{KGW98,EW88}

\begin{equation}
	\begin{fmffile}{pindeltave}
	\fmfset{arrow_len}{3mm}
	\fmfset{arrow_ang}{15}
	\unitlength=1mm
	\vtop{\vskip-10mm \hbox{
	\begin{fmfgraph*}(15,20)
		\fmftop{nout,piout}
		\fmfbottom{din}
		\fmfforce{(0.3w,0.0h)}{din}
		\fmfforce{(0.3w,1.0h)}{nout}
		\fmfforce{(0.80w,0.80h)}{piout}
		\fmfforce{(0.3w,0.5h)}{int}
		\fmf{dbl_plain_arrow}{din,int}
		\fmf{fermion}{int,nout}
		\fmf{dashes,label=$q$,label.side=right}{int,piout}
		\fmfdot{int}
		\fmflabel{$\zF$}{piout}
		\end{fmfgraph*}
	}}
	\ = \ 
	- \frac{g_{\pi N \Delta}}{2 m_0} S_\zi  q_\zi\; T_\zF\ ,
	\qquad
	\begin{fmfgraph*}(15,5)
		\fmfleft{din}
		\fmfright{dout}
		\fmfforce{(0.0w,0.2h)}{din}
		\fmfforce{(1.0w,0.2h)}{dout}
		\fmf{dbl_plain_arrow,label=$k$,label.side=left}{din,dout}
		\end{fmfgraph*}
	\ = \ 
	\frac{i}{k^0 - m_\Delta + i 0^+}
	\end{fmffile}
\end{equation}

Here, the $S_\zi$ are spin matrices and the $T_\zF$ are isospin matrices. The index $\zi=1..3$ is a space coordinate. The spin tensor $S$ couples the nucleon Pauli spinor and the momentum vector $\vec q$ to the spin of the delta. Likewise, the isospin tensor $T$ maps both isospin of the pion and nucleon to the isospin of the delta. The intermediate state must carry exactly the same quantum numbers as the system of incoming particles.\footnote{The strong interaction conserves spin and isospin.} From Clebsch-Gordon algebra it follows \cite{KGW98,EW88}
\begin{equation}
	S_\zi \; S^\dagger_\zj = \frac{2}{3} \delta_{\zi \zj} - \frac{i}{3} \varepsilon_{\zi \zj \zk} \sigma_\zk \ , \qquad
	T_\zG \; T^\dagger_\zF = \frac{2}{3} \delta_{\zG \zF} - \frac{i}{3} \varepsilon_{\zG \zF \zH} \lambda_\zH \ 
	\end{equation}
Since we are working in an SU(2) framework, $\lambda_\zH$ are simply the Pauli matrices. The amplitude of the ``direct'' process fig. \ref{fig-piNpiNprocess} (2a) is set up as
\begin{equation}
	T^\text{dir}_{\zG \zF}(q,q') = i \chi'^\dagger
	\left(-  \frac{g_{\pi N \Delta}}{2 m_0} S_\zi q'_\zi\, T_\zG \right )\frac{i}{(p^0+q^0)-m_\Delta}
	\left(-  \frac{g_{\pi N \Delta}}{2 m_0} S_\zj q_\zj\, T_\zF \right )^\dagger \chi
	\end{equation}
with $\chi$ and $\chi'$ denoting the Pauli spinors of the nucleon in the initial and final state, respectively.
From the propagator, the diagram acquires an energy denominator $\nobreak[(p^0+\nobreak q^0)-\nobreak m_\Delta]^{-1} = [\omega-\nobreak\Delta]^{-1}$. In the crossed process fig. \ref{fig-piNpiNprocess} (2b), the energy denominator becomes $[-\nobreak \omega -\nobreak \Delta]^{-1}$. The amplitudes of diagrams (2a) and (2b) add up to
\begin{equation} \begin{array}{lclll}
	\displaystyle T_{\zG \zF} = \frac{g_{\pi N \Delta}^2}{18 m_0^2} \frac{q'_\zi q_\zj}{\Delta^2 - \omega^2}\ \chi'^\dagger \big( &
		4 \Delta\, \delta_{\zi \zj}\, \delta_{\zG \zF} &
		- 2 \omega\, i \varepsilon_{\zi \zj \zk} \sigma_\zk\, \delta_{\zG \zF} \\
	&	- 2 \omega \, \delta_{\zi \zj}\, i \varepsilon_{\zG \zF \zH} \lambda_\zH &
		+ \Delta \, i \varepsilon_{\zi \zj \zk} \sigma_\zk\,  i \varepsilon_{\zG \zF \zH} \lambda_\zH
	& \big)	\chi
	\end{array} \end{equation}
We focus on the first term in the above sum. It constitutes a non-spin-flip isoscalar P-wave.\footnote{It is proportional to ${\vec q\,}'\cdot \vec q$, which makes it a P-wave in the partial wave decomposition, and it leaves spin and isospin invariant.} Thus we identify the $P_1^+$ delta contribution at threshold $\omega=m_\pi$ to be 
\begin{equation}
	P_1^{+(\Delta)} = \frac{2 g_{\pi N \Delta}^2}{9 m_0^2} \frac{\Delta}{\Delta^2 - m_\pi^2}
	= \frac{2 g_{\pi N \Delta}^2}{9 m_0^2\, \Delta } \left( 1 + \frac{m_\pi^2}{\Delta^2} - ... \right)
	\label{eq-p1plusdelta}
	\end{equation}
B$\chi$PT encodes the effect of the delta excitation in contact interactions like fig. \ref{fig-piNpiNprocess} (3). Possible candidates for interaction terms with two external pion lines can be found in $\mathcal{L}_{\pi N}^{(2)}$, $\mathcal{L}_{\pi N}^{(4)}$, etc. The amplitude of contact terms from $\mathcal{L}_{\pi N}^{(2)}$ reads
\begin{align}
	& \frac{2 c_3}{(f_\pi^0)^2}\ \bar u(p')\, q' \cdot q\, u(p) \; \delta_{\zG \zF} 
	+ \frac{c_2}{m_0^2 (f_\pi^0)^2}\ \bar u(p')\, \left( q' \cdot p\; q \cdot p + q' \cdot p'\; q \cdot p' \right) u(p)	\; \delta_{\zG \zF} \nonumber \\
	\xrightarrow{\text{static limit}} 
	& \chi'^\dagger \left[ \frac{2(c_2 + c_3)}{(f_\pi^0)^2} \omega^2 - \frac{2 c_3 }{(f_\pi^0)^2} {\vec q\,}'\cdot \vec{q}\; \right] \chi\ \delta_{\zG \zF}
	\label{eq-chiralcontact}
	\end{align}
where we have used $\displaystyle u(p) = \begin{pmatrix}\chi \\ \frac{\vec \sigma \cdot \vec p}{2 m_0} \chi \end{pmatrix}$ .\\
The second term in the static limit, proportional to $c_3$, is again an isoscalar non-spin-flip P-wave.\footnote{Naturally, we find the contribution $- 2 c_3 / (f_\pi^0)^2$ also in the HB$\chi$PT result $P_1^{+(\leq 2)}$ of eq. (\ref{eq-p1plus}).} 
This term must generate the entire isospin-scalar non-spinflip P-wave amplitude $P_1^{+(\Delta)}$ of the delta.

Here we come across an explanation for our observed discrepancies. The LEC $c_3$ should be determined (if it were possible) in the limit $m_\pi \rightarrow 0$. Using B$\chi$PT to order $p^3$, a hypothetical pion nucleon scattering P-wave analysis executed in the chiral limit would then measure $c_3$ precisely\footnote{Let us assume here that the delta mass shift $\Delta$ is only weakly dependent on $m_\pi$, and that diagrams \ref{fig-piNpiNprocess} (2a), (2b) constitute the dominant contribution to $P_1^+$.}, while a real world  experiment carried out at $m_\pi = m_\pi^\text{phys}$ yields a \emph{modified} value $\tilde c_3$. Comparing eq. (\ref{eq-p1plusdelta}) and (\ref{eq-chiralcontact}), we find
\begin{equation}
	 c_3 = - \frac{(f_\pi^0)^2 \, g_{\pi N \Delta}^2 }{9 m_0^2\, \Delta }, \quad
	 \tilde c_3 = - \frac{(f_\pi^0)^2 \, g_{\pi N \Delta}^2}{9 m_0^2 } \frac{\Delta}{\Delta^2 - (m_\pi^\text{phys})^2}, \quad
	 \rightarrow \quad 
	 \frac{\tilde c_3}{c_3} = \frac{\Delta^2}{\Delta^2 - (m_\pi^\text{phys})^2} \approx 1.3
\end{equation}
Of course, the $m_\pi$-dependence of $P_1^{+(\Delta)}$ can be encoded by B$\chi$PT as well, but this would require terms from $\mathcal{L}_{\pi N}^{(4)}$. The large number of coupling constants appearing in a full fourth order calculation make it difficult to produce a reliable fit to data \cite{FM00}. Moreover, the higher order terms would be large.
This is due to the fact that $\Delta$ is much smaller than $\Lambda_\chi$, so that the corrections of order $m_\pi^2/\Delta^2$ in eq. \ref{eq-p1plusdelta} do not fit properly into our power counting scheme.

\begin{figure}[tbh]
	\centering

	\unitlength=1mm
	\begin{fmffile}{deltaloopnn}
	\fmfset{arrow_len}{3mm}
	\fmfset{arrow_ang}{15}

	\setlength{\savearraycolsep}{\tabcolsep}
	\setlength{\tabcolsep}{15pt}
	\begin{tabular}{cccc}

		\begin{fmfgraph*}(10,30)
		\fmftop{nout}
		\fmfbottom{nin}
		\fmfforce{(0.5w,0.0h)}{nin}
		\fmfforce{(0.5w,1.0h)}{nout}
		\fmfforce{(0.5w,0.3h)}{int1}
		\fmfforce{(0.5w,0.7h)}{int2}
		\fmf{fermion}{nin,int1}
		\fmf{dbl_plain,label=$p-q$,label.side=left}{int1,int2}
		\fmf{fermion}{int2,nout}
		\fmf{dashes,right,label=$q$,label.side=right}{int1,int2}
		\fmfdot{int1,int2}
		\end{fmfgraph*}
		&
		\hspace{10mm}
		&	
		\begin{fmfgraph*}(18,30)
		\fmftop{nout1,nout2}
		\fmfbottom{nin1,nin2}
		\fmfforce{(0.1w,0.0h)}{nin1}
		\fmfforce{(0.9w,0.0h)}{nin2}
		\fmfforce{(0.1w,1.0h)}{nout1}
		\fmfforce{(0.9w,1.0h)}{nout2}
		\fmfforce{(0.1w,0.3h)}{int11}
		\fmfforce{(0.1w,0.7h)}{int12}
		\fmfforce{(0.9w,0.3h)}{int21}
		\fmfforce{(0.9w,0.7h)}{int22}
		\fmf{fermion}{nin1,int11}
		\fmf{dbl_plain}{int11,int12}
		\fmf{fermion}{int12,nout1}
		\fmf{fermion}{nin2,int21}
		\fmf{plain}{int21,int22}
		\fmf{fermion}{int22,nout2}
		\fmf{dashes}{int11,int21}
		\fmf{dashes}{int12,int22}
		\fmfdot{int11,int12,int21,int22}
		\end{fmfgraph*}	
		&
		\begin{fmfgraph*}(18,30)
		\fmftop{nout1,nout2}
		\fmfbottom{nin1,nin2}
		\fmfforce{(0.1w,0.0h)}{nin1}
		\fmfforce{(0.9w,0.0h)}{nin2}
		\fmfforce{(0.1w,1.0h)}{nout1}
		\fmfforce{(0.9w,1.0h)}{nout2}
		\fmfforce{(0.1w,0.3h)}{int11}
		\fmfforce{(0.1w,0.7h)}{int12}
		\fmfforce{(0.9w,0.3h)}{int21}
		\fmfforce{(0.9w,0.7h)}{int22}
		\fmf{fermion}{nin1,int11}
		\fmf{dbl_plain}{int11,int12}
		\fmf{fermion}{int12,nout1}
		\fmf{fermion}{nin2,int21}
		\fmf{plain}{int21,int22}
		\fmf{fermion}{int22,nout2}
		\fmf{dashes}{int11,int22}
		\fmf{dashes}{int12,int21}
		\fmfdot{int11,int12,int21,int22}
		\end{fmfgraph*}	
				
		\vspace{5pt} \\
		(4) & & (5a) & (5b) 
		\end{tabular}
		\setlength{\tabcolsep}{\savearraycolsep}
		
	\end{fmffile}
	\caption{(4) nucleon self-energy contribution from the delta resonance, (5a) (5b) single delta excitation in $N N$ scattering.}
	\label{fig-deltares}
	\end{figure}
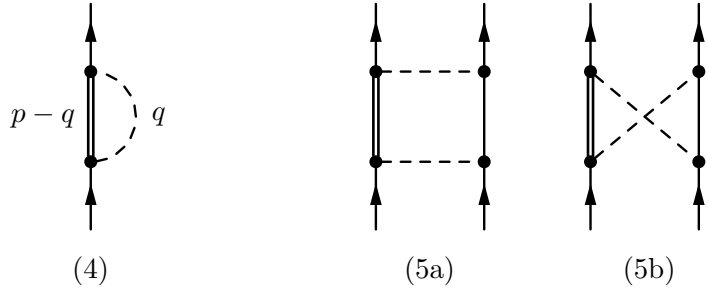

It is interesting to check what happens to the contribution associated with $c_3$ in our nucleon mass formula eq. (\ref{eq-massp4}).
Here, we must compare to a loop graph involving a delta-resonance in the intermediate state, see fig. \ref{fig-deltares} (4). Inserting a relativistic propagator for the pion, the amplitude reads
\begin{multline}
	i  \int \frac{d^4 q}{(2 \pi)^4} \left( \frac{- g_{\pi N \Delta}}{2 m_0} \right)^2 
	 S_\zi q_\zi T_\zF\; \frac{i}{q^2 - m_\pi^2 + i 0^+}\; T^\dagger_\zF q_\zj S^\dagger_\zj\; \frac{i}{(p^0-q^0) - m_\Delta + i 0^+}   = \\
	 \frac{g_{\pi N \Delta}^2}{3\,m_0^2} \int \frac{d^4 q}{(2 \pi)^4} \frac{i {\vec{q}\,}^2}{(q^0 + \omega - i 0^+)(q^0 - \omega + i 0^+)(q^0 + \Delta - i 0^+)}  = \\
	\frac{g_{\pi N \Delta}^2}{3\,m_0^2} \int \frac{d^3 q}{(2 \pi)^3} \frac{{\vec{q}\,}^2}{2 \omega (\omega + \Delta)}
	\label{eq-deltadiag}
	\end{multline}
In the last line, we have integrated with respect to $q^0$ in order to make the energy denominators visible. 
For comparison, the amplitude of the tadpole graph of fig. \ref{fig-massdiag} (b) created from the $c_3$-term is 
\begin{equation}
	\frac{3\,c_3}{(f_\pi^0)^2} \int \frac{d^3 q}{(2 \pi)^3} \frac{\omega^2 - {\vec q\,}^2}{2 \omega}
\end{equation}
The part proportional to $\vec{q}^2$ must generate the contribution of the delta. However, the $\omega$ dependence of the energy denominator in eq. (\ref{eq-deltadiag}) is different. Suppose for a moment we had obtained an energy denominator $[2\omega \Delta]^{-1}$ in eq. (\ref{eq-deltadiag}). Then we could encode the delta loop diagram perfectly in the tadpole graph, by setting the contribution to $c_3$ from the delta resonance to\footnote{For the numerical estimate, we use $g_{\pi N \Delta} \approx 28$, $m_0 \approx 0.88 \units{GeV}$.} 
\begin{equation}
	c_3^{(\Delta)} = -\frac{g_{\pi N \Delta}^2\, (f_\pi^0)^2}{9\, m_0^2\, \Delta} \approx -3.3 \units{GeV^{-1}}
	\end{equation}
The actual energy denominator $[2\omega (\omega+\Delta)]^{-1}$ makes higher order corrections necessary. Contrary to the case of $\pi N$ scattering, $\omega$ is \emph{added} to $\Delta$ in the denominator. Consequently, the contribution of the delta loop amplitude decreases with rising pion mass. We should expect that smaller values of $c_3$ work better for fits with the nucleon mass formula. This is indeed what we observe.

How do the effects of the delta dynamics show up in $NN$ scattering? According to ref. \cite{KGW98}, an interesting coincidence occurs for the isoscalar central and isovector tensor channel of the graphs \ref{fig-deltares} (4a) and (4b). Here, the energy denominators of the two graphs add up to the simple expression $2[\Delta\, \omega_1\, \omega_2]^{-1}$, where $\omega_1$ and $\omega_2$ are the energies of the exchanged pions. This combined energy denominator looks like a composition of two pion propagators and an energy independent delta propagator. Fixing $c_3$ from the amplitude of these graphs gives the same result in the chiral limit and at the physical point. Unfortunately, other channels of $NN$ scattering and contributions from double-delta excitations do not share this convenient feature. Nevertheless, it adds to the plausibility of the $NN$ result $c_3=-3.4\units{GeV^{-1}}$ from ref. \cite{EM02}, that the corresponding ``biased'' value $\tilde c_3 = -4.4\units{GeV^{-1}}$ 
is already in the range of results from $\pi N$ scattering analyses.\footnote{Compare e.g. to $\tilde c_3 = - 4.7 \pm 1 \units{GeV}$ from \cite{BPM00}}

As demonstrated in ref. \cite{HPW03}, attempts to fit lattice data of the axial vector coupling $g_A$ show even much more dramatically that the theory fails without inclusion of the delta fields. From the considerations above, it is easy to understand why there are observables which are affected particularly strongly by the delta. As soon as a process of the direct type (fig. \ref{fig-piNpiNprocess}~(2a)) is allowed by the symmetries, the corresponding energy denominator enhances the influence of the delta as $[m_\pi - \Delta]^{-1}$. With pion masses close to the delta mass, as we find them in lattice data, this factor explodes. In other words, ``the delta resonance goes in resonance.'' 

Even at the physical point, the effects of the delta dynamics can be quite significant numerically, as we saw in the case of pion nucleon scattering. Encoding them in large higher order contributions is problematic. 
 
The inescapable conclusion is that \textbf{delta degrees of freedom should always be included explicitly in Chiral Perturbation Theory with baryons}.

For the nucleon mass, the extension to a framework including delta fields has been done, and fits to lattice data are possible, see appendix \ref{sec-otherframew}.

\subsection{Constrained Fit with a Modified $c_3$}

For the present analysis, we use as a working approximation the modified value $\tilde{c_3} \equiv 1.3\, c_3$ as input to the pion nucleon scattering formulae, which we consequently denote $\tilde T^{+(\leq 3)}$ and $\tilde P^{+(\leq 3)}_1$.
Firstly, we reevaluate the parameter bounds permitted by empirical constraints. The result is shown in table \ref{tab-paramsemp2}.
Among these parameters, we select the set with the smallest $\chi^2$ with respect to the four
lattice points, see column (b) of table \ref{tab-paramsemp2}.

\begin{table}[hbt]
	\caption{Parameter values allowed by empirical constraints.}
	\label{tab-paramsemp2}
	\centering
	\vspace{10pt}
\begin{tabular}{ll||cc|r@{.}l}
	& & \multicolumn{2}{l|}{(a) constraints alone} & \multicolumn{2}{l}{(b) lattice fit} \\
	\hline
	$m_0$ & $(\mathrm{GeV})$ &                            0.871 .. 0.897 &            &  0&882   \\
	$c_1$ & $(\mathrm{GeV^{-1}})$ &                       -1.19 .. -0.68 &            & -0&91    \\
	$c_2$ & $(\mathrm{GeV^{-1}})$ &                        2.2 .. 4.0 &               &  2&95    \\
	$c_3$ & $(\mathrm{GeV^{-1}})$ &                       -4.2 .. -3.3 &              & -3&65    \\
	$e_1^{(4)}(1 \units{GeV})$ & $(\mathrm{GeV^{-3}})$ &  -10 .. 10 & input           &  1&88    \\
	$g_A^0$ & &                                            1.1 .. 1.3 & input         &  1&1     \\ 
	$f_\pi^0$ & $(\mathrm{GeV})$ &                         0.0862 .. 0.0924  & input  &  0&0924  \\
	\hline
	$\tilde T^{+(\leq 3)}$ & $(\mathrm{GeV^{-1}})$ &      -0.23 .. -0.01 & input      & -0&01    \\
	$\tilde P_1^{+(\leq 3)}$ & $(\mathrm{GeV^{-3}})$  &    1006 .. 1082 & input       & \multicolumn{2}{l}{1006} \\
	$\sigma_N$ & $(\mathrm{GeV})$ &                        0.037 .. 0.053 & input     & 0&050    \\
	\hline
	$\chi^2$ & & &                                                                    & 0&31    \\
	\end{tabular} \par 
\end{table}


Note that most input constraints in column (b) of table \ref{tab-paramsemp2} are \terminol{active}, i.e. the best fit to lattice lies on the border of the region we permit. An exception is the sigma-term $\sigma_N$, which comes out beautifully at $0.050\units{GeV}$, very close to the value obtained in our previous fit, see table \ref{tab-paramsb}. There is no reason to be concerned about the fact that the constraints are active. The four lattice points are simply far too little input to determine the seven parameters entering the nucleon mass formula. Therefore, the constraints provide essential additional input, and we see them actively influence the fit result. The drawback of a constrained fit of the above kind is that it becomes much more difficult to estimate errors, which we therefore have not done. $c_3$ now assumes a value quite close to the $NN$ scattering result $c_3 = -3.4\units{GeV}$. This is only possible, because thanks to our delta correction, $T^+$ and $P_1^+$ are evaluated with the modified value $\tilde c_3 = 1.3 (-3.65)\units{GeV} = -4.7\units{GeV}$.

\begin{figure}[htb]
  \begin{center}
    \includegraphics*{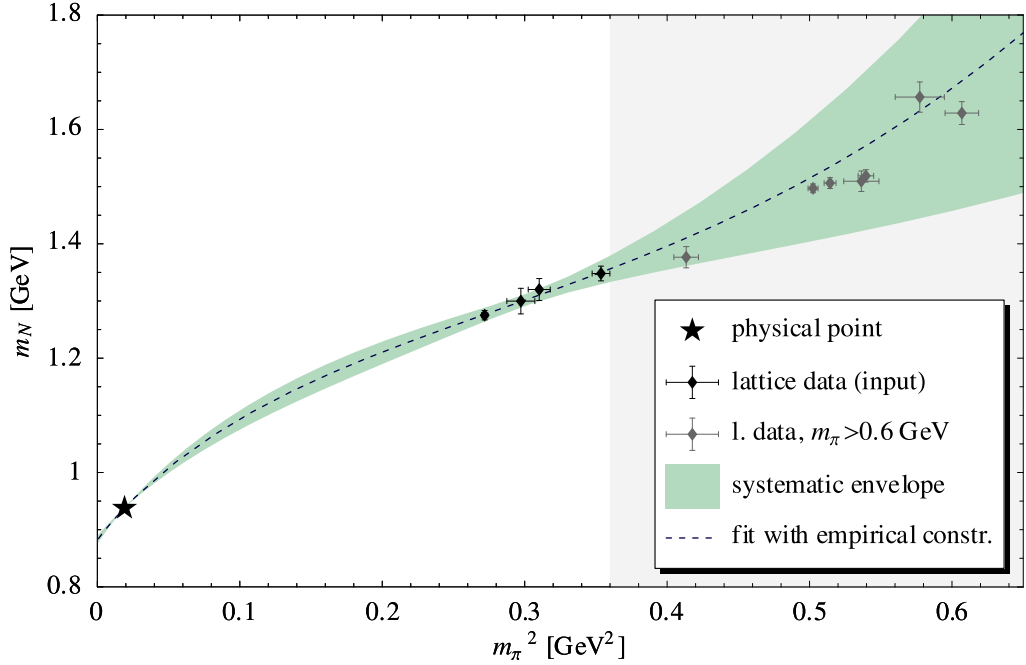}
    \caption{Fit to lattice data with empirical constraints $\tilde T^{+(\leq 3)}$, $\tilde P_1^{+(\leq 3)}$ and $\sigma_N$. For comparison, we plot again the systematic envelope of figure \ref{fig-band}.}
    \label{fig-constrfit}
  \end{center}
\end{figure}

The quality of the curve produced with the new parameters is formidable. The dashed line in fig. \ref{fig-constrfit} passes through all the lattice data points, see fig. \ref{fig-constrfit}. Moreover, the curve lies right in the center of the systematic envelope of our original setup in section \ref{sec-infvolband}. 

In conclusion, the introduction of strong constraints from $\pi N$ scattering leaves our interpolation curve practically invariant, if we take effects of the delta resonance into account. We thus obtain a consistent picture and have good reasons to be confident about our results.

\section{Finite Volume Effects}
\label{sec-voleff}

\subsection{Basics of the Finite Volume Correction}

Lattice calculations are carried out on a box of rather small side length, 
which causes deviations in the calculated observables. 
Chiral Effective Field Theory
can describe the effects of a finite simulation volume on the nucleon mass.
This way, more lattice data become accessible to $\chi$PT. \par
We still assume the time dimension to be so large that we can take it as
infinite. Based on the restriction of pion loop momenta 
to the discrete values permitted by periodic
boundary conditions, calculations for the mass difference caused
by finite simulation volumes done in \cite{K04} yield the following
contributions at order $p^3$ and $p^4$
\begin{eqnarray}
	\Delta^{(3)}m_N &=&
		\frac{3 (g_A^0)^2 m_0 m_\pi^2}{16 \pi^2 (f_\pi^0)^2} 
		\int_0^\infty dx \sum_{\vec{n}\in\mathbb{Z}^3\backslash\{\vec{0}\}} 
		K_0 \left(L |\vec{n}| \sqrt{m_0^2 x^2 + m_\pi^2 (1-x)}\right) 
		\label{eq-deltamvolp3} \\
	\Delta^{(4)}m_N &=& \frac{3 m_\pi^4 }{4 \pi^2 (f_\pi^0)^2} \sum_{\vec{n}\in\mathbb{Z}^3\backslash\{\vec{0}\}} 
		\left( (2 c_1- c_3) \frac{K_1(L |\vec{n}| m_\pi)}{L |\vec{n}| m_\pi} + 
		c_2 \frac{K_2(L |\vec{n}| m_\pi)}{(L |\vec{n}| m_\pi)^2} \right) 
		\label{eq-deltamvolp4}
	\end{eqnarray}
where $K_0$ and $K_1$ are modified Bessel functions. \par
We notice that these formulae introduce no new
parameters, and that $c_1$, $c_2$ and $c_3$ appear in linear combinations
different from those in the infinite volume formula (\ref{eq-massp4}).
Thus volume dependence can provide an entirely new kind of constraint on the fit parameters.
Originally rated as an undesirable artefact, finite volume effects be exploited to gain essential new statistical information.
We explore this possibility by fitting lattice data to the volume corrected nucleon mass
\begin{eqnarray}
	m_N^{(\text{FV})}(m_\pi,L) &\equiv& m_N^{(\leq 4)}(m_\pi) + \Delta^{(3)}m_N(m_\pi,L) + \Delta^{(4)}m_N(m_\pi,L)
	\label{eq-massvol}
	\end{eqnarray} \par

\subsection{Fast Implementation of the Finite Volume Correction}

In the course of fitting and statistical analysis, $m_N^{(\text{FV})}(m_\pi,L)$ must be evaluated numerically a large number of times. Therefore it is necessary to have a fast implementation of the volume correction formula. First of all, we cut off the sums at $\vec{n}^2=8$ and $\vec{n}^2=6$ in formula (\ref{eq-deltamvolp3}) and (\ref{eq-deltamvolp4}), respectively.
The integral in formula (\ref{eq-deltamvolp3}) must be determined numerically.
We speed up its evaluation with our own problem specific implementation.  \par

First of all, only the length $|\vec n|$ enters the sum over $\vec n$. Thus for each integer $k \equiv {\vec n}^2$, we count the number $N_k$ of thee dimensional vectors $\vec n$ with integer coordinates
and store these coefficients in a look-up table. The integrals in (\ref{eq-deltamvolp3}) are split into two parts
\begin{align}
	I_k & \equiv \int_{s_k}^\infty dx\, K_0(L \sqrt{k} \sqrt{m_0^2 x^2 + m_\pi^2(1-x)}) \\
	R_k & \equiv \int_{s_k}^\infty dx\, K_0(L \sqrt{k} \sqrt{m_0^2 x^2 + m_\pi^2(1-x)}) 
	\end{align}
$I_k$ is evaluated numerically. The cutoff $s_k$ has to be chosen so that the tail of the integral $R_k$ is negligible. To estimate $R_k$, consider the asymptotic expansion
\begin{equation}
	K_0(z) \rightarrow \sqrt{\frac{\pi}{2 z}}\; e^{-z}\qquad \text{for } z \rightarrow \infty
	\label{eq-besselasymp}
\end{equation}
For $z \geq \pi/2$ we have $K_0(z) \leq e^{-z}$. Furthermore, we approximate
\begin{equation}
	\sqrt{m_0^2 x^2 + m_\pi^2 (1-x)} \approx m_0 x \qquad
	\text{if } x \gtrsim 1 \text{ and } x\gg \frac{m_\pi}{m_0}
	\label{eq-bessargappr}
	\end{equation}
The latter condition is superfluous in our case. If the requirements are met, then
\begin{align}
	R_k & \approx 
	\int_{s_k}^\infty dx\, K_0(L \sqrt{k} m_0 x) \nonumber \\
	& \leq \int_{s_k}^\infty dx\, \exp(- L \sqrt{k} m_0 x) 
	= \frac{1}{L \sqrt{k} m_0} \exp(- L \sqrt{k} m_0 s_k )
	\end{align}
$R_k$ must be much smaller than the final integral, which we estimate to be of the order
\begin{equation}
	\tilde I_k \approx \int_0^\infty dx\, \sqrt{\frac{\pi}{2 z}}\; e^{-z} = \frac{\pi}{\sqrt{2k}\; L\; m_0 }
	\end{equation}
where we have made use of the asymptotic behavior eq. (\ref{eq-besselasymp}) and approximated the argument of the Bessel function according to (\ref{eq-bessargappr}) by $z \equiv L \sqrt{k} m_0 x$. Demanding $R_k \leq 10^{-4} \tilde I_k$ and enforcing the validity conditions of our approximations $x \gtrsim 1$, $z \geq \pi/2$ gives
\begin{equation}
	s_k = \mmax \left\{ \frac{-1}{L\; \sqrt{k}\; m_0} \ln \left(L\; \sqrt{k}\; m_0\; 10^{-4}\; \tilde I_k \right)\;,\
		1\;,\ \frac{\pi}{2\; L\; \sqrt{k}\; m_0}
		\right\} 
	\end{equation}
Now the $I_k$ are evaluated using Simpson integration with 11 sampling points each and summed up.
\begin{equation}
	\Delta^{(3)}m_N \approx
	\frac{3 (g_A^0)^2 m_0 m_\pi^2}{16 \pi^2 (f_\pi^0)^2} 
	\sum_{k=0}^8 N_k I_k
	\label{eq-deltamvolp3a} 
	\end{equation}

\subsection{Numerical Results}

Accepting lattice data for pion masses up to $0.65\units{GeV}$, we extend
our lattice input data to the first 10 points in table \ref{tab-latticedat}, 
providing us with two volume groups for which simulations in smaller lattices
exist.
The pion mass $m_\pi$ entering formula (\ref{eq-massvol}) is the mass of the pion
in the infinite volume. Within the volume groups of table \ref{tab-latticedat},
an infinite volume pion mass estimate can be taken 
from the pion mass in the largest volume, underlined in the table.
Ref. \cite{OLS05} demonstrates that this procedure is sensible.
We also copy the error of the pion mass from the largest volume. This is only
a working assumption and not entirely correct because in our $\chi^2$ analysis we treat the
pion masses of lattice points within a volume group as statistically independent quantities. \par
A weak point of our preceding study of large volume data was fixing $c_3$ to a value
still under debate and neglecting its uncertainty.
With finite volume data as a new source of information,
we can now release $c_3$ and determine it from the fit. Also, we accommodate our
systematic uncertainty about $c_2$ in a range from $3.1\units{GeV^{-1}}$ to $3.3\units{GeV^{-1}}$.
This range encompasses results from several HB$\chi$PT fits at $\mathcal{O}(p^3)$
to experimental $\pi N$ scattering data, see table 4 in \cite{FM00}. 

\begin{figure}[htb]
	\centering
	\includegraphics[width=\textwidth]{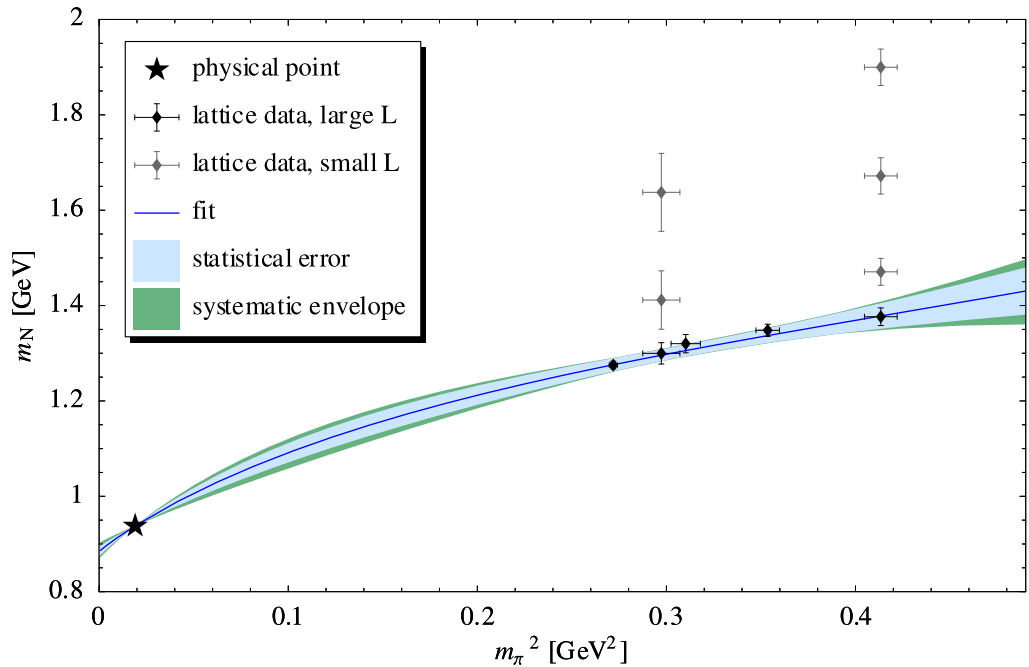}
	\caption{Fit to finite volume lattice data. All the lattice points shown are used as input
	for the fit. The points drawn as black diamonds are the calculations on the largest lattice
	of each volume group and can practically be treated as infinite volume results. The lattice points
	represented by brown triangles exhibit a sizable finite volume effect. 
	They are drawn at pion masses copied from the
	''large volume'' result of their group. The fit function, the
	68\% statistical error band and the systematic envelope are plotted in the infinite volume limit.}
	\label{fig-volband}
\end{figure}

\begin{figure}[htb]
	\centering
	\includegraphics[width=0.48\textwidth]{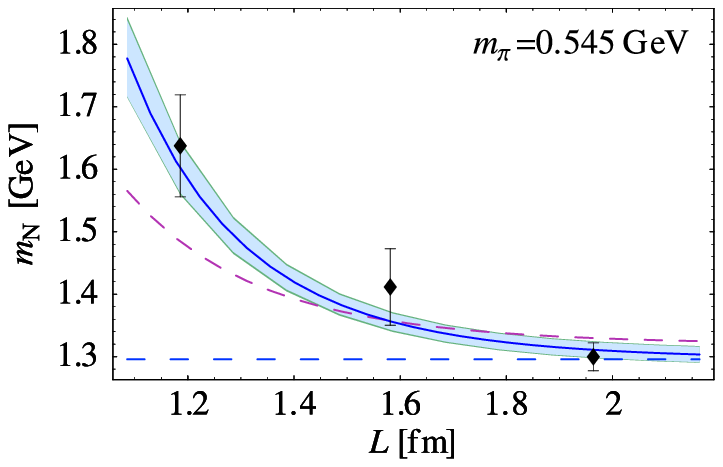} \hfill
	\includegraphics[width=0.48\textwidth]{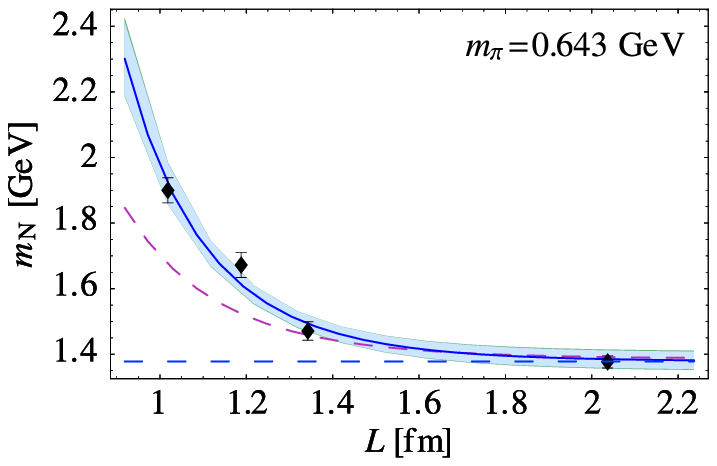}
	\caption{$L$-dependence of the fit function and error bands at the pion masses of
	the two volume groups. The shapes and colors of the lattice points correspond to those
	of fig. \ref{fig-volband}. The horizontal dashed line is the infinite volume limit
	of the fit function. The bent dashed lines represent the $\mathcal{O}(p^3)$ function, where
	$e_1^{(3)}(\lambda)$ has been determined from a secondary fit to data.}
	\label{fig-voll}
\end{figure}

\begin{figure}[htb]
	{\centering
	\includegraphics[width=\textwidth]{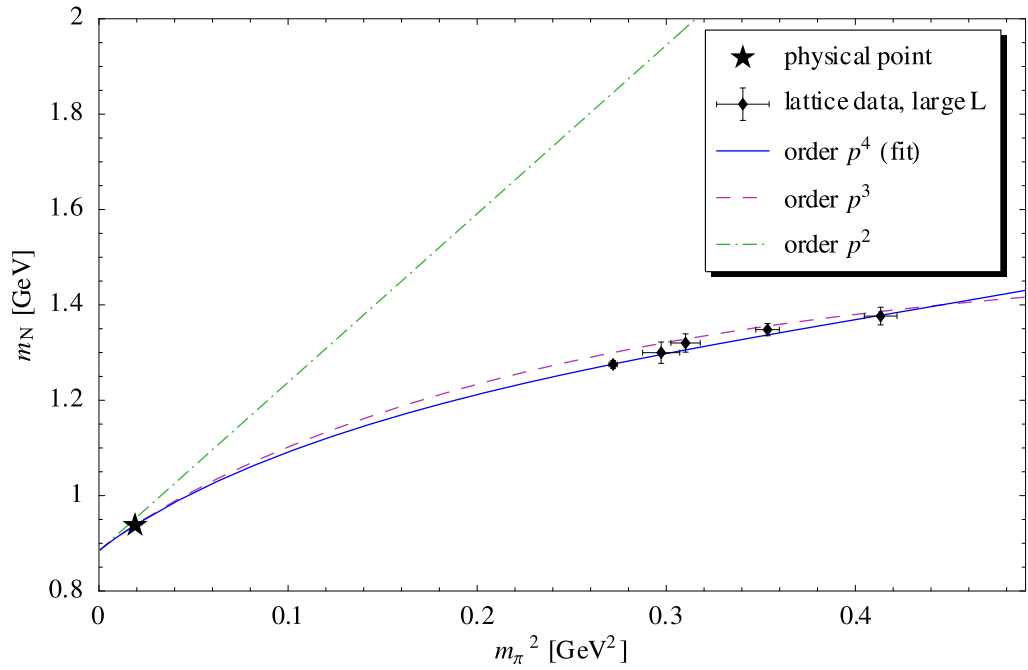}
	}
	\caption{Study of convergence:
	nucleon mass predictions at order $p^2$,  $p^3$ and $p^4$,
	using the parameters determined from the finite volume fit.
	At order $p^3$, $e_1^{(3)}(\lambda)$ cannot be imported from $p^4$ and needs to be determined from a secondary fit to the data.
	}
	\label{fig-volconv}
\end{figure}

The results of our calculations are summarized in table \ref{tab-params} and figs. \ref{fig-volband} and
\ref{fig-voll}. Looking at the latter figure, we rate convergence of our finite volume correction acceptable.
Notably, $c_3$ comes out in a range $-4.3 .. -1.4 \units{GeV^{-1}}$, i.e.
for finite volume data, a value of $c_3$ low in magnitude and compatible with $N N$ scattering results of ref. \cite{EM02}
is preferable. As a consequence, convergence on the $m_\pi$ axis in fig. \ref{fig-volconv} appears to be optimal.

\begin{table}[htb]
	\caption{parameter values and observables for the finite volume fit}
	\label{tab-paramsvol}
	\centering
	\vspace{10pt}
\begin{tabular}{ll||cc|cc}
	& & \multicolumn{2}{l|}{(a) statistical error} & \multicolumn{2}{l}{(b) systematic envelope} \\
	\hline
	$e_1^{(4)}(1 \units{GeV})$ & $(\mathrm{GeV^{-3}})$  & 2.40 $\pm$ 0.36 & fitted & 0.8 .. 3.6 & fitted \\
	$c_3$ & $(\mathrm{GeV^{-1}})$                       & -2.9 $\pm$ 0.6 & fitted & -4.3 .. -1.4 & fitted \\
	$m_0$ & $(\mathrm{GeV})$                            & 0.884 $\pm$ 0.006 & fitted & 0.873 .. 0.898 & fitted \\
	$c_1$ & $(\mathrm{GeV^{-1}})$                       & -0.88 $\pm$ 0.09 & elim. & -1.03 .. -0.69 & elim. \\
	$g_A^0$ &                                           & 1.267 & fixed & 1.10 .. 1.30 & scanned \\
	$f_\pi^0$ & $(\mathrm{GeV})$                        & 0.0924 & fixed & 0.0862 .. 0.0924 & scanned \\
	$c_2$ & $(\mathrm{GeV^{-1}})$                       & 3.2 & fixed & 3.10 .. 3.30 & scanned \\
	\hline
	$\chi^2/\text{d.o.f.}$ &                            & 0.75 & & 0.69 .. 0.82 & \\
	$\tilde T^{+(\leq 3)}$ & $(\mathrm{GeV^{-1}})$      & 5.1 $\pm$ 2.3 & & -2. .. 15. &  \\
	$\tilde P_1^{+(\leq 3)}$ & $(\mathrm{GeV^{-3}})$    & 690. $\pm$ 160. & & 180. .. 1190. & \\
	$\sigma_N$ & $(\mathrm{GeV})$                       & 0.0476 $\pm$ 0.0047 & & 0.036 .. 0.057 &  \\
	\hline
	$e_1^{(3)}(1 \units{GeV})$ & $(\mathrm{GeV^{-3}})$  & 2.54 & & & fitted \\
	\end{tabular} \par 
\end{table}

\begin{figure}[htb]
	\topalignbox{a) }\topalignbox{\includegraphics[width=0.45\textwidth]{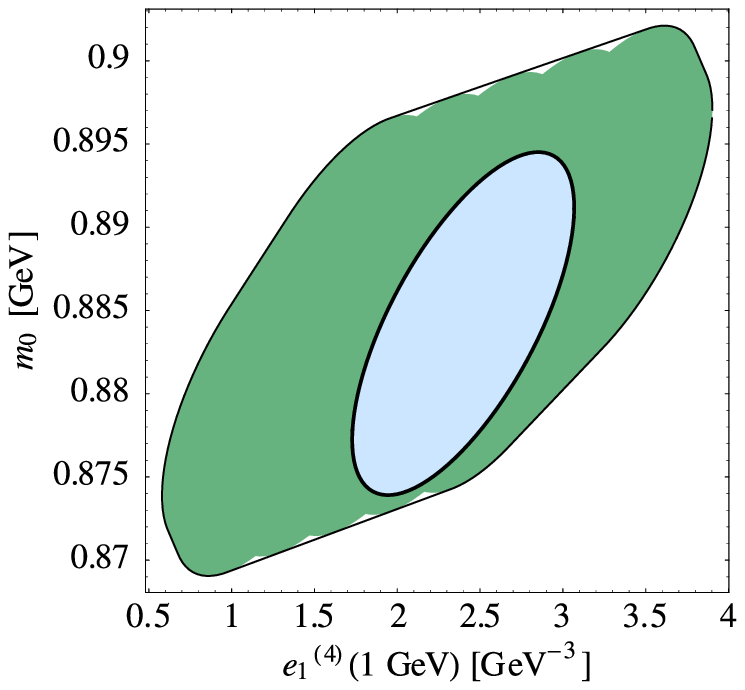}} 
	\hfill {}
	\topalignbox{b) }\topalignbox{\includegraphics[width=0.45\textwidth]{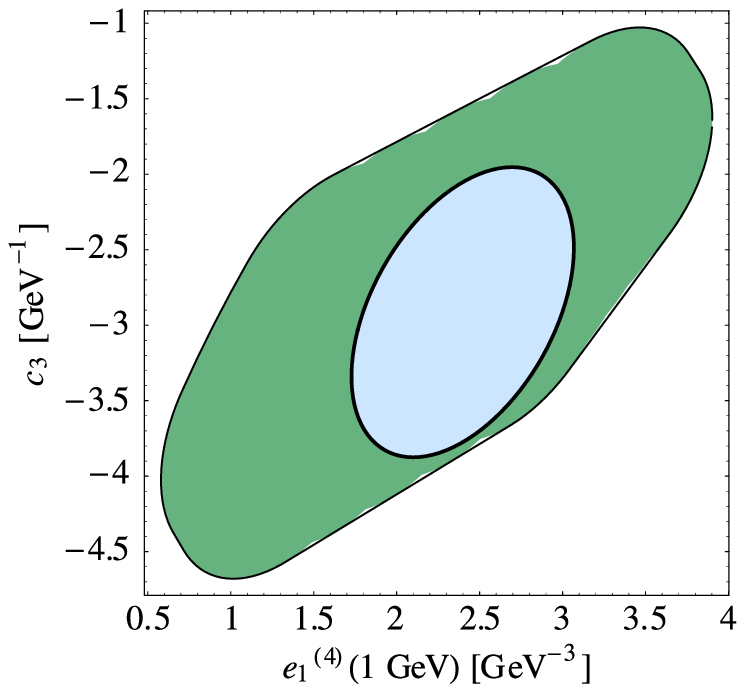}}
	\hfill {} \\
	\topalignbox{c) }\topalignbox{\includegraphics[width=0.45\textwidth]{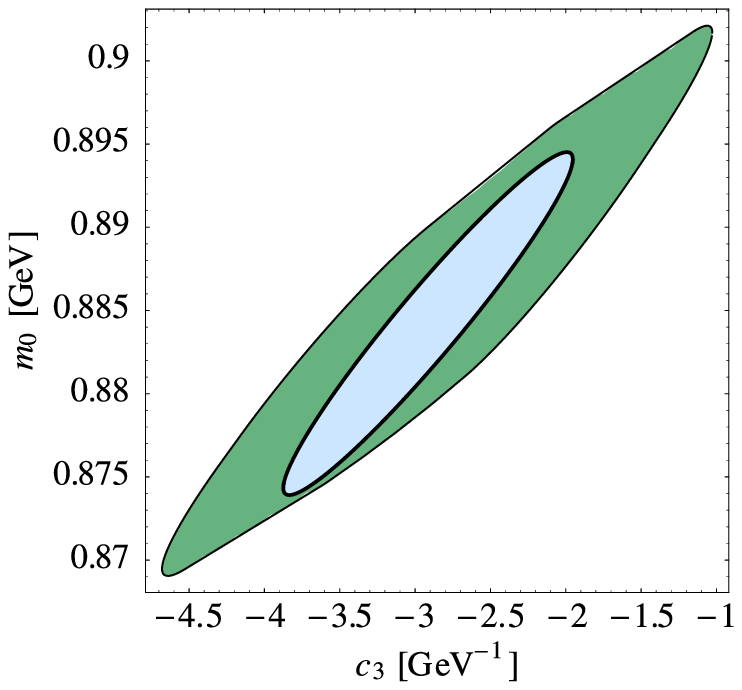}}
	\hfill {} 
	\topalignbox{d) }\topalignbox{\includegraphics[width=0.45\textwidth]{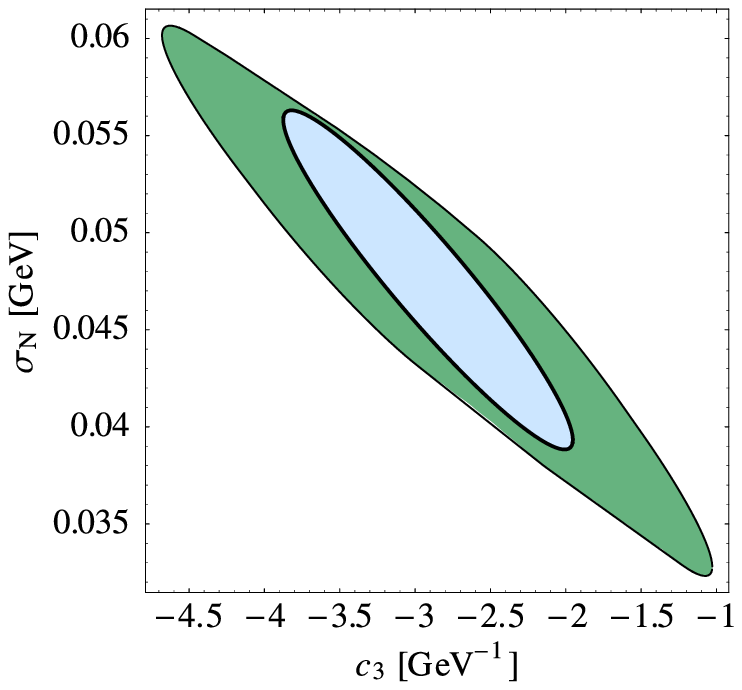}} 
	\hfill {}\\
	\topalignbox{e) }\topalignbox{\includegraphics[width=0.45\textwidth]{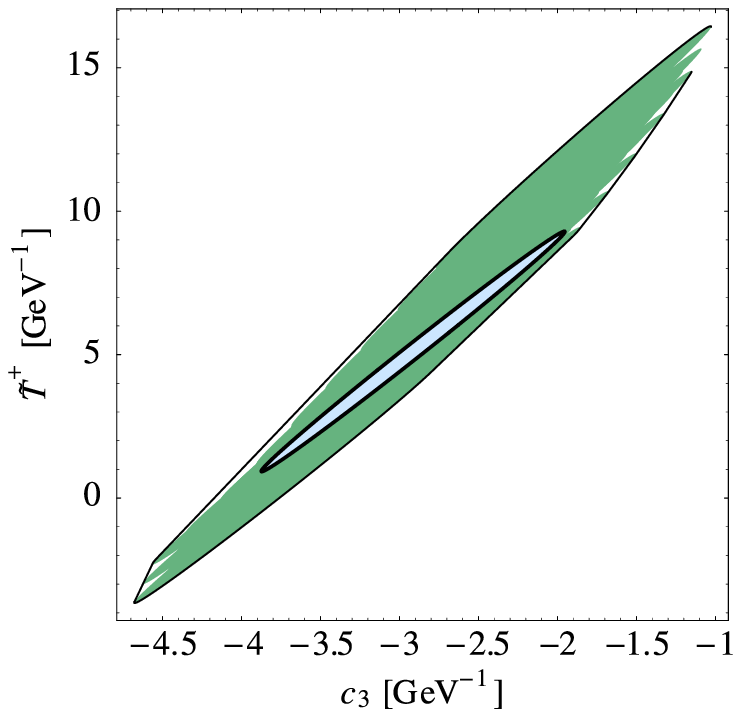}}
	\hfill {} 
	\topalignbox{f) }\topalignbox{\includegraphics[width=0.45\textwidth]{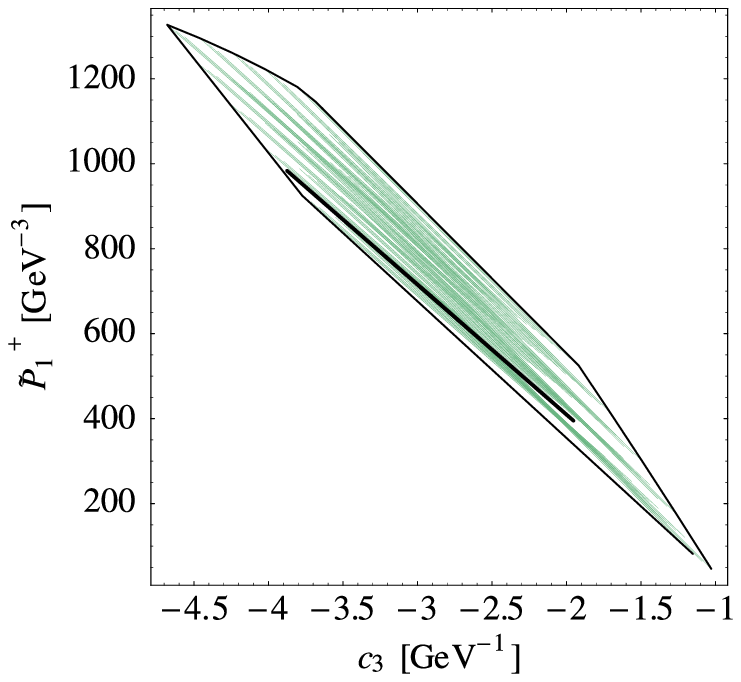}} 
	\hfill {}\\	
	\caption{$\CL=68\%$ 3-parameter confidence region for the finite volume fit. Color scheme as in fig. \ref{fig-confreg}. The systematic envelope reflects uncertainties in $g_A^0$, $f_\pi^0$ and $c_2$. Panes a), b), c) show projections on the different planes of fit parameter pairs. In panes d), e), f) the confidence region is translated to depict the statistical relation of $c_3$ to $\sigma_N$, $\tilde T^{+(\leq 3)}$ and $\tilde P_1^{+(\leq 3)}$.}
	\label{fig-volconfreg}
	\end{figure}

\begin{figure}[htb]
	\centering
	\includegraphics[width=\textwidth]{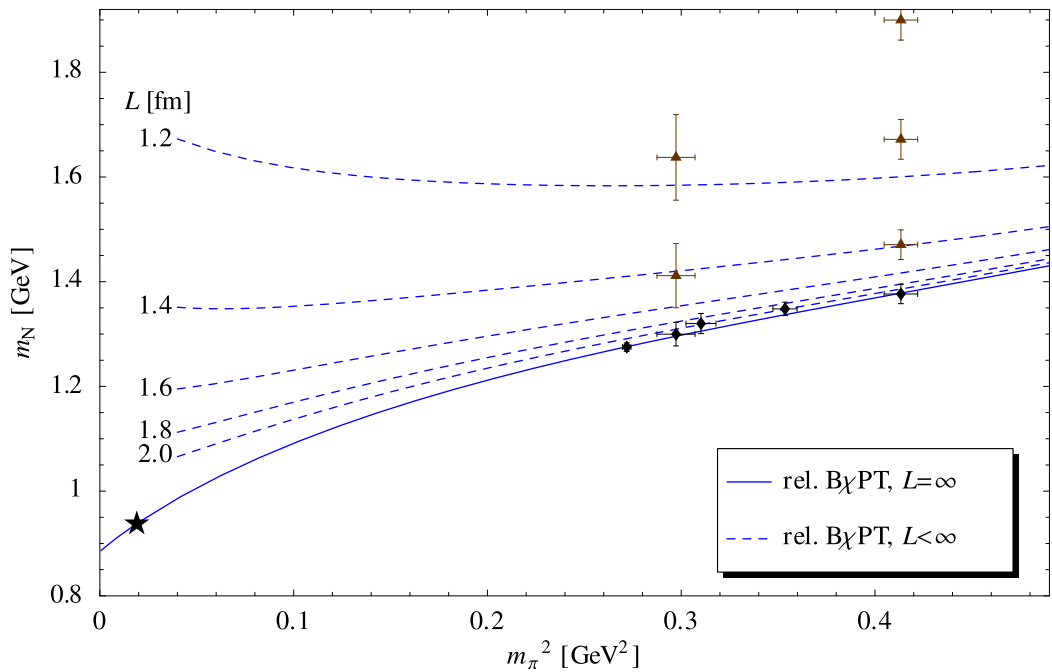} \\
	\caption{$m_N(m_\pi,L)$ evaluated at several box sizes $L$,
	with parameters from the fit to finite volume lattice data}
	\label{fig-lrangefv}
\end{figure}

\chapter{Chiral Extrapolation of Meson Mass Lattice Data}

\label{sec-mesonfit}

\newcommand{\ol}[1]{\overline{#1}}
\newcommand{\tf}[1]{\underline{#1}}
\newcommand{\tfol}[1]{\tf{\ol{#1}}}
\newcommand{\ren}[1]{#1^r}
\newcommand{\phys}{\text{phys}}

As long as excitations of multiplet states involving valence strange quarks are not in resonance during the process under study, we have the choice to pick either a two-flavor SU(2) or three-flavor SU(3) framework for our calculations. 
The two frameworks are related to each other via \terminol{matching conditions}: In going from SU(3) to SU(2), we find that couplings and masses in SU(2) contain contributions from the strangeness degrees of freedom. 
The MILC calculations \cite{Au04,Ber01} offer data on meson and baryon masses in fully dynamical three flavor lattice QCD. The quark masses employed are again isospin symmetric, $m_u = m_d\equiv \hat m$, and the strange quark mass $a m_s$ is placed at a value that reproduces roughly the physical strange quark mass $m_s^\text{phys}$. 

 
For MILC data, performed with staggered Fermions, the lattice renormalized quark masses directly enter as simulation parameters. This way, they are known to us precisely. One can employ them as input in a simultaneous fit to pion and kaon quark mass expansions from $\chi$PT.

The work done in ref. \cite{AuMeson} are meson and quark mass fits of great complexity. The authors apply \terminol{Staggered Chiral Perturbation Theory} to describe accurately the spurious states in the mass spectrum resulting from staggered Fermions. They address discretization errors and finite volume effects, and work at chiral order $p^6$. As a drawback, they have a large number of fit parameters. It is therefore interesting to compare to a relatively simple fit. Note however, that our analysis suffers from possibly large systematic errors in its input data.  

\section{Matching Conditions}
\label{sec-mesonmatching}

We employ the GMOR pion mass $\overline{m}_\pi = 2 B \hat m$ to parametrize the light quark mass $\hat m$. Therefore, it is vital to understand the link between SU(3) and SU(2) in the meson sector, before we address baryon masses.

With the renormalization invariant quantity $\hat m_K\equiv B m_s$, we write the SU(3) GMOR relations in the following form:

\begin{equation} 
	\renewcommand{\arraystretch}{2}
	\begin{array}{lclclcl}
	\ol{m}_\pi^2 & = & 2 B \hat m & & & &\\
	\ol{m}_K^2 & = & B \hat m + B m_s & = & \displaystyle \frac{1}{2} \ol{m}_\pi^2 + \hat{m}_K^2 & & \\
	\ol{m}_\eta^2 & = & \displaystyle \frac{2}{3} B \hat{m} + \frac{4}{3} B m_s 
		& = & \displaystyle  \frac{1}{3} \ol{m}_\pi^2 + \frac{4}{3} \hat{m}_K^2
		& = & \displaystyle - \frac{1}{3} \ol{m}_\pi^2 + \frac{4}{3} \ol{m}_K^2  
	\end{array}
	\renewcommand{\arraystretch}{1.0}
	\label{eq-GMOR3}
\end{equation}

With the definitions
\begin{equation}
	\ren{J}_3(\lambda) \equiv 2 \ren{L}_8(\lambda) - \ren{L}_5(\lambda)\ ,\qquad
	\ren{J}_4(\lambda) \equiv 2 \ren{L}_6(\lambda) - \ren{L}_4(\lambda)
	\end{equation}
\begin{equation}
	\mu_P \equiv \frac{\ol{m}_P^2}{2(4 \pi f_\pi^0)^2} \ln \frac{\ol{m}_P^2}{\lambda^2}
	\qquad \text{for any }P
	\end{equation}
the SU(3) pion and kaon mass to order $p^4$ read \cite{GL85}
\begin{align}
	m_\pi^2 & = \ol{m}_\pi^2 \left[1 - \frac{1}{3} \mu_\eta +  \frac{16 (\ol{m}_\pi^2 + \hat{m}_K^2)}{(f_\pi^0)^2} \ren{J}_4(\lambda) + \frac{8\, \ol{m}_\pi^2}{(f_\pi^0)^2} \ren{J}_3(\lambda) + \mu_\pi \right] \equiv F_1(\overline{m}_\pi^2,\hat m_K^2)
	\label{eq-pionsu3}\\
	m_K^2 & = \left( \frac{1}{2}\ol{m}_\pi^2 + \hat{m}_K^2 \right) \left[ 1 + \frac{2}{3} \mu_\eta  +  \frac{16 (\ol{m}_\pi^2 + \hat{m}_K^2)}{(f_\pi^0)^2} \ren{J}_4(\lambda) + \frac{4\ol{m}_\pi^2+8\hat{m}_K^2}{(f_\pi^0)^2} J_3^r(\lambda)  \right] \nonumber \\
	& \equiv F_2(\overline{m}_\pi^2,\hat m_K^2)
	\label{eq-kaonsu3}
	\end{align}
We should recover the SU(2) pion mass formula if we expand for $\hat m \ll m_s$ or, equivalently, $\ol{m}_\pi^2 \ll \hat{m}_K^2$:
\begin{align}
	m_\pi^2 & = \ol{m}_\pi^2 \left[1 + \frac{\hat{m}_K^2}{(4 \pi f_\pi^0)^2} \left( 16(4 \pi)^2 \ren{J}_4(\lambda) - \frac{2}{9} \ln \frac{ 4 \hat{m}_K^2}{3 \lambda^2}  \right) \right] \nonumber \\
	& + \frac{\ol{m}_\pi^4}{(4 \pi f_\pi^0)^2} 
	\left[ 8 (4 \pi )^2 \ren{J}_3(\lambda) + 16 (4 \pi )^2 \ren{J}_4(\lambda) - \frac{1}{18} \left( 1 + \ln \frac{ 4 \hat{m}_K^2}{3 \lambda^2} \right)
	+ \frac{1}{2} \ln \frac{\ol{m}_\pi^2}{\lambda^2} \right] \nonumber \\ & + \mathcal{O}({m_{\pi,K}}^6)
	\label{eq-mpimatchingform}
	\end{align}
 
Let us underline all quantities in the SU(2) framework. They can effectively absorb contributions from strangeness. The SU(2) GMOR relation simply reads $\tfol{m}_\pi^2 = 2\, \tf{B}\, \tf{\hat{m}}$. Chiral perturbation theory does not give us matching conditions for $\tf{B}$ and $\tf{\hat{m}}$ individually, since the quark masses $\tf{\hat{m}}$ are renormalized within a framework outside the scope of $\chi$PT. At this point, we cling to the concept of ref. \cite{GL85}, \emph{choosing} $\tf{\hat{m}} = \hat{m}$. \emph{We could not do this, if we ever wanted to compare to two-flavor lattice quark mass data, because then SU(2) lattice regularization would impose the quark mass concept to be used for $\tf{\hat{m}}$.} Comparing the term proportional to $\ol{m}_\pi^2$ in eq. (\ref{eq-mpimatchingform}) to the right hand side of GMOR, one finds \cite{GL85}
\begin{equation}
	\tfol{m}_\pi^2 = 2\,\tf{B}\,\hat{m} = \frac{\tf{B}}{B}\, \ol{m}_\pi^2\ ,
	\qquad
	\frac{\tf{B}}{B} = 1 + \frac{\hat{m}_K^2}{(f_\pi^0)^2} \left( 16 \ren{J}_4(\lambda) - \frac{2}{9 (4 \pi )^2} \ln \frac{ 4 \hat{m}_K^2}{3 \lambda^2} \right)
	\end{equation}
In the following, the correction $\tf{f}_\pi^0 = f_\pi^0 + \mathcal{O}(\hat m_K^2)$ can be ignored at our working order.
Comparing eq. (\ref{eq-mpimatchingform}) to the SU(2) pion mass expansion eq. (\ref{eq-pionmasssu2}), one reads off \cite{GL85}
\begin{equation}
	\tf{\elll}_3^r(\lambda) = 4 \ren{J}_3(\lambda) + 8 \ren{J}_4(\lambda) - \frac{1}{36 (4 \pi)^2 } \left( 1 + \ln \frac{ 4 \hat{m}_K^2}{3 \lambda^2} \right) + \mathcal{O}({m_{\pi,K}}^2)
	\label{eq-l3fromsu3}
	\end{equation}
	
\section{Fit Procedure}

The renormalization scheme changes for each set of lattice simulation parameters, mainly depending on the lattice spacing $a$. In general, quark mass parameters are not mutually comparable for the different simulations. Thanks to the farsighted choice of simulation parameters, a large number of MILC calculations has been performed at the same lattice spacing. Here, we shall pick a data set from lattice results for $a \approx 0.12\units{fm}$; the quark masses therein belong to the same renormalization scheme. For our low precision goal, the tiny residual scheme dependence on the sea quark masses is negligible, compare ref. \cite{AuMeson}. 

As input we take $a/r_1$, $a \hat m$, $a m_s$, $a m_\pi$, and $a m_K$ for 5 data points with pion masses from $0.257\units{GeV}$ to $0.610\units{GeV}$, see table \ref{tab-latticedatmeson}. Note that the data available to us are those of the \emph{lightest} taste of each particle\footnote{Many thanks to Dr. Claude Bernard for bringing this to my attention, and for giving advice on conversion to physical units.}, as tabularized in refs. \cite{Au04,Ber01}. The strange quark mass on the lattice has been kept fixed at $a m_s=0.05$ in our data set.
The conversion of a mass $m_X$ to physical units is done according to 
\begin{equation} 
	m_X = \frac{(a m_X)}{(a/r_1) \; r_1} \hbar c 
	\end{equation}
where we take $r_1 = 0.324(4)\units{fm}$, the value specified in ref. \cite{Au04} for the lattices with $a \approx 0.12\units{fm}$. The error $\Delta r_1 = 0.004\units{fm}$ in $r_1$ is systematic, i.e. it does not fluctuate for each lattice point. We do not include it in our formulation of $\chi^2$. Instead, we treat it together with the uncertainty in $f_\pi^0$ and carry out the analysis for several values within the error bounds. Ref. \cite{Au04} offers ``smoothed'' values for $r_1/a$. We prefer to use the unsmoothed values, because any statistical fluctuations should be accounted for in our fit to chiral perturbation theory.

The fitted parameters are $B$, $J_3^r(\lambda)$ and $J_4^r(\lambda)$. Resubstituting again $\ol{m}_\pi \rightarrow 2B\hat{m}$ and $\hat{m}_K \rightarrow B m_s$, we have two fit constraints for each lattice point from eq. (\ref{eq-pionsu3}) and (\ref{eq-kaonsu3}). For each lattice point, the constraints take the following form (in symbolical notation):

\setlength{\savearraycolsep}{\arraycolsep}
\setlength{\arraycolsep}{0pt}
\renewcommand{\arraystretch}{1.5}
\begin{equation} \begin{array}{lrlcllr}
	\text{``}\quad & f_{1}( a \hat m, a  m_s,  \overline{a/r_1} + & \delta(a/r_1) ; B, J_3^r, J_4^r) &\quad \approx \quad &\overline{a m_\pi} &\ \pm\  \Delta(a m_\pi) & \quad \text{''} \\
	\text{``} & f_{2}( a \hat m, a  m_s,  \overline{a/r_1} + & \delta(a/r_1) ; B, J_3^r, J_4^r) & \approx &\overline{a m_K} &\ \pm\  \Delta(a m_K) & \text{''} \\
	\text{``} & & \delta(a/r_1) & \approx & 0 &\ \pm\  \Delta(a/r_1) & \text{''}
	\end{array}
	\end{equation}
\setlength{\arraycolsep}{\savearraycolsep}
\renewcommand{\arraystretch}{1.0}

Here $\overline{a m_\pi}$, $\overline{a/r_1}$ and $\overline{a m_K}$ denote the central values of the lattice calculation results, and the quantities preceded by $\Delta$ are one-standard-deviation errors.
The auxiliary parameter $\delta(a/r_1)$ can be eliminated using the formalism of section \ref{sec-elimlinconst}, assuming $f_1$ and $f_2$ to be approximately linear in $\delta(a/r_1)$.

\section{Fit Results}

\begin{figure}[h]
	\centering
	\includegraphics[width=\textwidth]{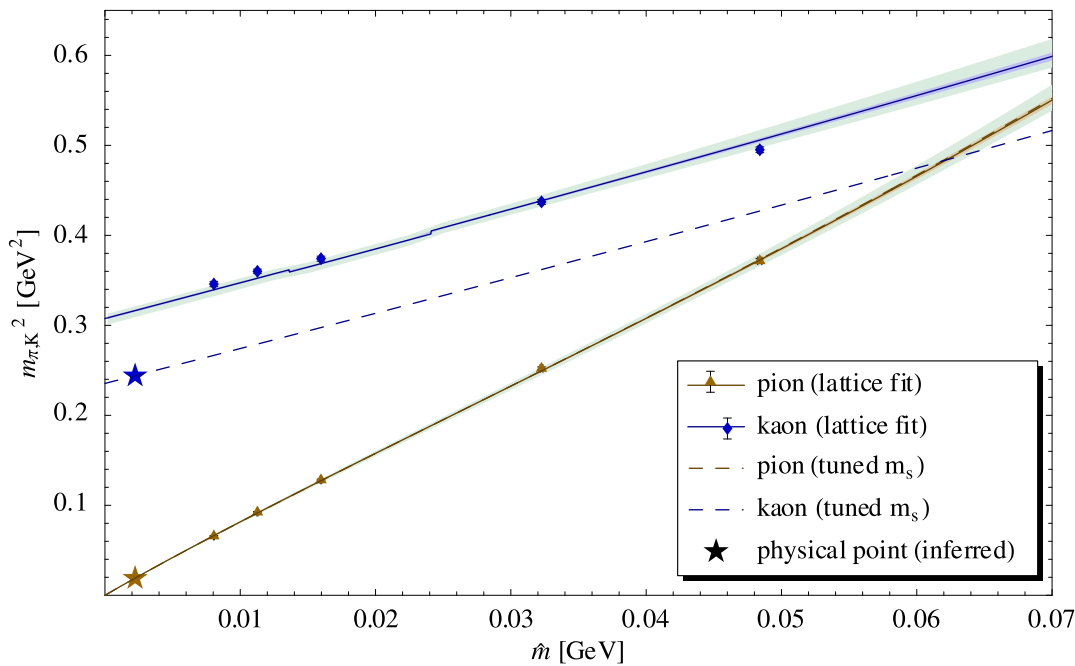} \\
	\caption{Simultaneous fit to MILC data for pion and kaon masses $m_\pi$,$m_K$ versus light lattice quark mass $\hat m$. The inner bands (here barely visible) represent the statistical error for $f_\pi^0 = f_\pi^\text{phys}$, the green outer bands also take into account uncertainties about $f_\pi^0$ and the lattice scale $r_1$.}
	\label{fig-mesonfit}
\end{figure}

\begin{figure}[h]
	\centering
	\includegraphics[width=\textwidth]{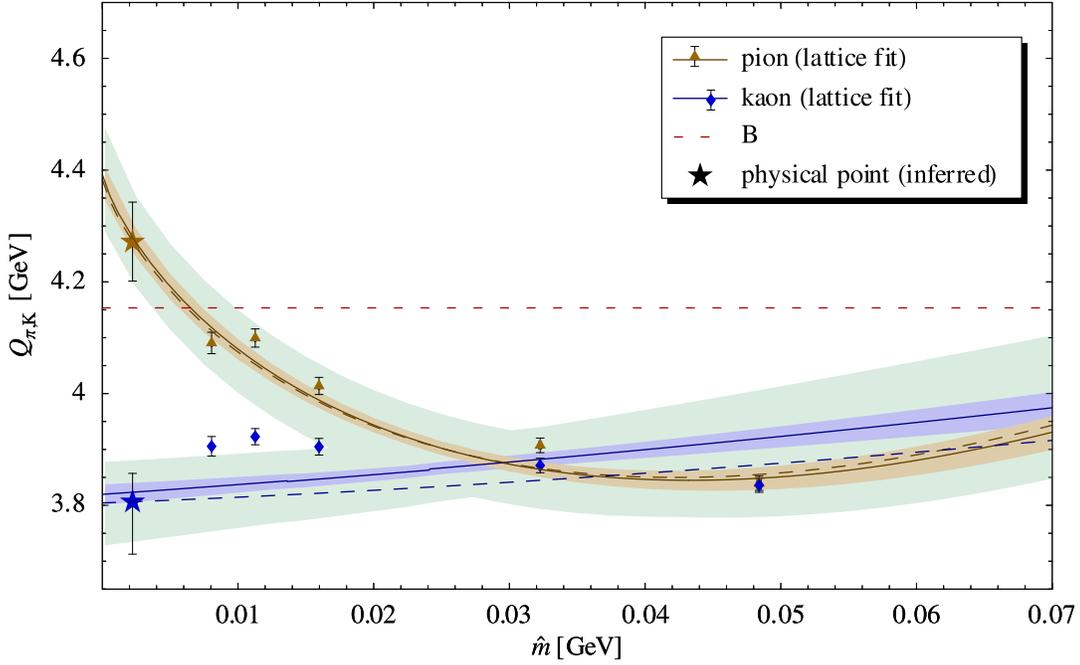} \\
	\caption{Simultaneous fit to MILC data for pion and kaon masses $m_\pi$ versus light lattice quark mass $\hat m$. Plotting the quantities $Q_\pi$, $Q_K$ permits better judgement of the quality of the fit. The inner bands represent the statistical error for $f_\pi^0 = f_\pi^\text{phys}$, the green outer bands also take into account uncertainties about $f_\pi^0$ and the lattice scale $r_1$.}
	\label{fig-mesonfitrel}
\end{figure}

The solid lines and the error bands in figs. \ref{fig-mesonfit} and \ref{fig-mesonfitrel} show the input data points and our fit curves expressed as functions of $\hat m$ and $m_s$. The steps in the fit curves, visible in particular in the kaon curve, are due to the fact that we take $m_s$ from the closest lattice point for the plot. Even though $a m_s = 0.05$ is constant, $m_s$ varies slightly with $a/r_1$. The location of the lattice points has been converted to physical units using the central value of $r_1$. The inner error bands correspond to statistical uncertainty for $f_\pi^0=f_\pi^\text{phys}$ and $r_1$ at its central value. The light green outer error bands arise from systematic errors, namely from performing the analysis with 5 different values of $f_\pi^0$ in the range (\ref{eq-gafpiinterv}), and with $r_1$ taking on its minimal and maximal values on the error bounds.  In fig. \ref{fig-mesonfit}, we plot $m_\pi^2$ and $m_K^2$ versus $\hat m$. The GMOR relations, which predict a linear rise of $m_\pi^2$ and $m_K^2$, are astonishingly well fulfilled. One can hardly notice the slight deviation of our fit curves from linearity. To make these deviations visible, we plot in fig.  \ref{fig-mesonfitrel} the quantities
\begin{equation}
	Q_\pi \equiv \frac{m_\pi^2}{2 \hat m}\ , \qquad
	Q_K  \equiv \frac{m_K^2}{\hat m + m_s}
	\end{equation}

\begin{table}[h]
	\caption{Parameter values and calculated quantities for the meson fit to MILC data (errors not reliable!)}
	\label{tab-paramsmeson}
	\centering
	\vspace{10pt}
\begin{tabular}{ll||cc|cc}
	& & \multicolumn{2}{l|}{(a) statistical error} & \multicolumn{2}{l}{(b) systematic envelope} \\
	\hline
	$B$ & $(\mathrm{GeV})$                      &	4.153 $\pm$ 0.026 & fitted & 4.04 .. 4.23 & fitted \\
	$J_3(\lambda=1\units{GeV}) \times 10^{3}$ & &	-0.0136 $\pm$ 0.0043 & fitted & -0.028 .. 0.019 & fitted \\
	$J_4(\lambda=1\units{GeV}) \times 10^{3}$ & &	0.021 $\pm$ 0.006 & fitted & 0.011 .. 0.039 & fitted \\
	$f_\pi^0$ & $(\mathrm{GeV})$                &	0.0924 & fixed & 0.0862 .. 0.0924 & scanned \\
	$r_1$ & $(\mathrm{fm})$                     &	0.324 & fixed & 0.3200 .. 0.3280 & scanned \\
	\hline                                        
	$\chi^2$/d.o.f &                            &	82 & & 75 .. 136 & \\
	\hline                                        
	$\overline{\elll}_3$ &                      &	3.882 $\pm$ 0.032 & & 3.75 .. 3.92 & \\
	$\ul{B}/B$ &                                &	1.0553 $\pm$ 0.0025 & & 1.050 .. 1.074 & \\
	$\hat m^\text{phys}$ & $(\mathrm{MeV})$     &	2.229 $\pm$ 0.009 & & 2.192 .. 2.267 & \\
	$m_s^\text{phys}$ & $(\mathrm{MeV})$        &	61.88 $\pm$ 0.19 & & 61.0 .. 63.5 & \\
	$\hat m_K^\text{phys}$ & $(\mathrm{GeV})$   &	0.5070 $\pm$ 0.0009 & & 0.5059 .. 0.5086 & \\
	\end{tabular} \par 
\end{table}

For our fit, we obtain a $\chi^2$/d.o.f. of $82$, which corresponds to an $8 \sigma$ discrepancy. Obviously, the input uncertainties are underestimated. This is no surprise. We have neglected significant artefacts from the finite simulation volume, from discretization, and certainly we need to distrust our perturbative expansion of order $p^4$ at larger quark masses. As a result, the statistical and systematic errors and error bands displayed are definitely not realistic. Nevertheless, they reflect the sensitivity of the fit results to the data. When excluding, for example, the point of lowest $\hat m$, or when modifying the weights of the points, we see some slight changes in the fit curves, visible especially in fig. \ref{fig-mesonfitrel}. However, some general features remain invariant, in particular the long arc describing the pion mass deviation from GMOR in figs. \ref{fig-mesonfitrel} and \ref{fig-mesonfitrpi}.

\section{Extrapolating to Physical Conditions}
	
Can we locate the physical point \cite{PDBook}
\begin{equation}
	m_\pi^\text{phys} = 0.138 \units{GeV}\ , \qquad
	m_K^\text{phys} = 0.494 \units{GeV}
	\end{equation}
in our plots?
We know that the value $a m_s = 0.05$ is only a guess, placing the strange quark mass somewhere in the vicinity of its physical value. Once we have extracted $B$, $J_3$ and $J_4$, it is easy to extrapolate to the true strange quark mass. To that end, we need to invert equations (\ref{eq-pionsu3}) and (\ref{eq-kaonsu3}). An approximation at working order can be formed by substituting $\hat m_K^2 \rightarrow m_K^2 - m_\pi^2 /2 $ and $\overline{m}_\pi^2 \rightarrow m_\pi^2$ in the square brackets and loop terms of eqs. (\ref{eq-pionsu3}) and (\ref{eq-kaonsu3}). Solving for $\hat m_K^2$ and $\overline{m}_\pi$, we obtain 
\begin{align}
	\overline{m}_\pi^2 & = m_\pi^2 + \frac{m_\pi^2}{(4 \pi f_\pi^0)^2} \Bigg[
	- 8 (4 \pi)^2 (  2 J_4^r(\lambda) m_K^2 + J_4^r(\lambda) m_\pi^2  + J_3^r(\lambda) m_\pi^2) \nonumber \\
	& + \left( \frac{2}{9} m_K^2 - \frac{1}{18} m_\pi^2 \right) \ln \frac{ 4 m_K^2 - m_\pi^2 }{3 \lambda^2} - \frac{m_\pi^2}{2} \ln \frac{m_\pi^2}{\lambda^2} \Bigg] = 2 B \hat m \equiv G_1(m_\pi^2,m_K^2) 
	\label{eq-mbarpisu3}\\
	\hat m_K^2 & =\frac{2 m_K^2 - m_\pi^2}{2} + \frac{1}{(4 \pi f_\pi^0)^2} \Bigg[
	 4 (4 \pi)^2 ( J_3^r(\lambda) m_\pi^2 + J_4^r(\lambda) m_\pi^2 - 2 J_3^r(\lambda) m_K^4 - 4 J_4^r(\lambda) m_K^4 ) \nonumber \\
	& +  \frac{m_\pi^2 - 16 m_K^2}{36} \ln \frac{ 4 m_K^2 - m_\pi^2 }{3 \lambda^2} + \frac{m_\pi^2}{4} \ln \frac{m_\pi^2}{\lambda^2} \Bigg] = B m_s \equiv G_2(m_\pi^2,m_K^2)
	\label{eq-mhatksu3}
	\end{align}
We can assess the quality of the approximation by resubstituting the results into the original formulae. 
In particular, evaluating $F_2(G_1(m_\pi^2,m_K^2),G_2(m_\pi^2,m_K^2))$ at $m_\pi^\phys$ and $m_K^\phys$, we should obtain $(m_K^\phys)^2$ again, but in fact, the resulting kaon mass is $1.5\units{MeV}$ smaller than the input. This error is also an indication of the accuracy of the perturbative expansion. We would like to have a more accurate inverse, because otherwise visible spurious offsets between extrapolated quantities would appear in the plots. We define residuals
\begin{align}
	R_1(m_\pi^2,m_K^2) & \equiv F_1(G_1(m_\pi^2,m_K^2),G_2(m_\pi^2,m_K^2)) - m_\pi^2 \nonumber \\
	R_2(m_\pi^2,m_K^2) & \equiv F_2(G_1(m_\pi^2,m_K^2),G_2(m_\pi^2,m_K^2)) - m_K^2 
	\end{align}
The iteratively corrected expressions
\begin{equation}
	\tilde G_i(m_\pi^2,m_K^2) \equiv G_i(m_\pi^2-R_1(m_\pi^2,m_K^2),m_K^2-R_2(m_\pi^2,m_K^2))
\end{equation}
perform better by three orders of magnitude for the kaon. They still constitute differentiable \texttt{Mathematica} expressions of the fit parameters, which can be processed by our error analysis tools.\footnote{Numerical evaluations at intermediate steps with simple error propagation are forbidden, because they lead to ``double counting'' of errors.} From $G_i(\;(m_\pi^\phys)^2,\;(m_K^\phys)^2\;)$ we calculate $\hat m^\phys$ and $m_s^\phys$ as shown in table \ref{tab-paramsmeson}. Note that they are lattice regularized. Now we also know where to place the markers of the physical point in our plots. Finally, we are able to plot as dashed lines the light quark mass dependence of pion and kaon mass for $m_s^\phys$.

\section{Making the Connection to SU(2)}

\begin{figure}[h]
	\centering
	\includegraphics[width=\textwidth]{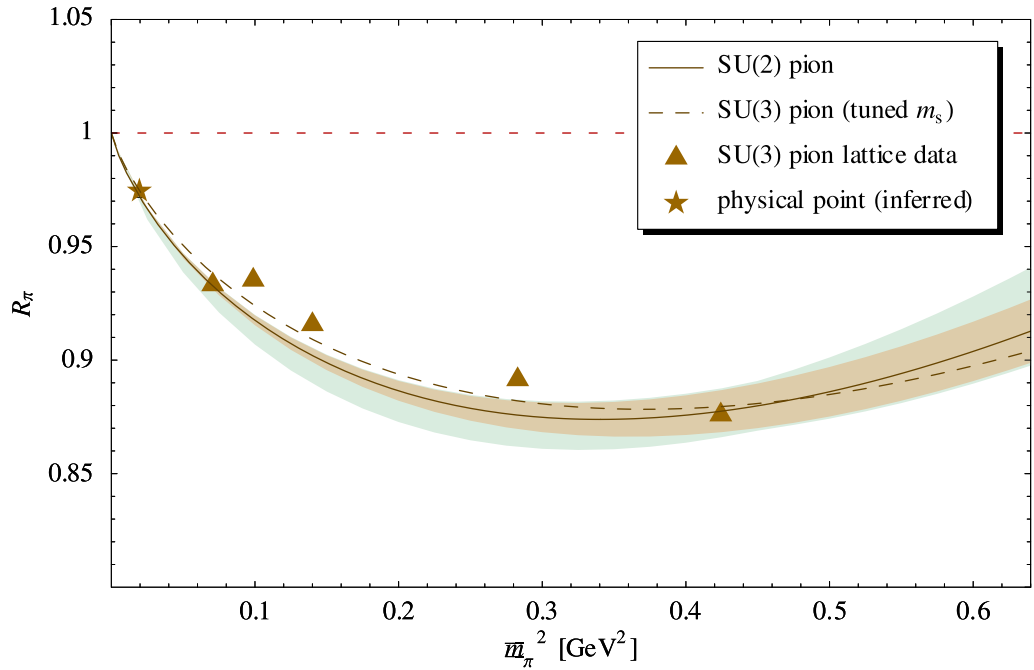} \\
	\caption{The SU(2) GMOR correction factor $R_\pi$ as inferred from the simultaneous SU(3) meson fit. The solid line and error bands are plotted using the SU(2) formula (\ref{eq-pionmasssu2}). The inner band represents the statistical error for $f_\pi^0 = f_\pi^\text{phys}$, the green outer bands also take into account uncertainties about $f_\pi^0$ and the lattice scale $r_1$. The dashed curve represents the SU(3) prediction using the tuned strange quark mass. (For the pion, the effect of this tuning is tiny.) }
	\label{fig-mesonfitrpi}
\end{figure}

The fit results can be transferred to the SU(2) framework using the matching conditions of section \ref{sec-mesonmatching}. The value of central interest is $\overline{\elll}_3$, see section \ref{sec-highordpi}. Even though the errors in table \ref{tab-paramsmeson} are definitely underestimated, a value of 
\begin{equation}
	\boxed{\overline{\elll}_3 \approx 4}
	\end{equation}
seems to be a stable result. This is in good agreement with the value $2.9 \pm 2.4$ from ref. \cite{GL84}. In fig. \ref{fig-mesonfitrpi}, we plot the SU(2) translation of our fit result in the same way as in fig. \ref{fig-lueschpl}. Mind that the location of the lattice data shown in this plot depend on the results of our fit. The curve indicates that in the range of available lattice data, $m_\pi^2 \approx 0.9\; \ol{\ul{m}}_\pi^2$, i.e. the SU(2) Gell-Mann - Oakes - Renner relations need roughly a $10\%$ correction.

\section{Comparing to Other Works}

\begin{figure}[h]
	\centering
	\includegraphics[width=.5\textwidth]{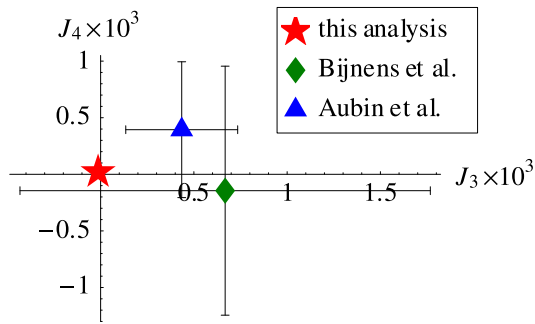} \\
	\caption{The values of $J_3^r$ and $J_4^r$ at the renormalization scale $\lambda=1\units{GeV}$ as calculated in the study at hand (without error bar), by Bijnens et al. \cite{Bij95} and Aubin et al. \cite{AuMeson}.}
	\label{fig-comparej3j4}
\end{figure}

Converting all values to our renormalization scale $\lambda=1\units{GeV}$ using
\begin{align}
	J_3^r(\lambda) & = J_3^r(\lambda_0) - \frac{1}{6 (4 \pi)^2} \ln \frac{\lambda_0}{\lambda} &
	J_4^r(\lambda) & = J_4^r(\lambda_0) + \frac{1}{36 (4 \pi)^2} \ln \frac{\lambda_0}{\lambda}
	\label{eq-runj3j4}
	\end{align}
we take values for the $L_i^r$ from \cite{Bij95}. Adding errors linearly, one obtains
$J_3^r = (0.7 \pm 1.1) \times 10^3$, $J_4^r = (-0.1 \pm 1.1) \times 10^3$. The results presented in \cite{AuMeson} are $J_3^r = (0.43 \pm 0.10 \pm 0.20) \times 10^3$ and $J_4^r = (0.39 \pm 0.20 \pm 0.40) \times 10^3$. Using the matching condition (\ref{eq-l3fromsu3}) with $\hat m_K$ from eq. (\ref{eq-mhatksu3}), we find that this corresponds to $\overline{\elll}_3 = 0.8 \pm 1.1 \pm 2.6$. As illustrated in fig. \ref{fig-comparej3j4}, these values are in good agreement with our results.\footnote{Our fit results are so close to the origin only for our particular choice of the renormalization scale $\lambda$. We checked that performing our study with $\lambda=2\units{GeV}$ and applying eq. (\ref{eq-runj3j4}) to convert back to $\lambda=1\units{GeV}$ produces exactly the same results. } We refrained from drawing an error bar for our result. A more detailed analysis would be needed to obtain a fair error estimate. We also find agreement with results for $L_5^r$ and $L_8^r$ from ref. \cite{ABT01}, which yield $J_3^r = (-0.58 \pm 0.7) \times 10^3$. 
	
\chapter{Conclusion}

\section{Summary}

In this work, we have thoroughly investigated the feasibility, reliability and perspectives of lattice extrapolations. Our main focus has rested on the nucleon mass. The following conclusions can be drawn:
\begin{itemize}
\item
As long as we fix a subset of parameters and constrain the fit curve to run through the physical point, we obtain a stable interpolant in the whole assumed range of validity of Chiral Perturbation Theory. The error band remains narrow from a pion mass of $0.6\units{GeV}$ down to the chiral limit. 
\item
The effect of uncertainties about the chiral limit values $g_A^0$ and $f_\pi^0$ is marginal for the interpolation curve, but appreciable for the extracted low energy constants.
\item
We have confidence that higher order effects are sufficiently suppressed. Unfortunately we cannot pursue a very stringent analysis at this point. This is due to the fact that the set of fourth order couplings subsumed in $e_1$ differs in the calculation to order $p^3$ and $p^4$.
\item
More indications for good convergence come from a numerical analysis replacing $g_A^0$ by the pion mass dependent $g_A$.
\item
With more lattice data becoming available over time, ab-initio extrapolations of lattice data come into reach. However, our nucleon mass formula alone will not be sufficient to make extrapolations purely from lattice data. The problem is inherently badly conditioned. Hope lies in simultaneous fits to a series of observables.
\item
Using $\chi$PT formulae of order $p^3$ for pion nucleon scattering, we observe discrepancies with empirical data. The problem has been identified as an effect of the delta resonance. \textbf{Future Chiral Perturbation Theory calculations with baryons should take the delta resonance into account explicitly}. We have presented a well-motivated workaround that can restore integrity in our particular scenario.
\item
Using a correction formula, we have been able to include lattice data for finite volume side lengths down to $L=1\units{fm}$. We have found that the gain in extractable information is substantial, because the finite volume dependence provides an entirely new kind of constraint.
\end{itemize}

In a second step, we have begun analyzing three-flavor lattice data. Chiral extrapolations rely on the pion mass to parametrize the quark mass. Using SU(3) pion and kaon mass formulae at chiral order $p^4$ simultaneously, we have fitted lattice data from MILC. Matching conditions have enabled us to infer SU(2) results.
\begin{itemize}
\item
Despite negligence of some sources of systematic errors, the fit formulae can be brought in satisfactory agreement with the lattice data.
\item
Effects of systematic discrepancies are reflected in a large $\chi^2$. The calculated errors are therefore underestimated.
\item
The extracted low energy constants are compatible with results from other groups.
\item
When performing simultaneous fits with nucleon mass data from SU(3) and SU(2) calculations, a relative factor $\ul{B}/B$ will have to be taken into account matching the quark mass parametrizations via the SU(3) and SU(2) pion masses. According to our fit, this amounts to a difference of about $6\%$. 
\item
We determine the SU(2) coupling constant $\overline{\elll}_3 \approx 3.9$.
\item
Present day lattice extrapolations ignoring corrections to the Gell-Mann~- Oakes - Renner relation underestimate the quark masses by roughly $10\%$.
\end{itemize}

\section{Outlook}

Even today, we still have rather incomplete knowledge of the
low-energy constants of Chiral Perturbation Theory as extracted from phenomenology.
The broad range of values for the low energy constants found in the literature stems
from many isolated studies on small sets of observables.
Yet effective field theories, such as Chiral Perturbation Theory, display their power in the 
description of many different phenomena with the same set of parameters.
Our success with the finite volume fit demonstrates the advantage
of a global analysis. The great task for the future will be a \textbf{simultaneous fit},
combining chiral formulae for many different observables in a single
$\chi^2$, applied to a broad range of statistically independent data from
experiment {\it and lattice}. The combined fit has the potential to 
produce precision results and reliable error information of a new quality.

\appendix

\chapter{Comparison to Other Frameworks}
\label{sec-otherframew}
 
The statistical error band allows us to compare our results to calculations performed within other theoretical frameworks.  

\begin{figure}[h]
 	\centering
 	\includegraphics{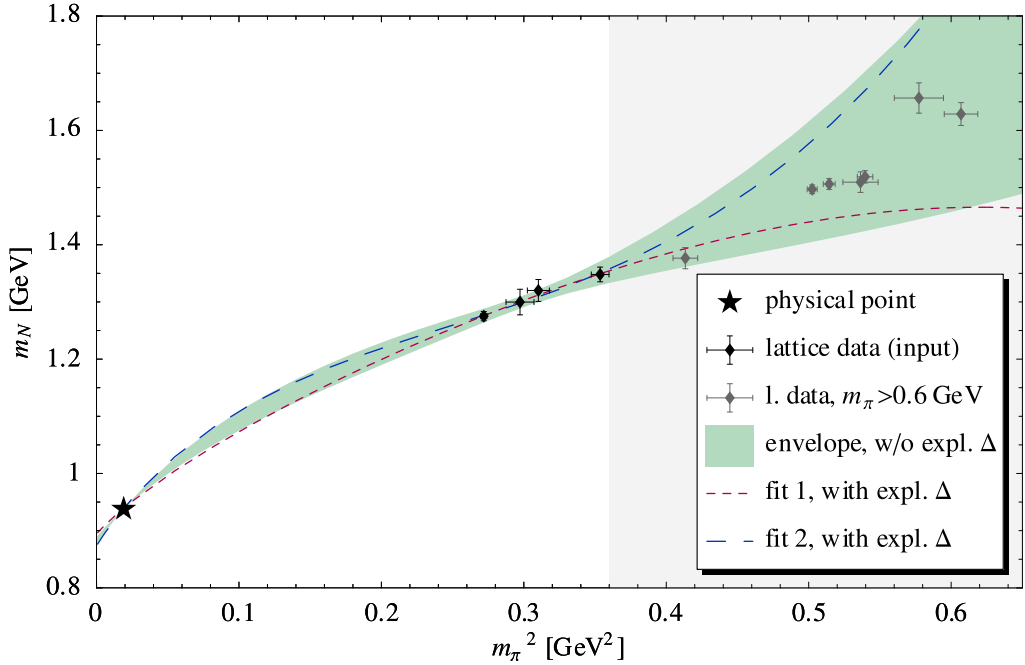}
 	\caption{Best fit curves based on formulae at order $\epsilon^3$ in manifestly covariant SSE, with $c_A=1.5$, $g_A^0=1.267$, $f_\pi^0=0.0924\units{GeV}$ and $\Delta=0.2711\units{GeV}$ as fixed input parameters \cite{PMHWW}. The red fine-dashed curve (fit~1) implements truncation at $\mathcal{O}(1/M_0)$. The blue long-dashed curve (fit~2) has been performed without truncation and with the decoupling of $\Delta\,(1232)$ only partially fulfilled. For comparison, we plot again the systematic envelope of figure \ref{fig-band}.}
 	\label{fig-deltaplot}
\end{figure}

We have seen that the influence of the delta resonance plays an important role in baryon chiral perturbation theory. 
Massimiliano Procura has calculated and applied nucleon mass formulae calculated within manifestly covariant SSE \cite{PMHWW}. 
Fig. \ref{fig-deltaplot} offers a glimpse at the results, and a comparison to our systematic envelope.
 
\begin{figure}[h]
 	\centering
 	\includegraphics{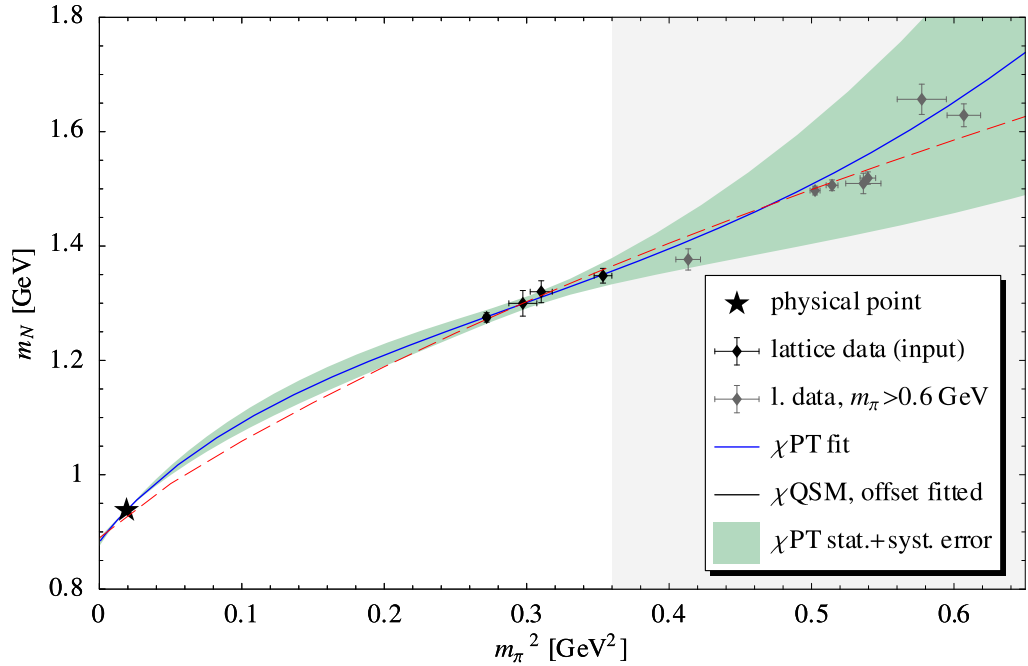}
 	\caption{$\chi$QSM result for $m_N(m_\pi)$ as obtained by P. Schweitzer et al. (red dashed line) \cite{Schweitzer,Goe05}. A subtraction of $0.321\units{GeV}$ is needed in order to account for the systematic overestimation of the nucleon mass in the soliton approach. For comparison, we plot our $\chi$PT best-fit curve and the systematic error band of fig. \ref{fig-band}. 
	}
 	\label{fig-cqsm}
\end{figure}

A very different approach to calculate the nucleon mass is the \terminol{Chiral Quark Soliton Model} (\terminol{$\chi$QSM}) \cite{Goe05,Diak97}. In contrast to $\chi$PT, this model should perform best at \emph{large} pion masses. Peter Schweitzer of the Ruhr-Universit{\"a}t Bochum has generously provided us with data material $m^{\chi\text{QSM}}_N(m_\pi)$ from a recent calculation of the nucleon mass within this model \cite{Schweitzer}. The results suffer from an offset, i.e. $m_N(m_\pi) = m^{\chi\text{QSM}}_N(m_\pi) \nobreak+ C$. From a simple fit to the four large volume lattice data points below $m_\pi=0.6\units{GeV}$, we have determined the offset to be $C=-0.321\units{GeV}$. The resulting curve, plotted next to our results, can be seen in fig. \ref{fig-cqsm}. The $\chi$QSM curve runs rather close to ours. The fact that it comes so close to the physical point gives a very optimistic outlook for the performance of the model.

\newpage	
\addtocounter{chapter}{1}
\addcontentsline{toc}{chapter}{\numberline{\Alph{chapter}}Lattice Data}
\thispagestyle{plain}
\textbf{\Huge Appendix \Alph{chapter} - Lattice Data}

\begin{table}[h!]
	\caption{Selection of two-flavor lattice data taken from \cite{K04} and \cite{OLS05}. }
	\label{tab-latticedat}
	\centering
	\vspace{10pt}
\begin{tabular}{rl|llrr|rr|c}
	no. & collab. & 
	$\beta$ &
	$\kappa$ &
	$a\ [\mathrm{fm}]$  & 
	$L\ [\mathrm{fm}]$ & 
	$m_\pi\ [\mathrm{GeV}]$ & 
	$m_N\ [\mathrm{GeV}]$ &
	large $L$ \\
	\hline \hline 
	19 & CP-PACS & 2.1 & 0.1382 & 0.111 & 2.68 & 0.5214(21) & 1.2751(82) & $\times$ \\
	\hline
	41 & JLQCD & 5.2 & 0.1355 & 0.098 & 1.96 & \ul{0.5453(91)} & 1.300(23) & $\times$ \\
	36 & JLQCD & 5.2 & 0.1355 & 0.099 & 1.58 & 0.560(16) & 1.412(62) &  \\
	31 & JLQCD & 5.2 & 0.1355 & 0.099 & 1.19 & 0.655(32) & 1.637(82) &  \\
	\hline
	8 & QCDSF & 5.25 & 0.13575 & 0.092 & 2.21 & 0.5570(70) & 1.320(20) & $\times$ \\
	\hline
	23 & CP-PACS & 2.2 & 0.1368 & 0.092 & 2.22 & 0.5946(53) & 1.348(13) & $\times$ \\
	\hline
	56 & \cite{OLS05} & 5.6 & 0.1575 & 0.085 & 2.04 & \ul{0.6429(68)} & 1.377(19) & $\times$ \\
	55 & \cite{OLS05} & 5.6 & 0.1575 & 0.084 & 1.34 & 0.660(12) & 1.471(29) &  \\
	54 & \cite{OLS05} & 5.6 & 0.1575 & $\sim$0.085 & 1.19 & 0.709(11) & 1.672(38) &  \\
	53 & \cite{OLS05} & 5.6 & 0.1575 & $\sim$0.085 & 1.02 & 0.832(22) & 1.900(39) &  \\
	\hline \
	\Small{\crossout{47}} & \Small{\cite{OLS05}} & \Small{5.5} & \Small{0.1596} & \Small{\textcolor{red}{\textbf{0.107}}} & \Small{1.71} & \Small{0.6793(71)} & \Small{1.410(18)} & \Small{$\times$, coarse!} \\
	\hline
	20 & CP-PACS & 2.1 & 0.1374 & 0.118 & 2.83 & 0.7088(25) & 1.4971(77) & $\times$ \\
	\hline
	14 & QCDSF & 5.29 & 0.1355 & 0.09 & 2.16 & \ul{0.7172(29)} & 1.5062(95) & $\times$ \\
	13 & QCDSF & 5.29 & 0.1355 & 0.09 & 1.44 & 0.7316(52) & 1.577(19) &  \\
	12 & QCDSF & 5.29 & 0.1355 & 0.087 & 1.04 & 0.826(12) & 1.963(30) &  \\
	\hline
	40 & JLQCD & 5.2 & 0.135 & 0.108 & 2.16 & \ul{0.7324(85)} & 1.509(18) & $\times$ \\
	35 & JLQCD & 5.2 & 0.135 & 0.108 & 1.73 & 0.7300(90) & 1.510(26) &  \\
	30 & JLQCD & 5.2 & 0.135 & 0.112 & 1.34 & 0.750(17) & 1.618(65) &  \\
	\hline
	24 & CP-PACS & 2.2 & 0.1363 & 0.095 & 2.29 & 0.7345(38) & 1.519(11) & $\times$ \\
	\hline
	51 & \cite{OLS05} & 5.6 & 0.157 & 0.091 & 1.46 & 0.746(12) & 1.533(29) & $\times$ \\
	\hline
	2 & UKQCD & 5.2 & 0.135 & 0.105 & 1.68 & 0.760(12) & 1.657(27) & $\times$ \\
	\hline
	\Small{\crossout{46}} & \Small{\cite{OLS05}} & \Small{5.5} & \Small{0.159} & \Small{\textcolor{red}{\textbf{0.114}}} & \Small{1.82} & \Small{0.7666(64)} & \Small{1.509(17)} & \Small{$\times$, coarse!} \\
	\hline
	7 & UKQCD & 5.25 & 0.1352 & 0.097 & 1.56 & 0.7791(76) & 1.629(20) & $\times$ \\
	\hline
	50 & \cite{OLS05} & 5.6 & 0.1565 & 0.095 & 1.51 & 0.823(12) & 1.637(27) & $\times$ \\
	\hline
	21 & CP-PACS & 2.1 & 0.1367 & 0.123 & 2.95 & 0.8304(33) & 1.6434(80) & $\times$ \\
	\hline
	39 & JLQCD & 5.2 & 0.1346 & 0.115 & 2.31 & 0.8388(98) & 1.650(20) & $\times$ \\
	\hline
	25 & CP-PACS & 2.2 & 0.1358 & 0.099 & 2.37 & 0.8389(41) & 1.652(12) & $\times$ \\
	\hline
	11 & QCDSF & 5.29 & 0.135 & 0.096 & 1.53 & 0.8681(64) & 1.721(15) & $\times$ \\
	\hline
	\Small{\crossout{45}} & \Small{\cite{OLS05}} & \Small{5.5} & \Small{0.158} & \Small{\textcolor{red}{\textbf{0.124}}} & \Small{1.99} & \Small{0.8795(82)} & \Small{1.631(31)} & \Small{$\times$, coarse!} \\
	\hline
	49 & \cite{OLS05} & 5.6 & 0.156 & 0.098 & 1.57 & 0.9002(70) & 1.719(16) & $\times$ \\
	\hline
	38 & JLQCD & 5.2 & 0.1343 & 0.121 & 2.41 & 0.9037(65) & 1.737(13) & $\times$ \\
	\hline
	6 & QCDSF & 5.25 & 0.1346 & 0.106 & 1.69 & 0.9207(51) & 1.768(14) & $\times$ \\
	\hline
	1 & QCDSF & 5.2 & 0.1342 & 0.123 & 1.96 & 0.9398(74) & 1.781(15) & $\times$ \\
	\hline
	9 & UKQCD & 5.26 & 0.1345 & 0.106 & 1.7 & 0.946(12) & 1.878(28) & $\times$ \\
	\hline
	22 & CP-PACS & 2.1 & 0.1357 & 0.13 & 3.12 & 0.9556(41) & 1.7980(85) & $\times$ \\
	\hline
	37 & JLQCD & 5.2 & 0.134 & 0.127 & 2.53 & 0.9598(74) & 1.801(15) & $\times$ \\
	\hline
	26 & CP-PACS & 2.2 & 0.1351 & 0.102 & 2.44 & 0.9694(45) & 1.809(17) & $\times$ \\
\end{tabular} \par
\end{table}

\begin{table}[htb]
	\caption{Selection of three-flavor meson lattice data taken from \cite{Ber01,Au04}. }
	\label{tab-latticedatmeson}
	\centering
	\vspace{10pt}
\begin{tabular}{r|ccccc|cc}
	no. & 
	$a \hat m$ &
	$a m_s$ &
	$a m_\pi$  & 
	$a m_K$ &
	$a/r1$ &
	$m_\pi\ (\mathrm{GeV})$ & 
	$m_\pi L$ \\
	\hline \hline 
	13 & 0.005 & 0.05 & 0.15938(16) & 0.36523(27) & 0.3782(16) & 0.2567(34) & 3.8 \\
	12 & 0.007 & 0.05 & 0.18881(19) & 0.37268(25) & 0.3783(13) & 0.3040(40) & 3.8 \\
	11 & 0.01 & 0.05 & 0.22421(12) & 0.38304(20) & 0.3814(14) & 0.3580(47) & 6.3 \\
	8 & 0.02 & 0.05 & 0.31125(16) & 0.40984(21) & 0.3775(12) & 0.5022(65) & 6.2 \\
	7 & 0.03 & 0.05 & 0.37787(18) & 0.43613(19) & 0.3775(12) & 0.6096(78) & 7.6 \\
	\end{tabular}
\end{table}

\bibliography{DA}
\bibliographystyle{mybibstyle} 

\chapter*{Danksagung}

Hiermit bedanke ich mich bei allen, die mich bei meiner Diplomarbeit unterst{\"u}tzt haben, insbesondere 
\begin{itemize}
\item
bei Prof. Wolfram Weise, der mich dieses Thema an seinem Lehrstuhl bearbeiten lie{\ss}. Seiner physikalischen Intuition entsprangen wertvolle pragmatische Ratschl{\"a}ge, die ganz wesentlich zum Gelingen dieser Arbeit beigetragen haben.
\item
bei Massimiliano Procura f{\"u}r die hervorragende und freundschaftliche Zusammen\-arbeit. Er stand mir bei Fragen jederzeit zur Verf{\"u}gung und gab mir wichtige Li\-teraturhinweise.
\item
bei Dr. Thomas Hemmert, der mir im Rahmen anregender Diskussionen Einblicke in die Bedenken, Vorschl{\"a}ge und Einsichten eines langj{\"a}hrigen Experten gew{\"a}hrte.
\item
bei Dr. habil. Norbert Kaiser und Dr. habil. Harald Grie{\ss}hammer f{\"u}r weiteren Expertenrat.
\item
bei allen, die mir beim Korrekturlesen geholfen haben: Bernhard M{\"u}ller, Massimiliano Procura, Prof. Wolfram Weise, Dr. Philipp H{\"a}gler, Rainer H{\"a}rtle und Monika Musch.
\item
bei Michael Thaler und Dr. Stefan Fritsch. Ihrer Administrationst{\"a}tigkeit ist es zu verdanken, dass es eine Freude war, mit den Computern des Lehrstuhls zu arbeiten. 
\item 
bei allen Mitarbeitern des Lehrstuhls T39 f{\"u}r das freundliche Arbeitsklima und die willkommenen Abwechslungen wie gelegentliches Frisbeewerfen, Baden oder Grillen.
\item
beim Freistaat Bayern f{\"u}r finanzielle F{\"o}rderung im Rahmen eines Stipendiums f{\"u}r besonders Begabte.
\item
bei meinen Eltern f{\"u}r ihre Unterst{\"u}tzung w{\"a}hrend des gesamten Studiums.
\end{itemize}

\end{document}